\documentclass[superscriptaddress, twocolumn,showpacs,
amssymb,amsmath,nobibnotes,aps,prd,
nofootinbib]{revtex4-1}
\pdfoutput=1
\usepackage{graphicx,subfigure,bm,color,psfrag,hyperref}
\usepackage{amsfonts}
\usepackage{lipsum}
\usepackage{mathtools}
\usepackage{verbatim}
\usepackage[normalem]{ulem}
\usepackage[dvipsnames]{xcolor}
\hypersetup{colorlinks,linkcolor={blue},citecolor={red},urlcolor={violet}}

\begin{document}

\title{Solving an Interacting Quintessence Model with a Sound Horizon Initial Condition and 
its Observational Constraints}

\author{Wenzhong Liu}
\email{lwz\_lwzlwz@163.com}
\affiliation{Department of Physics, Liaoning Normal University, Dalian 116029, China}
\affiliation{Department of Physics, Shenyang University, Shenyang 110096, China}

\author{Yabo Wu}
\email{ybwu61@126.com}
\affiliation{Department of Physics, Liaoning Normal University, Dalian 116029, China}

\author{Supriya Pan}
\email{supriya.maths@presiuniv.ac.in}
\affiliation{Department of Mathematics, Presidency University, 86/1 College Street, Kolkata 700073, India}
\affiliation{Institute of Systems Science, Durban University of Technology, PO Box 1334, Durban 4000, Republic of South Africa}

\author{Weiqiang Yang}
\email{d11102004@163.com}
\affiliation{Department of Physics, Liaoning Normal University, Dalian 116029, China}

\begin{abstract}
\noindent Astronomical observations suggest that the current standard $\Lambda$-Cold Dark Matter model in modern cosmology has some discrepancies when fitting the data during the whole expansion history of the universe. To solve the Hubble constant ($H_0$)  tension, usually an unknown mechanism is considered that shifts the sound horizon at the decoupling era. On the other hand, dynamical dark energy models are also considered to resolve the problems of the cosmological constant, and the additional degrees of freedom require initial conditions for a solution. In this article we have considered a coupled quintessence dark energy model with a special focus on its early-time behaviour. In our solution the initial conditions are naturally decided by setting the value of  the sound horizon at the recombination time, $\theta^*$. We find that during this process, $H_0$ could be derived and its value rises with the coupling strength of the interaction. We also performed the background and cosmic microwave background power spectrum analysis, and find that the existence of the interaction term affects the energy density during a narrow time interval range and shifts the early cosmic microwave background spectrum. We also constrained the parameter space of the underlying scenario using the markov chain monte carlo analysis. We find that the best-fit values of $H_0$ and $S_8$ are improved slightly for the interacting model, but not enough to release the tensions.
\end{abstract}

\maketitle

\section{Introduction}
\label{sec-introduction}

In the realm of modern cosmology, dark matter (DM) and dark energy (DE) are two most challenging problems. Currently, the most concordance model is $\Lambda$-Cold Dark Matter ($\Lambda$CDM) which assumes General Relativity (GR) as the universal theory of gravity. Within this model, the expansion of the universe today is dominated by a positive cosmological constant $\Lambda$ and cold dark matter (CDM). Although $\Lambda$CDM is well tested by an enormous number of astronomical observations, however, it suffers from several difficulties. Theoretically, it brings the fine-tuning problem~\cite{Weinberg:1988cp} and the coincidence problem~\cite{Zlatev:1998tr}. And during these years with the improvement of the observational precision, various discrepancies have emerged, including the  $5\sigma$ tension between the observed~\cite{Riess:2021jrx,Riess:2020fzl} and inferred~\cite{Aghanim:2018eyx} values of the Hubble constant $H_0 \equiv 100 h~{\rm km/s/Mpc}$~\cite{DiValentino:2020zio}, as well as the discrepancy between the cosmological and local determination of $S_8 \equiv \sigma_8 \sqrt{\Omega_m/3}$~\cite{DiValentino:2020vvd} that quantifies the root mean square density fluctuations when smoothed with a top-hat filter of radius $8h^{-1}/{\rm Mpc}~(\equiv \sigma_8)$ as a function of the present day value of the non-relativistic matter density parameter $\Omega_m$~\cite{DiValentino:2020zio,DiValentino:2020vvd}. In particular, the Hubble constant discrepancy is the significant one.
Assuming a flat $\Lambda$CDM model, 
the estimated value of $H_0$ by the cosmic microwave background (CMB) radiation by Planck Collaboration leads to $H_0 = 67.27 \pm 0.60~{\rm km/s/Mpc}$ at 68\% CL~\cite{Aghanim:2018eyx}, whereas the SH0ES collaboration finds a larger value of $H_0 = 73.2\pm  1.3~{\rm  km/s/Mpc}$~\cite{Riess:2020fzl}.  On the other hand, the flat $\Lambda$CDM-based Planck mission leads to $S_8 = 0.834 \pm 0.016$ (Planck TT,TE,EE+low E) \cite{Aghanim:2018eyx} while the weak lensing and galaxy clustering measurements lead different values on $S_8$, e.g.,  $S_8 = 0.775^{+0.026}_{-0.024}$ at 68\% CL from DES-Y3
\cite{DES:2021wwk}, $S_8 = 0.766^{+0.020}
_{-0.014}$ at 68\% CL by KiDS-1000 $\times$ {2dFLenS+BOSS}  \cite{Heymans:2020gsg}, $S_8 = 0.754^{+0.027}
{-0.029}$ at 68\% CL by KiDS-1000 PC$_{\ell}$ \cite{KiDS:2021opn} and $S_8 = 0.771^{+0.006}
{-0.032}$ at 68\% CL (see \cite{Abdalla:2022yfr} for more details).
These discrepancies may not be well explained by systematic errors ~\cite{Verde:2019ivm,Riess:2019qba,DiValentino:2020vnx} and thus many new cosmological models have been proposed to accommodate the data~\cite{DiValentino:2021izs,Perivolaropoulos:2021jda}. The above mentioned discrepancies have become a common test-ground to uncover properties of the dark sector.

In order to release these tensions one may consider replacing the cosmological constant by a dynamical DE. There are many proposals in this direction to release the $H_0$ tension \cite{DiValentino:2021izs,Perivolaropoulos:2021jda,Schoneberg:2021qvd} but the final proposal is yet to be discovered. One of the appealing proposals is to consider a non-gravitational interaction in the dark sector characterized by an energy-momentum exchange between between DM and DE~ \cite{Amendola:1999er}.
Phenomenologically, there is no objection in considering an interaction in the dark sector since the nature of both the dark components is not yet fully understood. The cosmological models allowing an interaction between the DM and DE, known as interacting DE (IDE) models, have been  widely investigated in the community  for their many interesting outcomes~\cite{Amendola:2003eq,Guo:2004xx,Cai:2004dk,Wang:2005ph,Barrow:2006hia,Setare:2007at,Feng:2007wn,Boehmer:2008av,Olivares:2008bx,Valiviita:2008iv,Wang:2008te,Gavela:2009cy,Valiviita:2009nu,LopezHonorez:2010esq,Chen:2010ws,Cao:2010fb,Clemson:2011an,He:2011qn,Harko:2012za,Sun:2013pda,Chimento:2013rya,Li:2013bya,Yang:2014gza,Yang:2014okp,Yang:2014hea,Li:2014eha,Tamanini:2015iia,Goncalves:2015eaa,Pan:2012ki,Yang:2016evp,Nunes:2016dlj,Yang:2017yme,Dutta:2017kch,DiValentino:2017iww,Santos:2017bqm,Yang:2017ccc,Yang:2017zjs,Yang:2017iew,Yang:2018euj,Yang:2018uae,Li:2019loh,Pan:2019gop,DiValentino:2019ffd,Cheng:2019bkh,DiValentino:2019ffd,DiValentino:2019jae,Pan:2020zza,DiValentino:2020leo,Pan:2020mst,DiValentino:2020kpf,Yang:2021hxg,Gariazzo:2021qtg,Gao:2021xnk,Guo:2021rrz,Chatzidakis:2022mpf,Zhao:2022ycr,Gao:2022ahg,Hou:2022rvk,Nunes:2022bhn,Zhai:2023yny,Li:2023fdk,Benisty:2024lmj,Halder:2024uao,Giare:2024smz,Giare:2024ytc,Halder:2024gag,Sabogal:2024yha,Ghedini:2024mdu} (see the reviews in this direction~ \cite{Bolotin:2013jpa,Wang:2016lxa,Wang:2024vmw}). The alleviation of the cosmic coincidence problem \cite{Amendola:1999er,Cai:2004dk,Pavon:2005yx,Huey:2004qv,delCampo:2008sr,delCampo:2008jx}, crossing of the phantom divide line \cite{Wang:2005jx,Sadjadi:2006qb,Pan:2014afa},  alleviation of the $H_0$ and $S_8$ tensions~\cite{Pan:2023mie,Yang:2022csz,Pan:2020bur,Pan:2019gop,Yang:2019uzo,Yang:2018uae,Pourtsidou:2016ico,An:2017crg,Kumar:2019wfs} have given a special importance to this class of cosmological models. However, modeling the DE sector is not unique, irrespective of the presence of interaction with other fields (e.g. CDM for instance) or not.  Among many possibilities, a scalar field $\phi$ is used to build a field theoretic description of the DE models. The simplest form of the DE in terms of the scalar field includes quintessence \cite{Tsujikawa:2013fta}.  In such quintessence models, the Lagrangian takes a simple canonical form with a self-interacting potential $V(\phi)$.  With some certain conditions, the field can present the so-called scaling or tracking behavior~\cite{Steinhardt:1999nw,Kiselev:2006jy,Roy:2013wqa,Gong:2014dia,Haro:2019peq,Urena-Lopez:2020npg,Alho:2023xel}. The coupling between the quintessence field and CDM has been investigated by many researchers in the community~\cite{Amendola:1999er,Amendola:2003eq,Xia:2009zzb,vandeBruck:2016hpz,Mifsud:2017fsy,VanDeBruck:2017mua,Liu:2019ygl,daFonseca:2021imp,Barros:2022bdv,Teixeira:2023zjt}.  
The presence of the coupling between the scalar field and CDM affects the evolution of the resulting interacting scenario at the background and perturbation levels.
At the level of background, the interaction may affect the cosmological parameters,
for instance, the equation of state parameter $w_{\phi}$ is affected even if the coupling is weak. Along the lines of the $H_0$ tension in the context of scalar field models (with or without interaction),
in recent years,
such models are also used to mimic an early DE (EDE)~\cite{Poulin:2018cxd,Sakstein:2019fmf,Niedermann:2019olb,Chudaykin:2020igl,Niedermann:2020dwg,Smith:2020rxx,Seto:2021xua,Agrawal:2019lmo}, which offer a potential solution to the $H_0$ tension. In an EDE model, the scalar field is usually assumed to be frozen initially~\cite{Karwal:2016vyq,Poulin:2018cxd,Poulin:2023lkg} and produces an energy injection around some time with $z>>1$. As a result the early sound horizon shrinks to get a larger $H_0$ today. Observational data from CMB and baryon acoustic oscillations (BAO) give tight constraint on the acoustic scale angle $100\theta^*$. The Planck 2018 leads to $100\theta^*=1.04110\pm 0.00031$ at 68\% CL~\cite{Aghanim:2018eyx}. Approximately, the peak spacing of the CMB temperature anisotropy spectra determines the angular size of the sound horizon at recombination by $\theta^*=\pi/\Delta {\ell}$~\cite{Knox:2019rjx}. On the other hand, the sound horizon $r_s^*$ is known as a ``standard ruler'' in a given model, which could be calculated if the early evolution is fully known. In a typical EDE scenario, $r_s^*$ suffers from a reduction due to the additional DE density, which in turn brings down the angular distance $D_A$ if $\theta^*$ is assumed to be fixed~\cite{Verde:2019ivm,Gomez-Valent:2020mqn,Wu:2020nxz,Jedamzik:2020zmd}, resulting in an enhancement to $H_0$. Inspired by these results, sound horizon could be used to set the initial conditions when considering the early evolution of an ordinary interacting quintessence model. In this paper we discuss the details of this method, and show the effects of the IDE model to the Hubble parameter after a direct accurate solution to the scalar field equations.

We focus on a typical quintessence scalar field model interacting with CDM in which  the quintessence field has an exponential potential and the interaction function depends both the CDM density and the quintessence field itself. And our purpose is to find the effects of the scalar field evolution, especially how the interacting scenario affects the evolution of the universe at the early stages, by properly setting the initial conditions considering the observational constrains on the sound horizon size. It is usually problematic to choose the proper initial conditions when we are dealing such kind of dynamical models. The initial values of $\phi$ and its derivative  must be given and different initial conditions on $\phi$ and its derivative will influence the behavior of the universe in the late stages to get different values of the present value of the CDM density and $H_0$, even if a tracking solution is achieved~\cite{Maziashvili:2023rjr,Alho:2023xel}. To avoid this ambiguity we use the set $\{\Omega_{ci}h^2,\Omega_{bi}h^2, \rho_{\phi}\}$ as the initial conditions where $\Omega_{c}h^2$ , $\Omega_{b}h^2$ are the physical CDM density and baryons density, respectively, in which $i$ is used to specify a fixed value of these parameters
and $\rho_{\phi}$ is the energy density of the quintessence field.  Notice that for the non-coupling case $\{\Omega_{ci},\Omega_{bi}\}$ could be found with their values today and  $\rho_{\phi}$ is decided by $\phi_{i}$, which varies arbitrarily, assuming the field is stationary at the beginning. When the system is solved we can get $100\theta^*$, since we can calculate both $r_s^*$ and $D_A$. Then we get the actual value of  $\rho_{\phi}$ with a shooting method for the given $100\theta^*$. On the other hand, iteration method should be applied to avoid a 2-dimension shooting if the scalar field couples with the cold dark matter. We find that a large coupling strength ($> 0.1$) 
may break the convergence of the iteration and make the system unsolvable. Thus, in this article we only consider the weak coupling situation, which still keeps quite a lot of generosity since the coupling is not expected to be much strong in most cases.  Notice that during such process, $h$ or $H_0$ is unlimited while $\theta^*$ is fixed. We find that the effects of the interaction is much larger than the change of the potential parameter.

The paper has been organized as follows. In section \ref{sec-2} we present the background and perturbation equations of the coupled quintessence scenarios. After that in section \ref{sec-3} we describe the observation datasets and the statistical methodology to constrain all the cosmological scenarios. Then in section \ref{sec-results} we discuss the observational constraints extracted out of all the scenarios considered. Finally, in section \ref{sec-discuss} we close this article with a brief summary of the results.

\section{Coupled Quintessence and the Initial Conditions}
\label{sec-2}

We consider that our universe is homogeneous and isotropic in the large scale and this description is well described by a spatially flat Friedmann-Lema\^{i}tre-Robertson-Walker (FLRW) metric
\begin{equation}
    ds^2=-dt^2+a(t)^2\delta_{ij}dx^{i}dx^{j},
\end{equation}
where $a(t)$ is the scale factor of the universe and $\delta_{ij}$ ($i,j = 1, 2, 3$) are the components of the metric tensor. In the synchronous gauge using the conformal time $\eta$ which is related to the scale factor and the cosmic time $t$ as, $d\eta=dt/a$, the metric takes this form:
\begin{equation}
    ds^2=a^2(\eta)(-d\eta^2+\delta_{ij}dx^{i}dx^{j}).
\end{equation}
We assume that the gravitational sector of the universe is described by GR and the matter sector is comprised of a quintessence scalar field, baryons, neutrinos, and CDM.
The Lagrangian of a  quintessence scalar field $\phi$ with the potential $V (\phi)$
is given by~\cite{Ratra:1987rm}
\begin{equation}
    \mathcal{L}=-\frac{1}{2}\partial_{\mu}\partial^{\mu}\phi-V(\phi).
\end{equation}
For the FLRW universe,
 energy density ($\rho_{\phi}$) and pressure ($p_{\phi}$) of the quintessence scalar field are
given by
\begin{eqnarray}
\rho_{\phi}=\frac{1}{2a^2}\phi'^2+V(\phi),\\
p_{\phi}=\frac{1}{2a^2}\phi'^2-V(\phi),
\end{eqnarray}
where prime stands for the derivative with respect to the conformal time $\eta$. Now, the first Friedmann equation in this context can be written as
\begin{equation}
    \mathcal{H}^2=\frac{8\pi G}{3}a^2\left(\rho_{\phi}+\rho_{c}+\rho_{b}+\rho_{\gamma}+\rho_{\nu}\right),
\end{equation}
where $\mathcal{H} = a'/a$ is the conformal Hubble parameter; $\rho_c$, $\rho_b$, $\rho_{\gamma}, \rho_{\nu}$ are respectively the energy density of CDM, baryons, radiation and neutrinos (one massive neutrino with a fixed mass $0.06$ eV and two massless neutrinos).
We consider that CDM interacts with the quintessence field during the evolution of the universe, while radiation, baryons and neutrinos do not take part in the interacting mechanism, e.g., baryons and  radiation follow their own conservation laws which lead to,
$\rho_{b}\propto (a/a_0)^{-3}$, $\rho_{\gamma}\propto (a/a_0)^{-4}$, where $a_0$ is the scale factor at present time. From now on, without any loss of generality, we set $a_0 =1$ and we work on the units so that $8 \pi G = 1$.
The conservation equations of the CDM and the quintessence scalar field in presence of an interaction between them follow
\begin{eqnarray}
    \rho_c'+3\mathcal{H}\rho_c=Q,\label{balance-cdm}\\
    \rho_{\phi}'+3\mathcal{H}(1 + w_{\phi}) \rho_{\phi} =-Q,\label{balance-de}
\end{eqnarray}
where $Q$ is the interaction function (also known as the interaction rate) between the dark sectors, $w_{\phi} = p_{\phi}/\rho_{\phi}$ is the equation of state of the quintessence scalar field. For $Q > 0$, the energy flow occurs from DE to CDM, while for $Q < 0$, the direction of energy flow is reversed (i.e. CDM to DE). One can see that the balance equations (\ref{balance-cdm}), (\ref{balance-de}) representing a coupled scenario can be expressed in terms of an uncoupled scenario by introducing the effective equation of state parameters for CDM, and quintessence scalar field, namely, $w_{c}^{\rm eff}$ and $w_{\phi}^{\rm eff}$, respectively, as follows

\begin{eqnarray}
&& \rho_c'+3\mathcal{H} \left(1+ w_{c}^{\rm eff}\right) \rho_c=0,\label{balance-eff-cdm}\\
&& \rho_{\phi}'+3\mathcal{H}(1 + w_{\phi}^{\rm eff}) \rho_{\phi}=0,\label{balance-eff-de}
\end{eqnarray}
where $w_{c}^{\rm eff}$, and $w_{\phi}^{\rm eff}$ are respectively given by
\begin{eqnarray}
    w_{c}^{\rm eff} =  - \frac{Q}{3 \mathcal{H} \rho_c}, \quad w_{\phi}^{\rm eff} = w_{\phi} + \frac{Q}{3 \mathcal{H} \rho_{\phi}}.
\end{eqnarray}
Thus, we see that in presence of an interaction characterized by the coupling function $Q$, the effective equation of state for CDM could be dynamical and the effective equation of state of the quintessence field may behave like a phantom fluid. Now, in order to proceed with the interacting dynamics, a specific form of $Q$ is essential. In this work we assume that the
interaction term $Q$ has the typical form \cite{Amendola:1999er,Xia:2009zzb,daFonseca:2021imp,Barros:2022bdv}

\begin{eqnarray}\label{model-Q}
Q=\beta\phi'\rho_c~,
\end{eqnarray}
in which the CDM energy density is coupled with the time-derivative of the field and $\beta$ is a constant dimensionless value measuring the strength of the interaction, known as the coupling parameter of the interaction model. In the field theory framework the coupling form of equation (\ref{model-Q}) suggests a Yukawa type interaction in the Lagrangian, resulting a long-range force only affecting the dark matter since the scalar field is assumed to have no direct interactions with the baryonic matter. Such a scenario implies that dark matter and baryons experience different accelerations in the galactic gravitational field and may violate the weak equivalent principle (WEP) locally. Analysis on the tidal tails\cite{Kesden_2006} has brought additional constrains on the coupling strength. To reconcile this problems it requires more details of the dark matter self interacting, and screening mechanisms such as the chameleon effect\cite{Khoury_2004} or Vainshtein screening \cite{VAINSHTEIN1972393} should be considered in dense environments while preserving cosmological-scale interactions.

 Notice that the sign of $\beta$ and $\phi'$ together decides the direction of the energy flow. The coupling becomes strong when $\phi$ evolves fast, and remains zero at the possible stationary points. Inserting $\rho_{\phi}$ and the interaction term (\ref{model-Q}) into the balance equation (\ref{balance-de}), 
we have the equivalent Klein-Gordon equation:
\begin{equation}
    \phi''+2\mathcal{H}\phi'+a^2\frac{dV}{d\phi}=-a^2\beta\rho_c~, \label{KGeq}
\end{equation}
and the effective equation of state parameters become
\begin{eqnarray}
  w_{c}^{\rm eff} =  - \frac{\beta \phi^{\prime}}{3 \mathcal{H}}, \quad w_{\phi}^{\rm eff} = w_{\phi} + \frac{\beta \phi^{\prime}}{3 \mathcal{H}} \times \frac{\rho_c}{\rho_{\phi}},
\end{eqnarray}
which explicitly show that the effective equation of state for CDM could be either positive or negative depending on the combined nature of $\beta$ and $\phi^{\prime}$ and in a similar fashion, if $\beta \phi^{\prime} < 0$,  $w_{\phi}^{\rm eff}$ could be less than $-1$ even if $w_{\phi} > -1$.  For $\beta  = 0$, one recovers the corresponding non-interacting scenario.

In the $\Lambda$CDM model, the set $\{h,\Omega_c h^2,\Omega_b h^2\}$ where $H_0 = 100 h $ km/s/Mpc, $\Omega_c$ and $\Omega_b$ are the density parameters for CDM and baryons, is enough to decide the background solution. Because, since in the $\Lambda$CDM structure, energy density of the cosmological constant, $\rho_{\Lambda}$, is a constant, we can get the initial conditions with all the parameter values today. However, in the dynamical DE model, $\rho_{\phi}$ varies with time and its current value depends on the initial value of the field, that means, $\phi_i$ and $\phi_i^{\prime}$ ($i$ corresponds to the value at a very early epoch). In this article, we use $a_i=10^{-8}$, corresponding to a redshift deep into the radiation era. And we also set $\phi_i^{\prime}=0$, assuming that the scalar field is at rest in the beginning. To set a proper value for $\phi_i$, we apply the shooting method. The starting value of $\phi_i$ is found by letting $\rho_{\phi}=\rho_{\Lambda}\approx 1.035\times 10^{-7}\text{Mpc}^{-2}$. Then by solving the Cauchy problem we can get $H(z)$, and perform the integrals to get $r_s^*$ and $D_A^*$ as follows
\begin{eqnarray}
    r_s^*=\int_{z_*}^{\infty}\frac{c_s(z)}{H(z)}dz,\\
    D_A^*=\int_{0}^{z_{*}}\frac{c}{H(z)}dz,
\end{eqnarray}

We use the given result of the angular size $\theta^{*}$ at recombination as the target to adjust the initial conditions. Here, we use $z^{*}$ rather than $z^{\rm drag}$ and the value $z^{*}=1091$ is given by Hu and Sugiyama~\cite{Hu:1995en}.
\begin{equation}
    \theta^*=\frac{r_s^*}{D_A^*},
\end{equation}
We adapt a similar process in the case that quintessence interacts with CDM. However, in this case not only $\rho_{\phi}$ is time varying, $\rho_c$ is not proportional to $a^{-3}$ as well. To avoid the mathematical unpromising of a 2-dimensional shooting, we use the perturbative iteration technique by assuming $\beta<<1$ as a weak coupling approximation, that means a small shift to $\Lambda$CDM. The perturbative expansion of $\rho_c$ and $\phi$ are written as
\begin{eqnarray}
  \rho_c &=& \rho^{(0)}_c+\rho_c^{(1)}+\rho_c^{(2)}+..., \\
  \phi &=& \phi^{(0)}+\phi^{(1)}+\phi^{(2)}+...,
\end{eqnarray}
where we have $\phi^{(n)}\sim\rho^{(n)}_{c}\sim O(\beta^{n})$ to define the magnitude of each term.  The unperturbed CDM energy density is proportional to $a^{-3}$, which is simply $\rho^ {(0)}_c=\rho_{c0}a^{-3}$. The zeroth-order term satisfies the non-interacting equations as below:
\begin{eqnarray}
    \rho'^{(0)}_{c}+3\mathcal{H}\rho^{(0)}_{c} &=& 0, \\
    {\phi''}^{(0)}+2\mathcal{H}{\phi'}^{(0)}+a^2\frac{dV}{d\phi} &=& 0.
\end{eqnarray}

\begin{figure}
\includegraphics[width=0.52\textwidth]{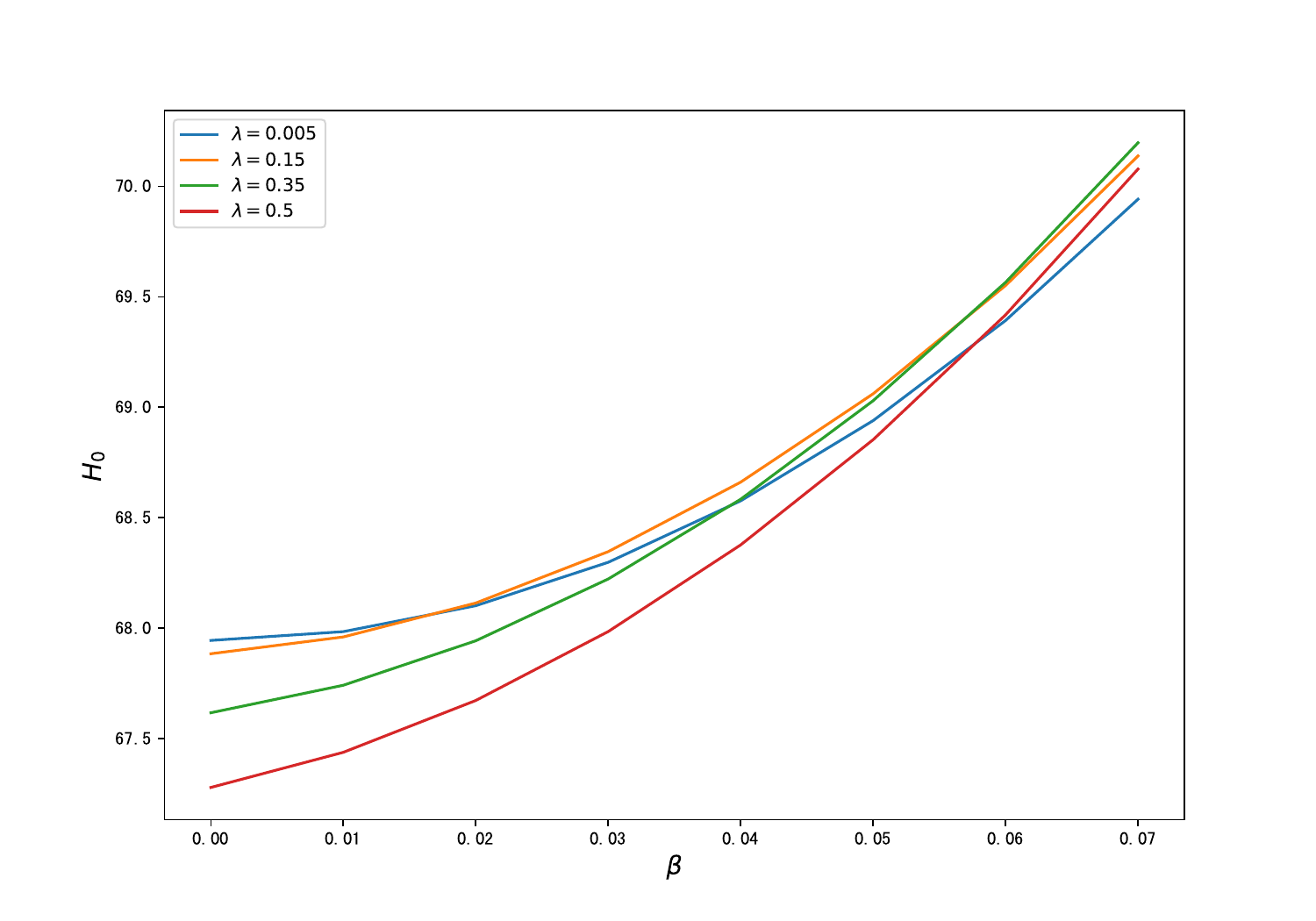}
\caption{The curves show the change of $H_{0}$ to the running values of coupling parameter $\beta$. Each curve corresponds to a certain value of $\lambda$, which appears in the potential $V(\phi)=V_0 \exp(-\lambda \phi)$. For all the curves $\theta^*$ is set to $1.0411$. }
\label{fig:H0}
\end{figure}

\begin{figure}
\includegraphics[width=0.52\textwidth]{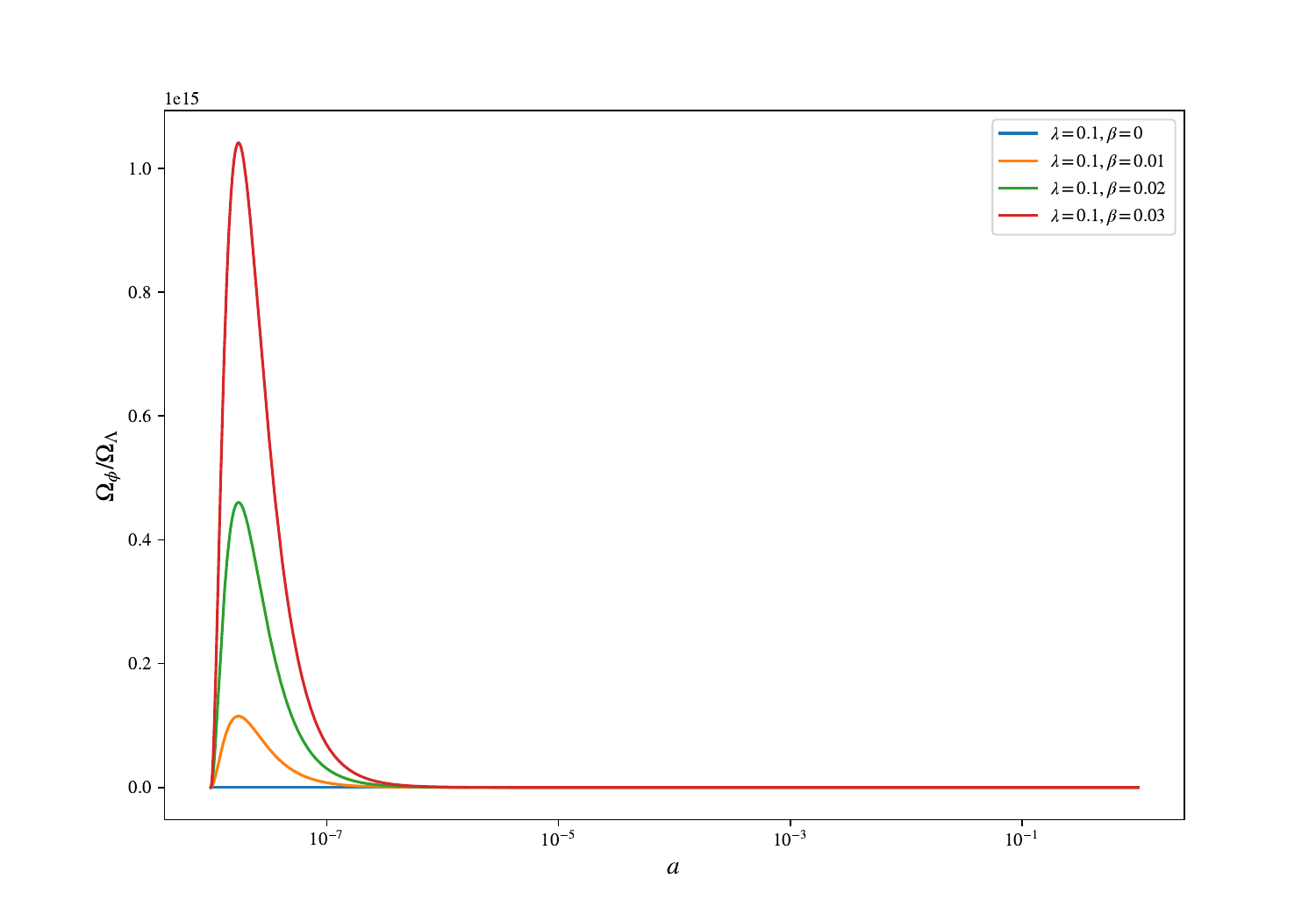}
\caption{The curves show the relative ratio $\Omega_{\phi}$ in a coupled quintessence model to $\Omega_{\Lambda}$ in a corresponding $\Lambda$CDM model for different values of $\lambda$ of the exponential potential $V(\phi)=V_0 \exp(-\lambda \phi)$ and the coupling parameter.  }
\label{fig:Omegaphi}
\end{figure}

The leading order and next leading order terms could be calculated out by solving the corresponding perturbative equations:
\begin{eqnarray}
    \rho'^{(1)}_{c}+3\mathcal{H}\rho^{(1)}_{c} &=& a^2\beta\rho^{(0)}_c~, \\
    {\phi''}^{(1)}+2\mathcal{H}\phi'^{(1)}+a^2\frac{dV}{d\phi} &=& -a^2\beta\rho^{(0)}_c~,\\
    \rho'^{(2)}_{c}+3\mathcal{H}\rho^{(2)}_{c} &=& a^2\beta\rho^{(1)}_c~, \\
    {\phi''}^{(2)}+2\mathcal{H}\phi'^{(2)}+a^2\frac{dV}{d\phi} &=& -a^2\beta\rho^{(1)}_c~.
\end{eqnarray}
And this sequence could be repeated iteratively until the solution converges. One can notice that the above calculations need a specific form of the potential. Although there are many choices of the potential in the literature, however, in this article we use the exponential potential
\begin{equation}
    V(\phi)=V_0\exp{(-\lambda\phi)},
\end{equation}
where $\lambda > 0$ is a dimensionless free parameter. 
Thus, we see that the proposed interacting quintessence scenario has two characteristic parameters, namely $\lambda$ which decides the slope of the potential and $\beta$ deciding the strength of the coupling.
The set $\{\Omega_c h^2, \Omega_{b}h^2,\theta^*,\lambda,\beta\}$ completely decides the evolution result. Notice that $H_0$ is also affected as well. And we show the varying of $H_0$  to the quintessence parameters in Fig.~\ref{fig:H0}. It is interesting to notice that the Hubble parameter could be enhanced with an increment of the coupling parameter $\beta$. The starting point of each curve corresponds to the non-interacting ($\beta=0$) case, and larger $\lambda$ in magnitude  tends to give smaller $H_0$ causing the $H_0$ tension worse. However, if the interacting term takes part in, curves with larger $\lambda$ will grow faster as $\beta$ gets larger. The curve with $\lambda=0.35$  allows $H_0>70$ for $\beta\sim 0.07$. In our calculation, the interaction of DM and DE brings quite large perturbation to the energy density evolution, and this perturbation is limited at very early time naturally. We can check this assertion by plotting the dimensionless density parameter of the field, which is shown in Fig.~\ref{fig:Omegaphi}. In this figure we see that for all the interacting cases $\Omega_{\phi}$ experiences a narrow pulser-shaped enhancement at the early time.


To compare with the CMB power spectrum we need to solve the linear perturbation. In the following we describe the evolution of the dark components at the level of perturbations. In order to proceed, we need a perturbed metric.
In the synchronous gauge~\cite{Ma:1995ey} and in conformal time, the metric is written as

\begin{equation}
    ds^2=a^2(\eta) \left[-d\eta^2+(\delta_{ij}+h_{ij})dx^idx^j \right],
\end{equation}
where $h_{ij}$ is the metric perturbation in the synchronous gauge. For the perturbations in the cosmological fluid, the dimensionless density contrast $\delta_i=\frac{\delta\rho_i}{\rho_i}$ describes the energy fluctuation of a given cosmological component $i$, and $\theta_i$ the velocity divergence with respect to the expansion.
For the scalar field $\theta_{\phi}$ could be directly written in terms of $\delta\phi$ as
\begin{equation}
    \theta_{\phi}=k^2\frac{\delta\phi}{\phi'},
\end{equation}
where $k$ is the Fourier-space wave number.
To solve the perturbation, it is more convenient to solve the field equations, i.e. the perturbed Klein-Gordon equation. By defining the time-derivative of $\phi$ as $v_{\phi}$, this equation could be written as 

\begin{eqnarray}
&& \delta\phi'=v_{\phi}, \label{perturb1} \\
&& v_{\phi}'=-2\mathcal{H}v_{\phi}-\frac{h'}{2}\phi'-k^2\delta\phi-a^2\delta\phi V_{\phi\phi}-\beta\rho_{c}a^2\delta_c \label{perturb2}
\end{eqnarray}
where $V_{\phi\phi}=\frac{d^2V}{d\phi^2}$. The density perturbation of the dark matter is also affected because of the interaction behaviour, and the dynamical equations of $\delta_c$ and $\theta_{c}$ would be shifted as follows
    \begin{eqnarray}
&& \delta_{c}'=-k\mathcal{Z}+\beta\delta\phi'-\theta_{c}~,\\
&& \theta_{c}'=-\frac{a'}{a}\theta_{c}-\beta\phi'\theta_{c}+k^2\beta\delta\phi~,
    \end{eqnarray}
where $\mathcal{Z}$ is defined as \cite{Ma:1995ey}:
\begin{equation}
    \mathcal{Z}= \left(\frac{\delta}{2k}+\eta_k \right)\frac{1}{aH}.
\end{equation}
To solve equations (\ref{perturb1}) and  (\ref{perturb2}) we need to find the initial conditions on $\delta\phi$ and $\delta\phi'$ rather than $\delta_{\phi}$ and $\theta_{\phi}$.
    In the synchronous gauge, the perturbed conformal fluid 4-velocity is written as $u^{\mu}=(1/a,v^{i}/a)$ (which is $[(1-\Psi)/a,v^i/a]$ in the conformal Newtonian gauge) \cite{Ma:1995ey}. So the gravitational potential does not appear in the expression of $\delta\rho_{\phi}$, that is
    \begin{equation}
        \delta\rho_{\phi}=\phi'\delta\phi'+V_{\phi}\delta\phi \quad \text{(Syn)}.
    \end{equation}

Thus, the perturbation variables of fluid and the field satisfy the following relations (we use synchronous gauge in all the calculations in this paper):
\begin{eqnarray}
&&\delta\phi=\frac{\phi'}{k^2}\theta_{\phi}~,\\
&&\delta\phi'=\rho_{\phi}\frac{\delta_{\phi}}{\phi'}-\frac{V_\phi\theta_{\phi}}{k}~.
\end{eqnarray}

\begin{figure}
\includegraphics[width=0.52\textwidth]{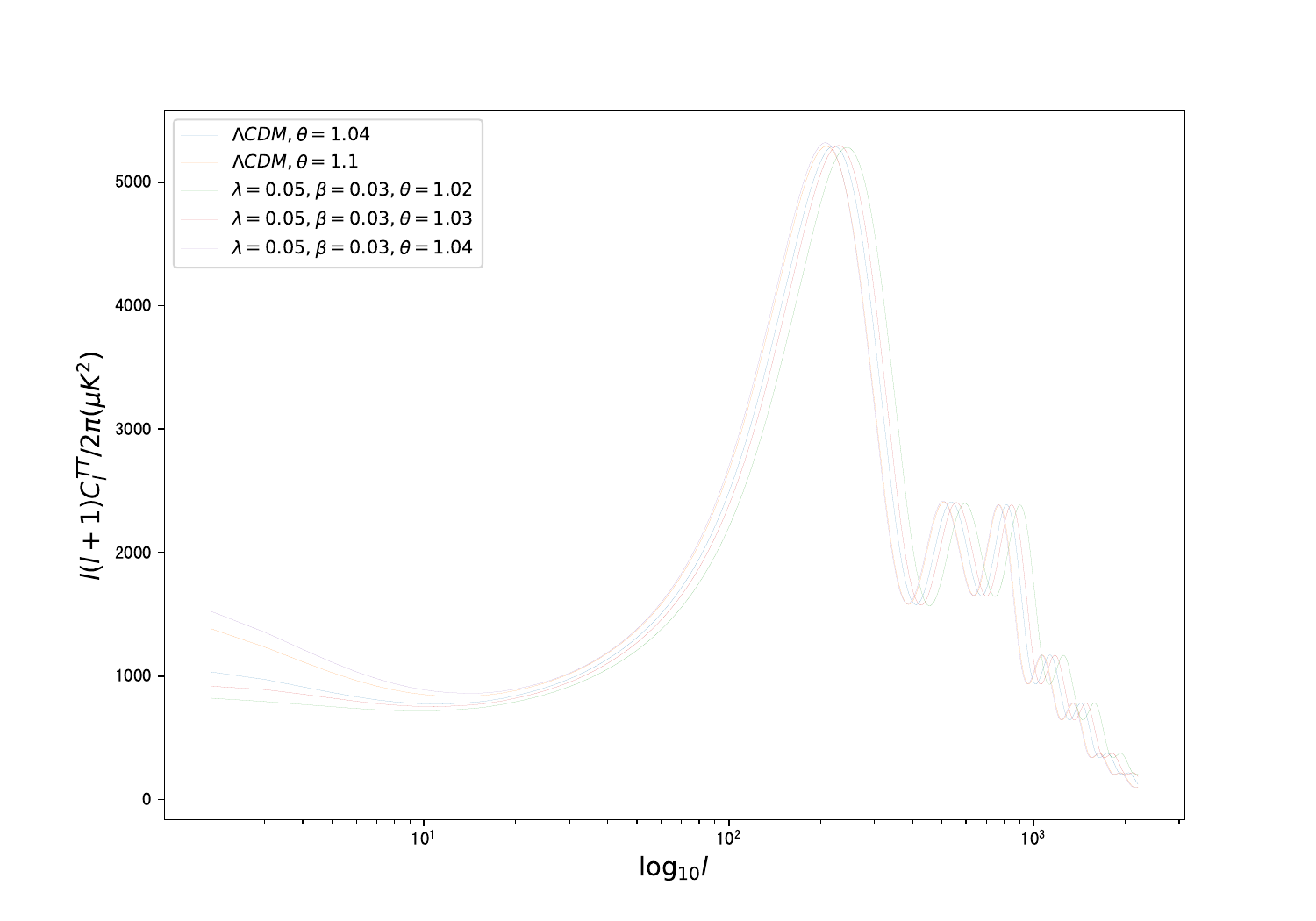}
\caption{The curves show the change in the CMB TT power spectrum caused due to the running values of $\theta^*$ and the coupling parameter $\beta$ for the potential $V(\phi)=V_0 e^{-\lambda\phi}$. }
\label{fig:TTtheta}
\end{figure}

\begin{figure}
\includegraphics[width=0.52\textwidth]{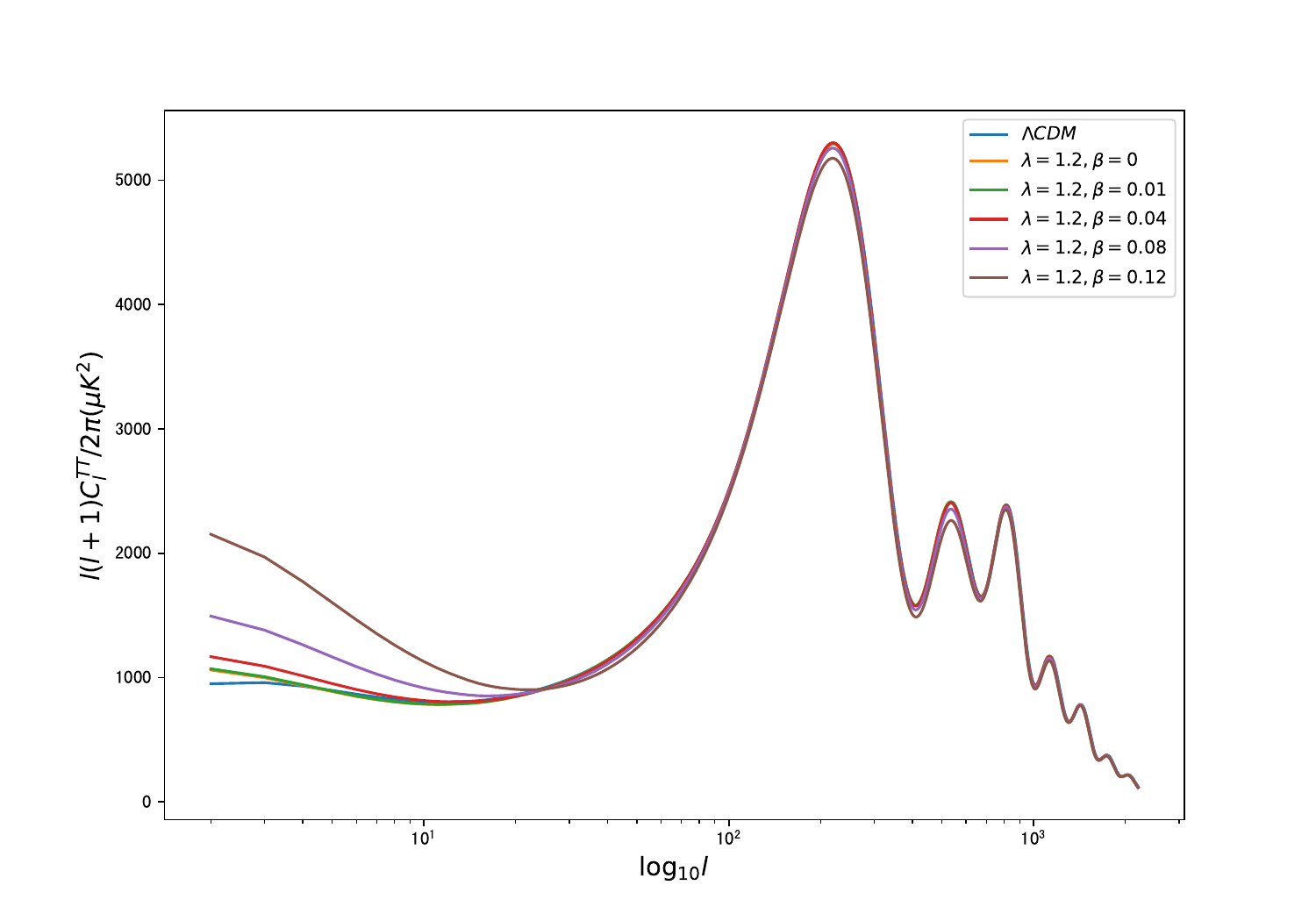}
\caption{The curves show the change in the CMB TT power spectrum for the running values of coupling parameter $\beta$, with $\theta^*$ chosen to be $1.0411$. For potential $V(\phi)=V_0 e^{-\lambda\phi}$. In this figure $\beta$ is set to be positive. }
\label{fig:TTbetaP}
\end{figure}

\begin{figure}
\includegraphics[width=0.52\textwidth]{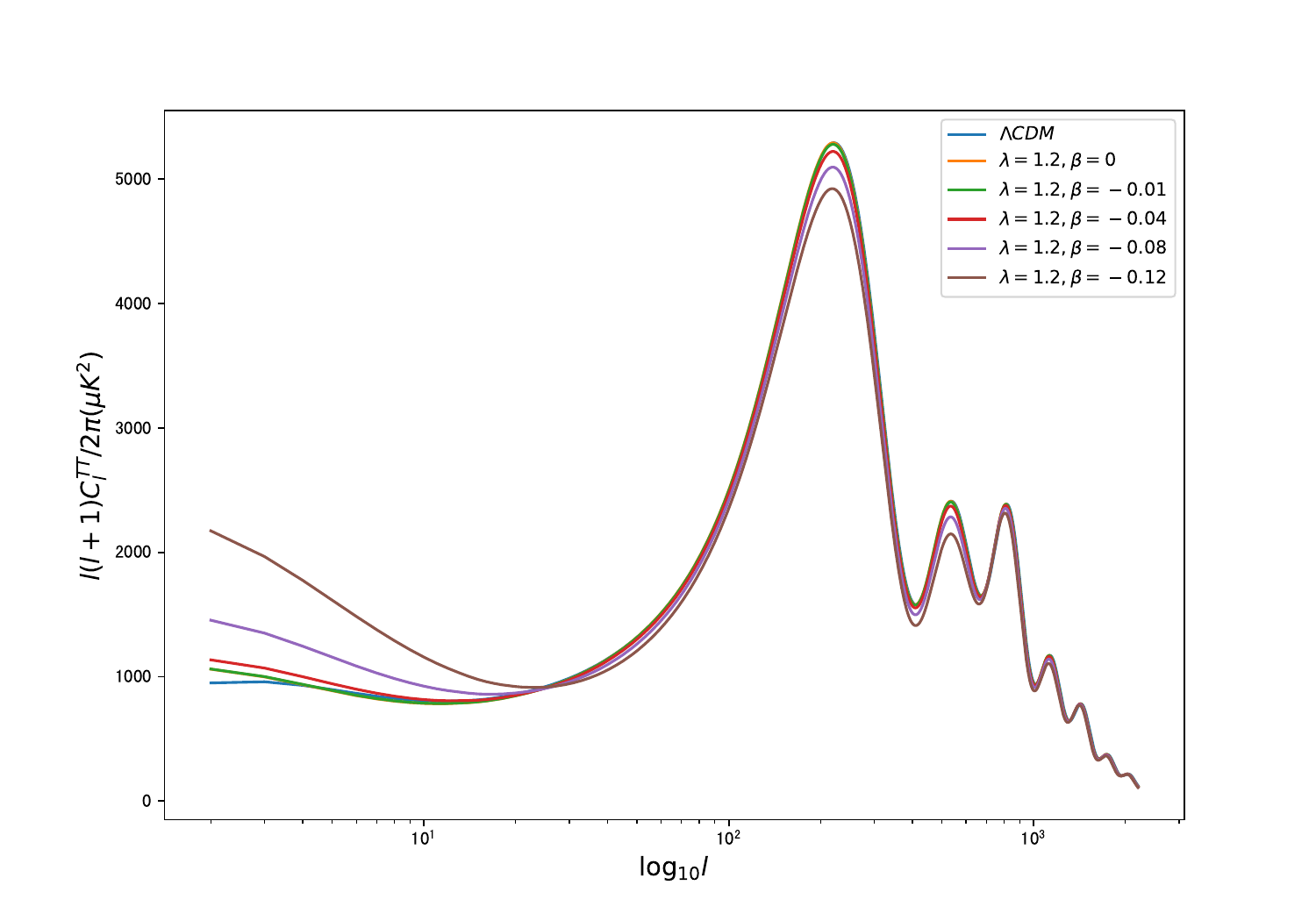}
\caption{The curves show the change in the CMB TT power spectrum for the running values of coupling parameter $\beta$, with $\theta^*$ chosen to be $1.0411$. For potential $V(\phi)=V_0 e^{-\lambda\phi}$. In this figure $\beta$ is set to be negative.  }
\label{fig:TTbetaN}
\end{figure}

We choose $\theta_{\phi}=\delta_{\phi}=0$, i.e. $\delta\phi=\delta\phi'=0$ as the initial conditions to get the CMB TT power spectrum. The results are graphically presented in Figs.~\ref{fig:TTtheta},~\ref{fig:TTbetaP} and ~\ref{fig:TTbetaN}.
In Fig.~\ref{fig:TTtheta} we can clearly see the displacement of the BAO peak positions as we shift the value of $\theta^*$, which is expected and makes it convenient for constraining the model with it. In Fig.~\ref{fig:TTbetaP} and Fig.~\ref{fig:TTbetaN} we plot the CMB TT spectra for different values of the coupling parameter $\beta$ taking a fixed value of $\lambda = 1.2$. Fig.~\ref{fig:TTbetaP} is for $\beta>0$ and Fig.~\ref{fig:TTbetaN} is for $\beta<0$. The results show that the power spectrum is affected on the low-l region, significantly. However, the peak values in the small scales are also shifted in the figure, suggesting the imprints caused by the coupling field. This is quite different with the non-interacting case of a dynamical quintessence DE and  we see that results are shifted if the sign of $\beta$ changes. However the angular power spectra exhibit the shifting differently on different scales. That is to say, at the low-l region $C_l$ is always shifted to larger values no matter $\beta$ is positive or negative, and the amplitudes are enhanced as the absolute values of $\beta$ rises. The coupling quintessence causes an temporary density enlargement at the early stages ($a<10^{-7}$) to increase the expansion rate, shrinking the sound horizon $r_s$. Since we set $\theta^*$ fixed the comoving angular diameter distance $D_{A}$ must also decrease, so $H_0$ will increase in both cases. So $\Omega_{c}$ decreases and $\Omega_{\phi}$ increases since $\Omega_{c}h^2$ is also fixed. The larger dark energy density at late stages will also enhance the ISW effect, as discussed in \cite{Weller_2003}, shown in Fig.\ref{fig:TTbetaP} and Fig.\ref{fig:TTbetaN}. Similar interpretation applies for Fig.\ref{fig:TTtheta}, where $D_A$ gets smaller as we manually increase $\theta^*$ while keeping $\lambda$ and $\beta$ unchanged (so $r_s$ unchanged), explaining the minor enhancement of low-l power spectra.

\section{Observational data and statistical methodology}
\label{sec-3}

In this section we describe the observational datasets that are used to constrain the proposed interacting quintessence scenario. In addition, we also constrain the uncoupled quintessence scenario enjoying the same potential of the quintessence field in order to compare with the interacting quintessence scenario.
We  use a series of cosmological observational datasets that are listed in the following.

\begin{itemize}
    \item \textbf{Cosmic microwave background (CMB) data:} We use the CMB data from Planck 2018 legacy release~\cite{Aghanim:2018eyx,Aghanim:2019ame}. In particular, we use Planck TT,TE,EE+low E~\cite{Aghanim:2018eyx,Aghanim:2019ame}.

    \item \textbf{Baryon acoustic oscillations (BAO)}: The BAO measurements from different astronomical surveys, e.g. 6dFGS~\cite{Beutler:2011hx}, SDSS-MGS~\cite{Ross:2014qpa}, and BOSS DR12~\cite{Alam:2016hwk} have bene used.

    \item \textbf{Pantheon sample from supernovae type Ia (SNIa)}: In addition, we include the Pantheon sample of 1048 SNIa distributed in the redshift interval $z \in [0.01, 2.3]$~\cite{Scolnic:2017caz}.

\end{itemize}

The scalar field acts as a dynamical DE and it affects the CMB power spectrum. To find the accurate solution we add a solving module in the Boltzmann code CAMB~\cite{Lewis:1999bs}. We solve the time evolution of the scalar field $\phi$ according to Klein-Gordon equation \ref{KGeq}. With the solution, we can get the energy densities of dark components as functions of time. The quintessence field and its coupling with the dark matter affects the CMB and large scale structure. To constrain the model we adopt Markov Chain Monte Carlo (MCMC) methods to sample the posterior distribution of the cosmological parameters.  We use a modified version of the  MCMC cosmological package  \texttt{CosmoMC}~\cite{Lewis:2002ah,Lewis:2013hha}, publicly available at \url{http://cosmologist.info/cosmomc/}, that is equipped with a convergence diagnostic based on the Gelman-Rubin criterion~\cite{Gelman:1992zz} and the package supports the Planck 2018 likelihood~\cite{Aghanim:2019ame}. We monitor the convergence of the generated MCMC chains using the standard $R$ parameter, requiring $R - 1 < 0.02$ for the MCMC chains to be considered as converged. The quintessence model span two different parameter spaces, a 6-dimensional space for the non-coupled case and a 7-dimensional space for the coupled case. The free parameters are:

\begin{displaymath}
\mathcal{P}_1 = \{\Omega_{b} h^2, \Omega_{\rm c} h^2, 100\theta_{MC}, \tau, n_{s}, {\rm{ln}}(10^{10} A_s), \lambda\},
\end{displaymath}

\begin{displaymath}
\mathcal{P}_2 = \{\Omega_{b} h^2, \Omega_{\rm c} h^2, 100\theta_{MC}, \tau, n_{s}, {\rm{ln}}(10^{10} A_s), \lambda,\beta\},
\end{displaymath}
where $\Omega_{b}$ and $\Omega_{\rm c}$ are respectively the baryons and CDM densities normalized to the critical density; $\theta_{MC}$  is an approximation of the ratio of the sound horizon to the angular diameter distance (which is adopted in \texttt{CosmoMC}~\cite{Lewis:2002ah,Lewis:2013hha} and is based on fitting formulae given in~\cite{Hu:1995en}); $\tau$ is the reionization optical depth; $n_{s}$ is the scalar spectral index; $A_s$ is the amplitude of the primordial scalar power spectrum; $\lambda$ is the slope of exponential potential of the quintessence field; and finally, $\beta$ is the interacting strength of the coupled case.
We impose flat priors on the free parameters as specified in Table~\ref{priors}.

\begin{table}
\begin{center}
\begin{tabular}{|c|c|}
\hline
Parameter                   & Prior\\
\hline
$\Omega_{b} h^2$            & $[0.005,0.1]$  \\
$\Omega_{\rm c} h^2$       & ~~~~~~~~$[0.001,0.99]$~~~~~~~~ \\
$100\theta_{MC}$             & $[0.5,10]$ \\
$\tau$                       & $[0.01,0.8]$ \\
$n_\mathrm{s}$               & $[0.7,1.3]$ \\
${\rm{ln}}(10^{10} A_s)$         & $[1.7, 5.0]$\\
$\lambda$                     & $[0,1]$  \\
$\beta$                     & $[-0.05,0.05]$  \\
\hline
\end{tabular}
\end{center}
\caption{Flat priors on various cosmological parameters of the non-interacting and interacting quintessence scenarios with the exponential potential.
}
\label{priors}
\end{table}

\section{Observational results}
\label{sec-results}

In this section we present the observational constraints on the quintessence model considering both the non-interacting and interacting cases.

\subsection{Non-interacting quintessence model}
\label{sec-results-no-interaction}

\begin{figure*}
	\includegraphics[width=0.85\textwidth]{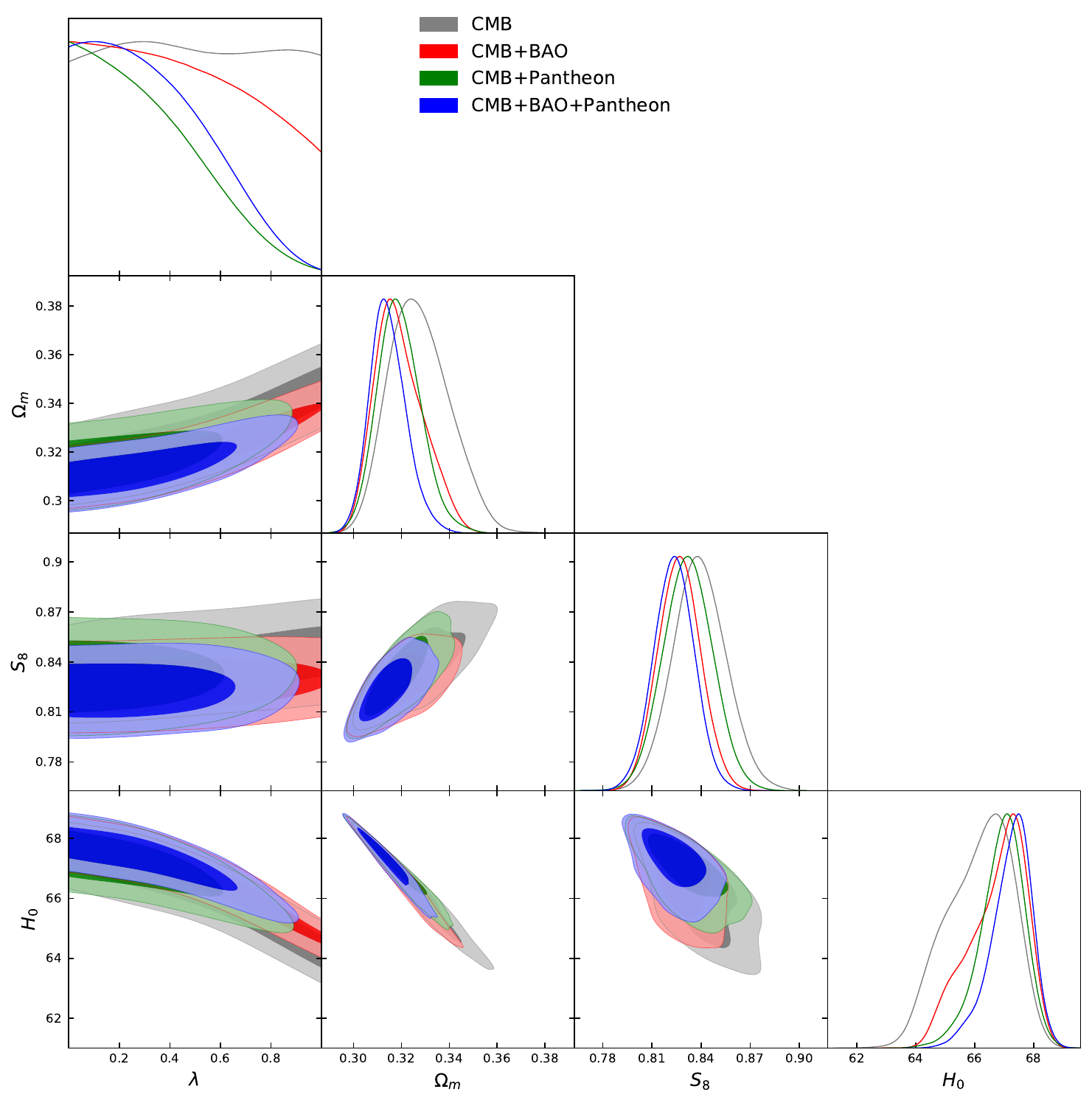}
	\caption{We show the 1-dimensional posterior distributions of some important parameters and the 2-dimensional joint contours at 68\% and 95\% CL between some of the model parameters  of the non-interacting quintessence scenario using various cosmological probes and their combinations. Here, $H_0$ and $r_{\rm drag}$ are given in the units [km/s/Mpc] and [Mpc],
 respectively. }
	\label{fig:quint1}
\end{figure*}

In Table~\ref{DEphi} we display the constraints at 68\% CL and at 95\% CL on the free cosmological parameters of the non-interacting quintessence model considering CMB, CMB+BAO, CMB+Pantheon and CMB+BAO+Pantheon. 
And the corresponding 2D contour plots are shown in Fig.~\ref{fig:quint1}. The results give a range of $0<\lambda<1$ to the index of the exponential potential for both CMB and BAO datasets. This is a weak limit since $[0,1]$ is the prior range of $\lambda$ when we start running the chains. However, a smaller upper bond of $\lambda$ could be acquired when Pantheon sample of SNIa is added in. The combination of CMB+BAO+Pantheon puts an upper limit of $\lambda$, e.g.  $0<\lambda<0.4462$ at 68\% CL and $0<\lambda<0.7427$ at 95\% CL. That is because of the fact that the quintessence DE density in the un-coupled case, falls quickly at early stages and does little effect to the CMB spectrum at the recombination era, while at the same time the initial condition is adjusted to fit the sound horizon of BAO in our calculations, as we discussed earlier. On the other hand, SNIa modeled by the late-stage accelerating expansion, which is relevant to the dynamics of the scalar field, and henceforth to the shape of the potential, limits the parameter $\lambda$. As the limit of $\lambda$ gets narrowed with the change of datasets, the central value of $H_0$ improves slightly, from $H_0 = 66.16$ km/s/Mpc at 68\% CL (for CMB alone) to $H_0  = 67.22$ km/s/Mpc at 68\% CL (CMB+BAO+Pantheon). That means, no effective reduction in the $H_0$ tension between Planck and SH0ES collaboration is found in this case. In fact, smaller $\lambda$ causes the model to approach towards the $\Lambda$CDM cosmology. We can see that the non-coupled scalar field with larger $\lambda$ makes the $H_0$ tension worse.

\begingroup
\squeezetable
\begin{center}
\begin{table*}                                                                                                                    \resizebox{\textwidth}{!}{
\begin{tabular}{ccccc}
\hline
Parameters & CMB & CMB+BAO & CMB+Pantheon& CMB+BAO+Pantheon  \\ \hline
$\Omega_{\rm c} h^2$ & $    0.1203_{-    0.0013-    0.0026}^{+    0.0014+    0.0027}$ & $    0.1191_{-    0.0011-    0.0020}^{+    0.0010+    0.0020}$ & $    0.1198_{-    0.0013-    0.0024}^{+    0.0013+    0.0025}$ & $    0.1190_{-    0.00095-    0.0019}^{+    0.0010+    0.0019}$\\

$\Omega_b h^2$ & $    0.02235_{-    0.00014-    0.00030}^{+    0.00015+    0.00029}$ & $    0.02243_{-    0.00013-    0.00026}^{+    0.00014+    0.00027}$ & $    0.02238_{-    0.00014-    0.00029}^{+    0.00014+    0.00029}$ & $    0.02244_{-    0.00014-    0.00022}^{+    0.00014+    0.00026}$ \\

$100\theta_{MC}$ & $    1.040_{-    0.00062-    0.0017}^{+    0.0011+    0.0013}$ & $    1.040_{-    0.00052-    0.0017}^{+    0.0010+    0.0013}$ & $    1.041_{-    0.00036-    0.0012}^{+    0.00064    0.0010}$ & $    1.041_{-    0.00036-    0.0013}^{+    0.00068+    0.00098}$\\

$\tau$ & $    0.05452_{-    0.0083-    0.015}^{+    0.0076+    0.017}$ & $    0.05631_{-    0.0081-    0.015}^{+    0.0076+    0.016}$ & $    0.05471_{-    0.0081-    0.016}^{+    0.0077+    0.016}$ & $    0.05629_{-    0.0086-    0.015}^{+    0.0076+    0.017}$\\

$n_s$ & $    0.9645_{-    0.0042-    0.0082}^{+    0.0044+    0.0086}$ & $    0.9671_{-    0.0038-    0.0075}^{+    0.0038+    0.0075}$ & $    0.9652_{-    0.0043-    0.0082}^{+    0.0042+    0.0083}$ & $    0.9674_{-    0.0037-    0.0072}^{+    0.0037+    0.0072}$\\

${\rm{ln}}(10^{10} A_s)$ & $    3.045_{-    0.017-    0.031}^{+    0.015+    0.034}$ & $    3.047_{-    0.016-    0.031}^{+    0.016+    0.033}$ & $    3.045_{-    0.016-    0.033}^{+    0.016+    0.034}$ & $    3.046_{-    0.016-    0.033}^{+    0.016+    0.034}$\\

$\lambda$ & $ <1 $ & $ <1 $ & $ <0.4090\,<0.7272 $ & $ <0.4462\,<0.7427 $  \\

\hline

$\Omega_m$ & $    0.3276_{-    0.015-    0.024}^{+    0.011+    0.026}$& $    0.3192_{-    0.012-    0.019}^{+    0.0085+    0.021}$ & $    0.3192_{-    0.010-    0.018}^{+    0.0084+    0.019}$ & $    0.3145_{-    0.0084-    0.015}^{+    0.0068+    0.016}$\\

$\sigma_8$ & $    0.8033_{-    0.010-    0.023}^{+    0.013+    0.020}$ & $    0.8017_{-    0.0094-    0.023}^{+    0.013+    0.021}$ & $    0.8067_{-    0.0084-    0.019}^{+    0.010+    0.018}$ & $    0.8044_{-    0.0082-    0.020}^{+    0.010+    0.018}$ \\

$H_0\,[{\rm km/s/Mpc}]$ & $   66.16_{-    0.97-    2.2}^{+    1.4+    2.0}$ & $   66.77_{-    0.73-    2.1}^{+    1.3+    1.7}$ & $   66.92_{-    0.63-    1.6}^{+    0.85+    1.5}$ & $   67.22_{-    0.54-    1.6}^{+    0.81+    1.3}$\\

\hline

\end{tabular}          }
\caption{68\% and 95\% CL constraints on the free parameters (above the horizontal line) and the derived ones (below the horizontal line) of the non-interacting quintessence model with the exponential potential have been summarized for various cosmological probes.  }\label{DEphi}
\end{table*}
\end{center}
\endgroup

\begingroup
\squeezetable
\begin{center}
\begin{table*}                                                                                                                    \resizebox{\textwidth}{!}{
\begin{tabular}{ccccc}

\hline\hline
Parameters & CMB & CMB+BAO & CMB+Pantheon& CMB+BAO+Pantheon  \\ \hline
$\Omega_{\rm c} h^2$ & $    0.1196_{-    0.0014-    0.0032}^{+    0.0017+    0.0032}$ & $    0.1190_{-    0.0010-    0.0024}^{+    0.0013+    0.0021}$ & $    0.1191_{-    0.0014-    0.0032}^{+    0.0017+    0.0029}$ & $    0.1188_{-    0.0010-    0.0022}^{+    0.0012+    0.0020}$\\

$\Omega_b h^2$ & $    0.02236_{-    0.00015-    0.00030}^{+    0.00016+    0.00030}$ & $    0.02239_{-    0.00014-    0.00028}^{+    0.00014+    0.00029}$ & $    0.02239_{-    0.00015-    0.00028}^{+    0.00015+    0.00028}$ & $    0.02230_{-    0.00014-    0.00027}^{+    0.00014+    0.00028}$ \\

$100\theta_{MC}$ & $    1.040_{-    0.00061-    0.0017}^{+    0.0011+    0.0014}$ & $    1.040_{-    0.00051-    0.0017}^{+    0.0010+    0.0013}$ & $    1.040_{-    0.00040-    0.0012}^{+    0.00071    0.0011}$ & $    1.041_{-    0.00038-    0.0012}^{+    0.00066+    0.0010}$\\

$\tau$ & $    0.05477_{-    0.0084-    0.015}^{+    0.0074+    0.016}$ & $    0.05503_{-    0.0081-    0.015}^{+    0.0082+    0.015}$ & $    0.05556_{-    0.0085-    0.015}^{+    0.0076+    0.016}$ & $    0.05519_{-    0.0076-    0.015}^{+    0.0077+    0.016}$\\

$n_s$ & $    0.9648_{-    0.0044-    0.0085}^{+    0.0042+    0.0091}$ & $    0.9664_{-    0.0038-    0.0075}^{+    0.0037+    0.0074}$ & $    0.9663_{-    0.0042-    0.0087}^{+    0.0047+    0.0083}$ & $    0.9668_{-    0.0036-    0.0075}^{+    0.0042+    0.0068}$\\

${\rm{ln}}(10^{10} A_s)$ & $    3.046_{-    0.016-    0.033}^{+    0.015+    0.034}$ & $    3.045_{-    0.016-    0.031}^{+    0.016+    0.032}$ & $    3.047_{-    0.016-    0.031}^{+    0.016+    0.034}$ & $    3.045_{-    0.016-    0.032}^{+    0.016+    0.032}$\\

$\lambda$ & $ <1 $ & $ <0.6244\,<1 $ & $ <0.4578\,<0.7902 $ & $ <0.4672\,<0.7833 $  \\

$\beta$ & $    0.002441_{-    0.023-    0.040}^{+    0.023+    0.041}$ & $    0.002647_{-    0.022-    0.042}^{+    0.027+    0.041}$ & $    0.003270_{-    0.022-    0.053}^{+    0.029+    0.047}$ & $    0.005448_{-    0.020-    0.044}^{+    0.030+    0.039}$\\

\hline

$\Omega_m$ & $    0.3220_{-    0.016-    0.026}^{+    0.013+    0.028}$& $    0.3166_{-    0.013-    0.020}^{+    0.0096+    0.022}$ & $    0.3135_{-    0.011-    0.022}^{+    0.012+    0.022}$ & $    0.3118_{-    0.0088-    0.017}^{+    0.0077+    0.017}$\\

$\sigma_8$ & $    0.8020_{-    0.0092-    0.028}^{+    0.014+    0.024}$ & $    0.8007_{-    0.0080-    0.028}^{+    0.014+    0.024}$ & $    0.8045_{-    0.0086-    0.022}^{+    0.012+    0.021}$ & $    0.8038_{-    0.0080-    0.021}^{+    0.010+    0.018}$ \\

$H_0\,[{\rm km/s/Mpc}]$ & $   66.60_{-    1.1-    2.4}^{+    1.4+    2.2}$ & $   67.00_{-    0.81-    2.1}^{+    1.3+    1.8}$ & $   67.35_{-    0.90-    1.9}^{+    0.90+    1.9}$ & $   67.47_{-    0.60-    1.7}^{+    0.82+    1.5}$\\

\hline\hline

\end{tabular}          }
\caption{68\% and 95\% CL constraints on the free parameters (above the horizontal line) and the derived ones (below the horizontal line) of the interacting quintessence model with the exponential potential have been summarized for various cosmological probes.  }\label{DEphicoup}
\end{table*}
\end{center}
\endgroup
\begin{figure*}
	\includegraphics[width=0.85\textwidth]{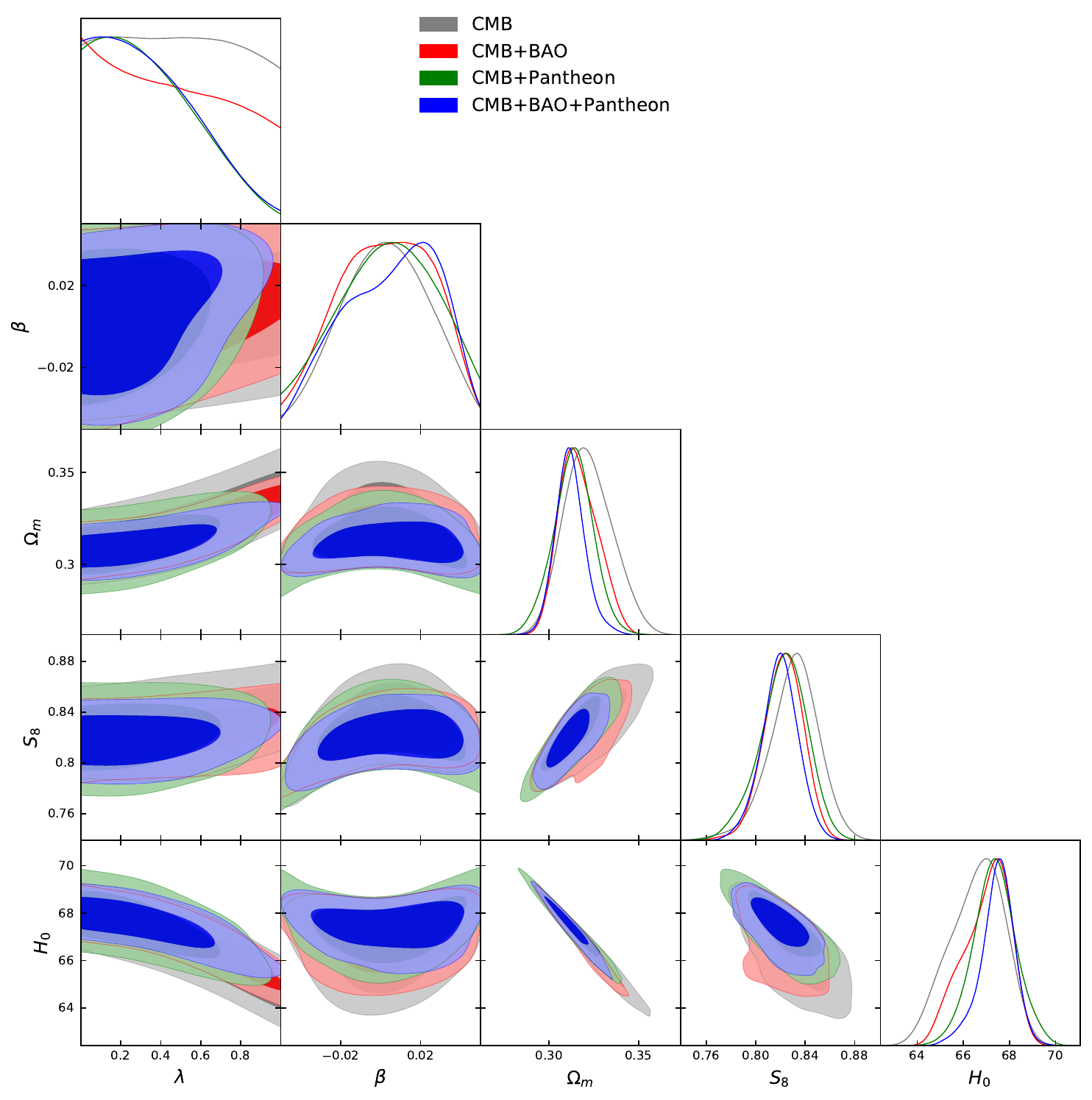}
	\caption{We show the 1-dimensional posterior distributions of some important parameters and the 2-dimensional joint contours at 68\% and 95\% CL between some of the model parameters  for the non-interacting quintessence scenario using various cosmological probes and their combinations. $H_0$ and $r_{\rm drag}$ are given in [km/s/Mpc] and [Mpc], respectively. }
	\label{fig:quint2}
\end{figure*}

\subsection{Interacting quintessence model}
\label{sec-results-interaction}

In this part we summarize the observational constraints on the interacting quintessence model. The parameter space in this case is extended because of the inclusion of the coupling parameter $\beta$. In Table~\ref{DEphicoup} we show the constraints at 68\%CL and at 95\% CL on the free and derived cosmological parameters. The corresponding 2D contour plots are shown in Fig.~\ref{fig:quint2}.

In this case, the quintessence potential parameter $\lambda$ exhibits similar characteristics to what we have observed in the context of non-interacting case. We got an unconstrained range of $\lambda$ for CMB. If BAO data are added to CMB, the 68\% CL upper limit on $\lambda$ becomes 0.6244, while 95\% CL upper limits is not changed. When Pantheon sample is considered, the upper limit of $\lambda$ shrinks significantly, to 0.4578 (at 68\% CL) and 0.7902 (at 95\% CL) for CMB+Pantheon, and to 0.4672 (68\% CL) and 0.7833 (95\% CL) for CMB+BAO+Pantheon. Notably, the central values and ranges of $H_0$ are also improved in the interacting scenario. The combined dataset CMB+BAO+Pantheon leads to $H_0 = 67.47_{-0.60}^{+0.82}$ km/s/Mpc at 68\% CL, which is quite close to the $\Lambda$CDM-based Planck value $H_ 0 =  67.27_{-0.60}^{+0.60}$ km/s/Mpc at 68\% CL~\cite{Aghanim:2018eyx}. We note that the constraints on $H_0$ should be influenced by the choice of the potential of the scalar field. In this article we have considered the exponential potential, while more interesting forms need to be tested aiming to see how the choices of the quintessence potential influence the $H_0$ tension.

Focusing on the coupling parameter $\beta$, we find that
the results are quite interesting. Firstly, we note that all the datasets are sensitive to $\beta$ and provide its tight constraints.
As we discussed before, the $\phi$-CDM interacting model can shift the energy density of the quintessence at the early stages of the universe around the decoupling era, causing the CMB power spectrum and BAO sound horizon to be shifted as well. Secondly, according to the statistical results, $\beta$ is required to be very small, approximately $|\beta|<0.05$ at 95\% CL. This is consistent with our assumption that the interaction between the scalar field and the dark matter follows a weak coupling dynamics. And the ranges are similar for all the datasets. On the other hand, it is obvious that within the constraining range of $\beta$, it could be positive and negative as well, and looks visually symmetrical about zero, hence, no strong indication of an interaction is suggested by any of the datasets considered here. However, after a careful observation,  we find that positive $\beta$ is slightly favoured for all the datasets, for instance, the central value of $\beta$ is 0.002441 for CMB and 0.005448 for CMB+BAO+Pantheon. Generally SNIa data have a tendency of giving a larger $\beta$ than CMB. Although the central values obtained in all the datasets are close to zero, however, the possibility of the interaction between the quintessence field and CDM cannot be excluded which is clearly visualized from Fig. \ref{fig:quint2}. We further comment that the constraints on the coupling parameter is also dependent on the choice of the coupling function. As the intrinsic nature of the interaction is reflected through its influence on the early stage of the universe, hence, other kind of interaction models should be investigated along the lines of research presented in this work.

\section{Bayesian Evidence Analysis}
We also did calculations of the log-Bayesian evidence to compare the quintessence model (with interacting or not) with $\Lambda CDM$ model. For a given dataset $D$ and some model $M$ with a set of parameters $\Theta$, Bayesian evidence (or Marginal Likelihood) is defined as
\begin{equation}
    B=P(D|M)=\int\mathcal{L}(D|\Theta,M)\pi(\Theta|M)d\Theta
\end{equation}
Where $\mathcal{L}(D|\Theta,M)$ is the likelihood function at the given model parameters, and $\pi(\Theta|M)$ is the prior distribution of $\Theta$.
The posterior probability distribution function of the model is written as
\begin{equation}
  P(M|D)=\frac{P(D|M)P(M)}{P(D)}
\end{equation}
Because the marginal probability $P(D)$ depends only on the data, for two models $M_1$ and $M_2$ the ratio of their posterior probabilities could be written bellow:
\begin{equation}
    \frac{P(M_1|D)}{P(M_2|D)}=\frac{P(D|M_1)P(M_1)}{P(D|M_2)P(M_2)}=\frac{B_1}{B_2}\frac{P(M_1)}{P(M_2)}
\end{equation}
So the posterior probability is proportional to the Bayesian evidence $B_{i}$ of model $M_i$. The relative log-Bayesian evidence between model $M_i$ and $M_j$ is defined as:
\begin{equation}
    \ln{B_{ij}}=\ln{B_{i}}-\ln{B_{j}}
\end{equation}
Here $M_j$ is set to be the $\Lambda CDM$ model and $M_i$ for the quintessence model (with or without interaction). To calculate the Bayesian evidence for all the observational data we use the cosmological code MCEvidence\cite{Heavens:2017hkr,Heavens:2017afc}. During this processes only the MCMC chains are needed to extract the cosmological parameters using the observational datasets. Similar discussion also appears in\cite{Pan_2018,Yang_2019}. The log-Bayesian evidence $\ln{B_{ij}}$ measures the performance of model $M_i$ with respect to the reference model $M_j$. In Table\ref{tab:jeffreys} we display the revised Jeffreys scale\cite{Kass:1995loi} that quantifies the observational support of the underlying cosmological model. In Table\ref{tab:phiDEbayesian} and \ref{tab:phiIDEbayesian} we display the values of $\ln{B_{ij}}$ for all the observational datasets. From the results we find that for all the combinations observational data favour $\Lambda CDM$ over the quintessence. The non-interacting model is more favoured than the the interacting model.



\begin{table} [ht]
\begin{tabular}{ccc}
\hline
$\ln B_{ij}$ &  Strength of evidence for model ${M}_i$ \\ \hline
$0 \leq \ln B_{ij} < 1$ & Weak \\
$1 \leq \ln B_{ij} < 3$ & Definite/Positive \\
$3 \leq \ln B_{ij} < 5$ & Strong \\
$\ln B_{ij} \geq 5$ & Very strong \\
\hline\hline
\end{tabular}
\caption{Revised Jeffreys scale level to compare the underlying cosmological models. } \label{tab:jeffreys}
\end{table}

\begin{table} [ht]
\begin{center}
\begin{tabular}{ccc}
\hline
Dataset & $\ln B_{ij}$ & ~Strength of evidence\\

\hline
 CMB  & \(-1.1\) & Definite/Positive \\
 CMB+BAO  & \(-3.3\) & Strong \\
 CMB+Pantheon  & \(-5.1\) & Very Strong \\
 CMB+BAO+Pantheon  & \(-4.2\) & Strong \\
\hline
\end{tabular}
\caption{Relative log-Bayesian evidence of the non-interacting quintessence model to the reference model $\Lambda$CDM (labeled with $j$). The negative sign actually indicates that the reference model is favored. }
\label{tab:phiDEbayesian}
\end{center}
\end{table}


\begin{table} [ht]
\begin{center}
\begin{tabular}{ccc}
\hline

Dataset & $\ln B_{ij}$ & ~Strength of evidence\\

\hline
 CMB  & \(-1.9\) & Definite/Positive \\
 CMB+BAO  & \(-3.7\) & Strong \\
 CMB+Pantheon  & \(-7.6\) & Very Strong \\
 CMB+BAO+Pantheon  & \(-6.0\) & Very Strong \\
\hline
\end{tabular}
\caption{Relative log-Bayesian evidence of the interacting quintessence model to the reference model $\Lambda$CDM (labeled with $j$). The negative sign actually indicates that the reference model is favored. }
\label{tab:phiIDEbayesian}
\end{center}
\end{table}


\section{Summary and Conclusions}
\label{sec-discuss}

In modern cosmology, $H_0$ tension problem suggests some kind of contradiction between the early and late stages evolution of the universe in the standard model. If we use a dynamical DE to replace the cosmological constant, the initial conditions on the external degrees of freedom are always critical when we actually perform the calculations. And these initial conditions are fundamentally decided by the basic physics of these new fluid components or some kind of field at the very early stage of the universe. In this paper we tried to solve the evolution equations of a dynamical DE in a specific method in which the initial conditions are given by setting the value of sound horizon $\theta^*$ manually. The fine-tuning of the initial conditions is avoided by a shooting process to $\theta^*$, which is a well-defined observable parameter.
	
We considered a simple cosmological model in which DE is described by a quintessence scalar field with an exponential potential $V(\phi)=V_0\exp{-\lambda\phi}$ and it could interact with CDM in which the coupling term takes the typical form $Q=\beta\phi'\rho_c$, a well known interactiom form in the literature~\cite{Amendola:1999er,Xia:2009zzb,daFonseca:2021imp,Barros:2022bdv}. In order to understand the effects of the interaction, we have also considered the non-interacting quintessence model ($\beta = 0$) with the exponential potential.
For the non-interacting case, the background evolution exhibits a typical behavior of a thawing quintessence model and the CMB angular power spectrum is affected only in the low-$\ell$ region,and this effect is quite small in our parameter range. For the interacting case, the low-$\ell$ CMB angular power spectrum is shifted to a larger extent. The peak value for larger $\ell$ is also varied slightly. Remarkably, $H_0$ could also be shifted as $\beta$ varies with a constant $\lambda$. Stronger coupling causes a larger $H_0$ today. At the same time, $H_0$ becomes smaller for large $\lambda$, thus, a small $\lambda$ with a non-zero $\beta$ could enhance $H_0$ to 69 $-$ 71 km/s/Mpc as a typical value. In all these parameter space analysis, we keep the sound horizon $\theta^*$ to be fixed. We find that this indicates a similar mechanism as EDE because the interaction of the dark constituents naturally causes a narrow energy enhancement around the recombination stage. To examine this conclusion we performed MCMC analyses for different datasets considering both the non-interacting and interacting scenarios. The results are shown in Table. ~\ref{DEphi}, Fig. ~\ref{fig:quint1} and Table. ~\ref{DEphicoup}, Fig. ~\ref{fig:quint2}, respectively.

The characteristic parameters $\lambda$ and $\beta$ are set in range $0<\lambda<1$ and $-0.05<\beta<0.05$ as a uniform prior for the interacting case. Here, $\beta$ is chosen to be small as a weak coupling approximation, which is also in agreement with the constrained results. The conclusions are summarized as bellow:

    \begin{enumerate}

        \item  In both cases, an upper limit on $\lambda$  can be given only when the Pantheon sample is added to CMB and CMB+BAO. The lower limit zero  fits with the cosmological constant. We have $0<\lambda<0.4462$ at 68\% CL (CMB+BAO+Pantheon)for the non-interacting model and $0<\lambda<0.4672$ at 68\% CL (CMB+BAO+Pantheon) for the interacting model.

        \item  In the non-interacting case, the $H_0$ value decreases with increasing $\lambda$. This is reflected from the anti-correlation between $\lambda$ and $H_0$ as displayed in
        Fig. \ref{fig:quint1}). 
        However, according to the results as summarized in Table~\ref{DEphi}, there is no release on $H_0$ tension for the non-interacting model.

        \item  In the interacting case, the central value of $\beta$ is small and positive, which is consistent with the weak coupling assumption. Although we do not find any strong indication of a coupling between CDM and the quintessence, characterized by the coupling function $Q = \beta \phi^{\prime} \rho_c$, however, based on the existing results, there is no strong indication for ruling out the possibility of the interaction (see also Fig. \ref{fig:quint2}) even if the coupling parameter is very mild (for the most constraining dataset, namely, CMB+BAO+Pantheon, the central value of $\beta$ is $0.005448$).

        \item In the interacting case, we find no evidence of any significant release of $H_0$ tension either. However, the central value is enhanced within 68\% CL compared with the non-interacting case for all the cosmological probes employed in this work.
    \end{enumerate}

   In summary, we have examined an interacting quintessence with exponential potential with a new calculation in which the field initial condition is set by the sound horizon $\theta^*$. We find that the interacting behavior of the dark components could increase the DE density at the early stage of universe which could release the $H_0$ tension like an EDE. Also the $\phi$-CDM coupling shows some enhancement on $H_0$ than the non-interacting DE with the same field potential form, however, that is not enough to resolve the tension. This can motivate the scientific community to investigate the interacting quintessence models considering, first of all, the upcoming astronomical probes and secondly, by changing the potential of the scalar field as well as the interaction rate between the quintessence and the dark matter.

\section{Acknowledgments}
 Y. Wu's work is supported by the National Natural Science Foundation of China under Grant No. 12075109.
 W. Yang's work is supported by the National Natural Science Foundation of China under Grants No. 12175096, and Liaoning Revitalization Talents Program under Grant no. XLYC1907098.
SP acknowledges the financial support from the Department of Science and Technology (DST), Govt. of India under the Scheme  ``Fund for Improvement of S\&T Infrastructure (FIST)'' (File No. SR/FST/MS-I/2019/41).

\bibliography{biblio}

\begin{thebibliography}{154}%
\makeatletter
\providecommand \@ifxundefined [1]{%
 \@ifx{#1\undefined}
}%
\providecommand \@ifnum [1]{%
 \ifnum #1\expandafter \@firstoftwo
 \else \expandafter \@secondoftwo
 \fi
}%
\providecommand \@ifx [1]{%
 \ifx #1\expandafter \@firstoftwo
 \else \expandafter \@secondoftwo
 \fi
}%
\providecommand \natexlab [1]{#1}%
\providecommand \enquote  [1]{``#1''}%
\providecommand \bibnamefont  [1]{#1}%
\providecommand \bibfnamefont [1]{#1}%
\providecommand \citenamefont [1]{#1}%
\providecommand \href@noop [0]{\@secondoftwo}%
\providecommand \href [0]{\begingroup \@sanitize@url \@href}%
\providecommand \@href[1]{\@@startlink{#1}\@@href}%
\providecommand \@@href[1]{\endgroup#1\@@endlink}%
\providecommand \@sanitize@url [0]{\catcode `\\12\catcode `\$12\catcode
  `\&12\catcode `\#12\catcode `\^12\catcode `\_12\catcode `\%12\relax}%
\providecommand \@@startlink[1]{}%
\providecommand \@@endlink[0]{}%
\providecommand \url  [0]{\begingroup\@sanitize@url \@url }%
\providecommand \@url [1]{\endgroup\@href {#1}{\urlprefix }}%
\providecommand \urlprefix  [0]{URL }%
\providecommand \Eprint [0]{\href }%
\providecommand \doibase [0]{http://dx.doi.org/}%
\providecommand \selectlanguage [0]{\@gobble}%
\providecommand \bibinfo  [0]{\@secondoftwo}%
\providecommand \bibfield  [0]{\@secondoftwo}%
\providecommand \translation [1]{[#1]}%
\providecommand \BibitemOpen [0]{}%
\providecommand \bibitemStop [0]{}%
\providecommand \bibitemNoStop [0]{.\EOS\space}%
\providecommand \EOS [0]{\spacefactor3000\relax}%
\providecommand \BibitemShut  [1]{\csname bibitem#1\endcsname}%
\let\auto@bib@innerbib\@empty
\bibitem [{\citenamefont {Weinberg}(1989)}]{Weinberg:1988cp}%
  \BibitemOpen
  \bibfield  {author} {\bibinfo {author} {\bibfnamefont {S.}~\bibnamefont
  {Weinberg}},\ }\href {\doibase 10.1103/RevModPhys.61.1} {\bibfield  {journal}
  {\bibinfo  {journal} {Rev. Mod. Phys.}\ }\textbf {\bibinfo {volume} {61}},\
  \bibinfo {pages} {1} (\bibinfo {year} {1989})}\BibitemShut {NoStop}%
\bibitem [{\citenamefont {Zlatev}\ \emph {et~al.}(1999)\citenamefont {Zlatev},
  \citenamefont {Wang},\ and\ \citenamefont {Steinhardt}}]{Zlatev:1998tr}%
  \BibitemOpen
  \bibfield  {author} {\bibinfo {author} {\bibfnamefont {I.}~\bibnamefont
  {Zlatev}}, \bibinfo {author} {\bibfnamefont {L.-M.}\ \bibnamefont {Wang}}, \
  and\ \bibinfo {author} {\bibfnamefont {P.~J.}\ \bibnamefont {Steinhardt}},\
  }\href {\doibase 10.1103/PhysRevLett.82.896} {\bibfield  {journal} {\bibinfo
  {journal} {Phys. Rev. Lett.}\ }\textbf {\bibinfo {volume} {82}},\ \bibinfo
  {pages} {896} (\bibinfo {year} {1999})},\ \Eprint
  {http://arxiv.org/abs/astro-ph/9807002} {arXiv:astro-ph/9807002} \BibitemShut
  {NoStop}%
\bibitem [{\citenamefont {Riess}\ \emph {et~al.}(2022)\citenamefont {Riess}
  \emph {et~al.}}]{Riess:2021jrx}%
  \BibitemOpen
  \bibfield  {author} {\bibinfo {author} {\bibfnamefont {A.~G.}\ \bibnamefont
  {Riess}} \emph {et~al.},\ }\href {\doibase 10.3847/2041-8213/ac5c5b}
  {\bibfield  {journal} {\bibinfo  {journal} {Astrophys. J. Lett.}\ }\textbf
  {\bibinfo {volume} {934}},\ \bibinfo {pages} {L7} (\bibinfo {year} {2022})},\
  \Eprint {http://arxiv.org/abs/2112.04510} {arXiv:2112.04510 [astro-ph.CO]}
  \BibitemShut {NoStop}%
\bibitem [{\citenamefont {Riess}\ \emph {et~al.}(2021)\citenamefont {Riess},
  \citenamefont {Casertano}, \citenamefont {Yuan}, \citenamefont {Bowers},
  \citenamefont {Macri}, \citenamefont {Zinn},\ and\ \citenamefont
  {Scolnic}}]{Riess:2020fzl}%
  \BibitemOpen
  \bibfield  {author} {\bibinfo {author} {\bibfnamefont {A.~G.}\ \bibnamefont
  {Riess}}, \bibinfo {author} {\bibfnamefont {S.}~\bibnamefont {Casertano}},
  \bibinfo {author} {\bibfnamefont {W.}~\bibnamefont {Yuan}}, \bibinfo {author}
  {\bibfnamefont {J.~B.}\ \bibnamefont {Bowers}}, \bibinfo {author}
  {\bibfnamefont {L.}~\bibnamefont {Macri}}, \bibinfo {author} {\bibfnamefont
  {J.~C.}\ \bibnamefont {Zinn}}, \ and\ \bibinfo {author} {\bibfnamefont
  {D.}~\bibnamefont {Scolnic}},\ }\href {\doibase 10.3847/2041-8213/abdbaf}
  {\bibfield  {journal} {\bibinfo  {journal} {Astrophys. J. Lett.}\ }\textbf
  {\bibinfo {volume} {908}},\ \bibinfo {pages} {L6} (\bibinfo {year} {2021})},\
  \Eprint {http://arxiv.org/abs/2012.08534} {arXiv:2012.08534 [astro-ph.CO]}
  \BibitemShut {NoStop}%
\bibitem [{\citenamefont {Aghanim}\ \emph
  {et~al.}(2020{\natexlab{a}})\citenamefont {Aghanim} \emph
  {et~al.}}]{Aghanim:2018eyx}%
  \BibitemOpen
  \bibfield  {author} {\bibinfo {author} {\bibfnamefont {N.}~\bibnamefont
  {Aghanim}} \emph {et~al.} (\bibinfo {collaboration} {Planck}),\ }\href
  {\doibase 10.1051/0004-6361/201833910} {\bibfield  {journal} {\bibinfo
  {journal} {Astron. Astrophys.}\ }\textbf {\bibinfo {volume} {641}},\ \bibinfo
  {pages} {A6} (\bibinfo {year} {2020}{\natexlab{a}})},\ \Eprint
  {http://arxiv.org/abs/1807.06209} {arXiv:1807.06209 [astro-ph.CO]}
  \BibitemShut {NoStop}%
\bibitem [{\citenamefont {Di~Valentino}\ \emph
  {et~al.}(2021{\natexlab{a}})\citenamefont {Di~Valentino} \emph
  {et~al.}}]{DiValentino:2020zio}%
  \BibitemOpen
  \bibfield  {author} {\bibinfo {author} {\bibfnamefont {E.}~\bibnamefont
  {Di~Valentino}} \emph {et~al.},\ }\href {\doibase
  10.1016/j.astropartphys.2021.102605} {\bibfield  {journal} {\bibinfo
  {journal} {Astropart. Phys.}\ }\textbf {\bibinfo {volume} {131}},\ \bibinfo
  {pages} {102605} (\bibinfo {year} {2021}{\natexlab{a}})},\ \Eprint
  {http://arxiv.org/abs/2008.11284} {arXiv:2008.11284 [astro-ph.CO]}
  \BibitemShut {NoStop}%
\bibitem [{\citenamefont {Di~Valentino}\ \emph
  {et~al.}(2021{\natexlab{b}})\citenamefont {Di~Valentino} \emph
  {et~al.}}]{DiValentino:2020vvd}%
  \BibitemOpen
  \bibfield  {author} {\bibinfo {author} {\bibfnamefont {E.}~\bibnamefont
  {Di~Valentino}} \emph {et~al.},\ }\href {\doibase
  10.1016/j.astropartphys.2021.102604} {\bibfield  {journal} {\bibinfo
  {journal} {Astropart. Phys.}\ }\textbf {\bibinfo {volume} {131}},\ \bibinfo
  {pages} {102604} (\bibinfo {year} {2021}{\natexlab{b}})},\ \Eprint
  {http://arxiv.org/abs/2008.11285} {arXiv:2008.11285 [astro-ph.CO]}
  \BibitemShut {NoStop}%
\bibitem [{\citenamefont {Abbott}\ \emph {et~al.}(2022)\citenamefont {Abbott}
  \emph {et~al.}}]{DES:2021wwk}%
  \BibitemOpen
  \bibfield  {author} {\bibinfo {author} {\bibfnamefont {T.~M.~C.}\
  \bibnamefont {Abbott}} \emph {et~al.} (\bibinfo {collaboration} {DES}),\
  }\href {\doibase 10.1103/PhysRevD.105.023520} {\bibfield  {journal} {\bibinfo
   {journal} {Phys. Rev. D}\ }\textbf {\bibinfo {volume} {105}},\ \bibinfo
  {pages} {023520} (\bibinfo {year} {2022})},\ \Eprint
  {http://arxiv.org/abs/2105.13549} {arXiv:2105.13549 [astro-ph.CO]}
  \BibitemShut {NoStop}%
\bibitem [{\citenamefont {Heymans}\ \emph {et~al.}(2021)\citenamefont {Heymans}
  \emph {et~al.}}]{Heymans:2020gsg}%
  \BibitemOpen
  \bibfield  {author} {\bibinfo {author} {\bibfnamefont {C.}~\bibnamefont
  {Heymans}} \emph {et~al.},\ }\href {\doibase 10.1051/0004-6361/202039063}
  {\bibfield  {journal} {\bibinfo  {journal} {Astron. Astrophys.}\ }\textbf
  {\bibinfo {volume} {646}},\ \bibinfo {pages} {A140} (\bibinfo {year}
  {2021})},\ \Eprint {http://arxiv.org/abs/2007.15632} {arXiv:2007.15632
  [astro-ph.CO]} \BibitemShut {NoStop}%
\bibitem [{\citenamefont {Loureiro}\ \emph {et~al.}(2022)\citenamefont
  {Loureiro} \emph {et~al.}}]{KiDS:2021opn}%
  \BibitemOpen
  \bibfield  {author} {\bibinfo {author} {\bibfnamefont {A.}~\bibnamefont
  {Loureiro}} \emph {et~al.} (\bibinfo {collaboration} {KiDS, Euclid}),\ }\href
  {\doibase 10.1051/0004-6361/202142481} {\bibfield  {journal} {\bibinfo
  {journal} {Astron. Astrophys.}\ }\textbf {\bibinfo {volume} {665}},\ \bibinfo
  {pages} {A56} (\bibinfo {year} {2022})},\ \Eprint
  {http://arxiv.org/abs/2110.06947} {arXiv:2110.06947 [astro-ph.CO]}
  \BibitemShut {NoStop}%
\bibitem [{\citenamefont {Abdalla}\ \emph {et~al.}(2022)\citenamefont {Abdalla}
  \emph {et~al.}}]{Abdalla:2022yfr}%
  \BibitemOpen
  \bibfield  {author} {\bibinfo {author} {\bibfnamefont {E.}~\bibnamefont
  {Abdalla}} \emph {et~al.},\ }\href {\doibase 10.1016/j.jheap.2022.04.002}
  {\bibfield  {journal} {\bibinfo  {journal} {JHEAp}\ }\textbf {\bibinfo
  {volume} {34}},\ \bibinfo {pages} {49} (\bibinfo {year} {2022})},\ \Eprint
  {http://arxiv.org/abs/2203.06142} {arXiv:2203.06142 [astro-ph.CO]}
  \BibitemShut {NoStop}%
\bibitem [{\citenamefont {Verde}\ \emph {et~al.}(2019)\citenamefont {Verde},
  \citenamefont {Treu},\ and\ \citenamefont {Riess}}]{Verde:2019ivm}%
  \BibitemOpen
  \bibfield  {author} {\bibinfo {author} {\bibfnamefont {L.}~\bibnamefont
  {Verde}}, \bibinfo {author} {\bibfnamefont {T.}~\bibnamefont {Treu}}, \ and\
  \bibinfo {author} {\bibfnamefont {A.~G.}\ \bibnamefont {Riess}},\ }\href
  {\doibase 10.1038/s41550-019-0902-0} {\bibfield  {journal} {\bibinfo
  {journal} {Nature Astron.}\ }\textbf {\bibinfo {volume} {3}},\ \bibinfo
  {pages} {891} (\bibinfo {year} {2019})},\ \Eprint
  {http://arxiv.org/abs/1907.10625} {arXiv:1907.10625 [astro-ph.CO]}
  \BibitemShut {NoStop}%
\bibitem [{\citenamefont {Riess}(2019)}]{Riess:2019qba}%
  \BibitemOpen
  \bibfield  {author} {\bibinfo {author} {\bibfnamefont {A.~G.}\ \bibnamefont
  {Riess}},\ }\href {\doibase 10.1038/s42254-019-0137-0} {\bibfield  {journal}
  {\bibinfo  {journal} {Nature Rev. Phys.}\ }\textbf {\bibinfo {volume} {2}},\
  \bibinfo {pages} {10} (\bibinfo {year} {2019})},\ \Eprint
  {http://arxiv.org/abs/2001.03624} {arXiv:2001.03624 [astro-ph.CO]}
  \BibitemShut {NoStop}%
\bibitem [{\citenamefont {Di~Valentino}(2021)}]{DiValentino:2020vnx}%
  \BibitemOpen
  \bibfield  {author} {\bibinfo {author} {\bibfnamefont {E.}~\bibnamefont
  {Di~Valentino}},\ }\href {\doibase 10.1093/mnras/stab187} {\bibfield
  {journal} {\bibinfo  {journal} {Mon. Not. Roy. Astron. Soc.}\ }\textbf
  {\bibinfo {volume} {502}},\ \bibinfo {pages} {2065} (\bibinfo {year}
  {2021})},\ \Eprint {http://arxiv.org/abs/2011.00246} {arXiv:2011.00246
  [astro-ph.CO]} \BibitemShut {NoStop}%
\bibitem [{\citenamefont {Di~Valentino}\ \emph
  {et~al.}(2021{\natexlab{c}})\citenamefont {Di~Valentino}, \citenamefont
  {Mena}, \citenamefont {Pan}, \citenamefont {Visinelli}, \citenamefont {Yang},
  \citenamefont {Melchiorri}, \citenamefont {Mota}, \citenamefont {Riess},\
  and\ \citenamefont {Silk}}]{DiValentino:2021izs}%
  \BibitemOpen
  \bibfield  {author} {\bibinfo {author} {\bibfnamefont {E.}~\bibnamefont
  {Di~Valentino}}, \bibinfo {author} {\bibfnamefont {O.}~\bibnamefont {Mena}},
  \bibinfo {author} {\bibfnamefont {S.}~\bibnamefont {Pan}}, \bibinfo {author}
  {\bibfnamefont {L.}~\bibnamefont {Visinelli}}, \bibinfo {author}
  {\bibfnamefont {W.}~\bibnamefont {Yang}}, \bibinfo {author} {\bibfnamefont
  {A.}~\bibnamefont {Melchiorri}}, \bibinfo {author} {\bibfnamefont {D.~F.}\
  \bibnamefont {Mota}}, \bibinfo {author} {\bibfnamefont {A.~G.}\ \bibnamefont
  {Riess}}, \ and\ \bibinfo {author} {\bibfnamefont {J.}~\bibnamefont {Silk}},\
  }\href {\doibase 10.1088/1361-6382/ac086d} {\bibfield  {journal} {\bibinfo
  {journal} {Class. Quant. Grav.}\ }\textbf {\bibinfo {volume} {38}},\ \bibinfo
  {pages} {153001} (\bibinfo {year} {2021}{\natexlab{c}})},\ \Eprint
  {http://arxiv.org/abs/2103.01183} {arXiv:2103.01183 [astro-ph.CO]}
  \BibitemShut {NoStop}%
\bibitem [{\citenamefont {Perivolaropoulos}\ and\ \citenamefont
  {Skara}(2022)}]{Perivolaropoulos:2021jda}%
  \BibitemOpen
  \bibfield  {author} {\bibinfo {author} {\bibfnamefont {L.}~\bibnamefont
  {Perivolaropoulos}}\ and\ \bibinfo {author} {\bibfnamefont {F.}~\bibnamefont
  {Skara}},\ }\href {\doibase 10.1016/j.newar.2022.101659} {\bibfield
  {journal} {\bibinfo  {journal} {New Astron. Rev.}\ }\textbf {\bibinfo
  {volume} {95}},\ \bibinfo {pages} {101659} (\bibinfo {year} {2022})},\
  \Eprint {http://arxiv.org/abs/2105.05208} {arXiv:2105.05208 [astro-ph.CO]}
  \BibitemShut {NoStop}%
\bibitem [{\citenamefont {Sch\"oneberg}\ \emph {et~al.}(2022)\citenamefont
  {Sch\"oneberg}, \citenamefont {Franco~Abell\'an}, \citenamefont
  {P\'erez~S\'anchez}, \citenamefont {Witte}, \citenamefont {Poulin},\ and\
  \citenamefont {Lesgourgues}}]{Schoneberg:2021qvd}%
  \BibitemOpen
  \bibfield  {author} {\bibinfo {author} {\bibfnamefont {N.}~\bibnamefont
  {Sch\"oneberg}}, \bibinfo {author} {\bibfnamefont {G.}~\bibnamefont
  {Franco~Abell\'an}}, \bibinfo {author} {\bibfnamefont {A.}~\bibnamefont
  {P\'erez~S\'anchez}}, \bibinfo {author} {\bibfnamefont {S.~J.}\ \bibnamefont
  {Witte}}, \bibinfo {author} {\bibfnamefont {V.}~\bibnamefont {Poulin}}, \
  and\ \bibinfo {author} {\bibfnamefont {J.}~\bibnamefont {Lesgourgues}},\
  }\href {\doibase 10.1016/j.physrep.2022.07.001} {\bibfield  {journal}
  {\bibinfo  {journal} {Phys. Rept.}\ }\textbf {\bibinfo {volume} {984}},\
  \bibinfo {pages} {1} (\bibinfo {year} {2022})},\ \Eprint
  {http://arxiv.org/abs/2107.10291} {arXiv:2107.10291 [astro-ph.CO]}
  \BibitemShut {NoStop}%
\bibitem [{\citenamefont {Amendola}(2000)}]{Amendola:1999er}%
  \BibitemOpen
  \bibfield  {author} {\bibinfo {author} {\bibfnamefont {L.}~\bibnamefont
  {Amendola}},\ }\href {\doibase 10.1103/PhysRevD.62.043511} {\bibfield
  {journal} {\bibinfo  {journal} {Phys. Rev. D}\ }\textbf {\bibinfo {volume}
  {62}},\ \bibinfo {pages} {043511} (\bibinfo {year} {2000})},\ \Eprint
  {http://arxiv.org/abs/astro-ph/9908023} {arXiv:astro-ph/9908023} \BibitemShut
  {NoStop}%
\bibitem [{\citenamefont {Amendola}\ and\ \citenamefont
  {Quercellini}(2003)}]{Amendola:2003eq}%
  \BibitemOpen
  \bibfield  {author} {\bibinfo {author} {\bibfnamefont {L.}~\bibnamefont
  {Amendola}}\ and\ \bibinfo {author} {\bibfnamefont {C.}~\bibnamefont
  {Quercellini}},\ }\href {\doibase 10.1103/PhysRevD.68.023514} {\bibfield
  {journal} {\bibinfo  {journal} {Phys. Rev. D}\ }\textbf {\bibinfo {volume}
  {68}},\ \bibinfo {pages} {023514} (\bibinfo {year} {2003})},\ \Eprint
  {http://arxiv.org/abs/astro-ph/0303228} {arXiv:astro-ph/0303228} \BibitemShut
  {NoStop}%
\bibitem [{\citenamefont {Guo}\ \emph {et~al.}(2005)\citenamefont {Guo},
  \citenamefont {Cai},\ and\ \citenamefont {Zhang}}]{Guo:2004xx}%
  \BibitemOpen
  \bibfield  {author} {\bibinfo {author} {\bibfnamefont {Z.-K.}\ \bibnamefont
  {Guo}}, \bibinfo {author} {\bibfnamefont {R.-G.}\ \bibnamefont {Cai}}, \ and\
  \bibinfo {author} {\bibfnamefont {Y.-Z.}\ \bibnamefont {Zhang}},\ }\href
  {\doibase 10.1088/1475-7516/2005/05/002} {\bibfield  {journal} {\bibinfo
  {journal} {JCAP}\ }\textbf {\bibinfo {volume} {05}},\ \bibinfo {pages} {002}
  (\bibinfo {year} {2005})},\ \Eprint {http://arxiv.org/abs/astro-ph/0412624}
  {arXiv:astro-ph/0412624} \BibitemShut {NoStop}%
\bibitem [{\citenamefont {Cai}\ and\ \citenamefont {Wang}(2005)}]{Cai:2004dk}%
  \BibitemOpen
  \bibfield  {author} {\bibinfo {author} {\bibfnamefont {R.-G.}\ \bibnamefont
  {Cai}}\ and\ \bibinfo {author} {\bibfnamefont {A.}~\bibnamefont {Wang}},\
  }\href {\doibase 10.1088/1475-7516/2005/03/002} {\bibfield  {journal}
  {\bibinfo  {journal} {JCAP}\ }\textbf {\bibinfo {volume} {03}},\ \bibinfo
  {pages} {002} (\bibinfo {year} {2005})},\ \Eprint
  {http://arxiv.org/abs/hep-th/0411025} {arXiv:hep-th/0411025} \BibitemShut
  {NoStop}%
\bibitem [{\citenamefont {Wang}\ \emph {et~al.}(2006)\citenamefont {Wang},
  \citenamefont {Lin},\ and\ \citenamefont {Abdalla}}]{Wang:2005ph}%
  \BibitemOpen
  \bibfield  {author} {\bibinfo {author} {\bibfnamefont {B.}~\bibnamefont
  {Wang}}, \bibinfo {author} {\bibfnamefont {C.-Y.}\ \bibnamefont {Lin}}, \
  and\ \bibinfo {author} {\bibfnamefont {E.}~\bibnamefont {Abdalla}},\ }\href
  {\doibase 10.1016/j.physletb.2006.04.009} {\bibfield  {journal} {\bibinfo
  {journal} {Phys. Lett. B}\ }\textbf {\bibinfo {volume} {637}},\ \bibinfo
  {pages} {357} (\bibinfo {year} {2006})},\ \Eprint
  {http://arxiv.org/abs/hep-th/0509107} {arXiv:hep-th/0509107} \BibitemShut
  {NoStop}%
\bibitem [{\citenamefont {Barrow}\ and\ \citenamefont
  {Clifton}(2006)}]{Barrow:2006hia}%
  \BibitemOpen
  \bibfield  {author} {\bibinfo {author} {\bibfnamefont {J.~D.}\ \bibnamefont
  {Barrow}}\ and\ \bibinfo {author} {\bibfnamefont {T.}~\bibnamefont
  {Clifton}},\ }\href {\doibase 10.1103/PhysRevD.73.103520} {\bibfield
  {journal} {\bibinfo  {journal} {Phys. Rev. D}\ }\textbf {\bibinfo {volume}
  {73}},\ \bibinfo {pages} {103520} (\bibinfo {year} {2006})},\ \Eprint
  {http://arxiv.org/abs/gr-qc/0604063} {arXiv:gr-qc/0604063} \BibitemShut
  {NoStop}%
\bibitem [{\citenamefont {Setare}(2007)}]{Setare:2007at}%
  \BibitemOpen
  \bibfield  {author} {\bibinfo {author} {\bibfnamefont {M.~R.}\ \bibnamefont
  {Setare}},\ }\href {\doibase 10.1088/1475-7516/2007/01/023} {\bibfield
  {journal} {\bibinfo  {journal} {JCAP}\ }\textbf {\bibinfo {volume} {01}},\
  \bibinfo {pages} {023} (\bibinfo {year} {2007})},\ \Eprint
  {http://arxiv.org/abs/hep-th/0701242} {arXiv:hep-th/0701242} \BibitemShut
  {NoStop}%
\bibitem [{\citenamefont {Feng}\ \emph {et~al.}(2007)\citenamefont {Feng},
  \citenamefont {Wang}, \citenamefont {Gong},\ and\ \citenamefont
  {Su}}]{Feng:2007wn}%
  \BibitemOpen
  \bibfield  {author} {\bibinfo {author} {\bibfnamefont {C.}~\bibnamefont
  {Feng}}, \bibinfo {author} {\bibfnamefont {B.}~\bibnamefont {Wang}}, \bibinfo
  {author} {\bibfnamefont {Y.}~\bibnamefont {Gong}}, \ and\ \bibinfo {author}
  {\bibfnamefont {R.-K.}\ \bibnamefont {Su}},\ }\href {\doibase
  10.1088/1475-7516/2007/09/005} {\bibfield  {journal} {\bibinfo  {journal}
  {JCAP}\ }\textbf {\bibinfo {volume} {09}},\ \bibinfo {pages} {005} (\bibinfo
  {year} {2007})},\ \Eprint {http://arxiv.org/abs/0706.4033} {arXiv:0706.4033
  [astro-ph]} \BibitemShut {NoStop}%
\bibitem [{\citenamefont {Boehmer}\ \emph {et~al.}(2008)\citenamefont
  {Boehmer}, \citenamefont {Caldera-Cabral}, \citenamefont {Lazkoz},\ and\
  \citenamefont {Maartens}}]{Boehmer:2008av}%
  \BibitemOpen
  \bibfield  {author} {\bibinfo {author} {\bibfnamefont {C.~G.}\ \bibnamefont
  {Boehmer}}, \bibinfo {author} {\bibfnamefont {G.}~\bibnamefont
  {Caldera-Cabral}}, \bibinfo {author} {\bibfnamefont {R.}~\bibnamefont
  {Lazkoz}}, \ and\ \bibinfo {author} {\bibfnamefont {R.}~\bibnamefont
  {Maartens}},\ }\href {\doibase 10.1103/PhysRevD.78.023505} {\bibfield
  {journal} {\bibinfo  {journal} {Phys. Rev. D}\ }\textbf {\bibinfo {volume}
  {78}},\ \bibinfo {pages} {023505} (\bibinfo {year} {2008})},\ \Eprint
  {http://arxiv.org/abs/0801.1565} {arXiv:0801.1565 [gr-qc]} \BibitemShut
  {NoStop}%
\bibitem [{\citenamefont {Olivares}\ \emph {et~al.}(2008)\citenamefont
  {Olivares}, \citenamefont {Atrio-Barandela},\ and\ \citenamefont
  {Pavon}}]{Olivares:2008bx}%
  \BibitemOpen
  \bibfield  {author} {\bibinfo {author} {\bibfnamefont {G.}~\bibnamefont
  {Olivares}}, \bibinfo {author} {\bibfnamefont {F.}~\bibnamefont
  {Atrio-Barandela}}, \ and\ \bibinfo {author} {\bibfnamefont {D.}~\bibnamefont
  {Pavon}},\ }\href {\doibase 10.1103/PhysRevD.77.103520} {\bibfield  {journal}
  {\bibinfo  {journal} {Phys. Rev. D}\ }\textbf {\bibinfo {volume} {77}},\
  \bibinfo {pages} {103520} (\bibinfo {year} {2008})},\ \Eprint
  {http://arxiv.org/abs/0801.4517} {arXiv:0801.4517 [astro-ph]} \BibitemShut
  {NoStop}%
\bibitem [{\citenamefont {Valiviita}\ \emph {et~al.}(2008)\citenamefont
  {Valiviita}, \citenamefont {Majerotto},\ and\ \citenamefont
  {Maartens}}]{Valiviita:2008iv}%
  \BibitemOpen
  \bibfield  {author} {\bibinfo {author} {\bibfnamefont {J.}~\bibnamefont
  {Valiviita}}, \bibinfo {author} {\bibfnamefont {E.}~\bibnamefont
  {Majerotto}}, \ and\ \bibinfo {author} {\bibfnamefont {R.}~\bibnamefont
  {Maartens}},\ }\href {\doibase 10.1088/1475-7516/2008/07/020} {\bibfield
  {journal} {\bibinfo  {journal} {JCAP}\ }\textbf {\bibinfo {volume} {07}},\
  \bibinfo {pages} {020} (\bibinfo {year} {2008})},\ \Eprint
  {http://arxiv.org/abs/0804.0232} {arXiv:0804.0232 [astro-ph]} \BibitemShut
  {NoStop}%
\bibitem [{\citenamefont {Wang}\ and\ \citenamefont
  {Zhang}(2008)}]{Wang:2008te}%
  \BibitemOpen
  \bibfield  {author} {\bibinfo {author} {\bibfnamefont {S.}~\bibnamefont
  {Wang}}\ and\ \bibinfo {author} {\bibfnamefont {Y.}~\bibnamefont {Zhang}},\
  }\href {\doibase 10.1016/j.physletb.2008.09.055} {\bibfield  {journal}
  {\bibinfo  {journal} {Phys. Lett. B}\ }\textbf {\bibinfo {volume} {669}},\
  \bibinfo {pages} {201} (\bibinfo {year} {2008})},\ \Eprint
  {http://arxiv.org/abs/0809.3627} {arXiv:0809.3627 [astro-ph]} \BibitemShut
  {NoStop}%
\bibitem [{\citenamefont {Gavela}\ \emph {et~al.}(2009)\citenamefont {Gavela},
  \citenamefont {Hernandez}, \citenamefont {Lopez~Honorez}, \citenamefont
  {Mena},\ and\ \citenamefont {Rigolin}}]{Gavela:2009cy}%
  \BibitemOpen
  \bibfield  {author} {\bibinfo {author} {\bibfnamefont {M.~B.}\ \bibnamefont
  {Gavela}}, \bibinfo {author} {\bibfnamefont {D.}~\bibnamefont {Hernandez}},
  \bibinfo {author} {\bibfnamefont {L.}~\bibnamefont {Lopez~Honorez}}, \bibinfo
  {author} {\bibfnamefont {O.}~\bibnamefont {Mena}}, \ and\ \bibinfo {author}
  {\bibfnamefont {S.}~\bibnamefont {Rigolin}},\ }\href {\doibase
  10.1088/1475-7516/2009/07/034} {\bibfield  {journal} {\bibinfo  {journal}
  {JCAP}\ }\textbf {\bibinfo {volume} {07}},\ \bibinfo {pages} {034} (\bibinfo
  {year} {2009})},\ \bibinfo {note} {[Erratum: JCAP 05, E01 (2010)]},\ \Eprint
  {http://arxiv.org/abs/0901.1611} {arXiv:0901.1611 [astro-ph.CO]} \BibitemShut
  {NoStop}%
\bibitem [{\citenamefont {Valiviita}\ \emph {et~al.}(2010)\citenamefont
  {Valiviita}, \citenamefont {Maartens},\ and\ \citenamefont
  {Majerotto}}]{Valiviita:2009nu}%
  \BibitemOpen
  \bibfield  {author} {\bibinfo {author} {\bibfnamefont {J.}~\bibnamefont
  {Valiviita}}, \bibinfo {author} {\bibfnamefont {R.}~\bibnamefont {Maartens}},
  \ and\ \bibinfo {author} {\bibfnamefont {E.}~\bibnamefont {Majerotto}},\
  }\href {\doibase 10.1111/j.1365-2966.2009.16115.x} {\bibfield  {journal}
  {\bibinfo  {journal} {Mon. Not. Roy. Astron. Soc.}\ }\textbf {\bibinfo
  {volume} {402}},\ \bibinfo {pages} {2355} (\bibinfo {year} {2010})},\ \Eprint
  {http://arxiv.org/abs/0907.4987} {arXiv:0907.4987 [astro-ph.CO]} \BibitemShut
  {NoStop}%
\bibitem [{\citenamefont {Lopez~Honorez}\ \emph {et~al.}(2010)\citenamefont
  {Lopez~Honorez}, \citenamefont {Reid}, \citenamefont {Mena}, \citenamefont
  {Verde},\ and\ \citenamefont {Jimenez}}]{LopezHonorez:2010esq}%
  \BibitemOpen
  \bibfield  {author} {\bibinfo {author} {\bibfnamefont {L.}~\bibnamefont
  {Lopez~Honorez}}, \bibinfo {author} {\bibfnamefont {B.~A.}\ \bibnamefont
  {Reid}}, \bibinfo {author} {\bibfnamefont {O.}~\bibnamefont {Mena}}, \bibinfo
  {author} {\bibfnamefont {L.}~\bibnamefont {Verde}}, \ and\ \bibinfo {author}
  {\bibfnamefont {R.}~\bibnamefont {Jimenez}},\ }\href {\doibase
  10.1088/1475-7516/2010/09/029} {\bibfield  {journal} {\bibinfo  {journal}
  {JCAP}\ }\textbf {\bibinfo {volume} {09}},\ \bibinfo {pages} {029} (\bibinfo
  {year} {2010})},\ \Eprint {http://arxiv.org/abs/1006.0877} {arXiv:1006.0877
  [astro-ph.CO]} \BibitemShut {NoStop}%
\bibitem [{\citenamefont {Chen}\ \emph {et~al.}(2011)\citenamefont {Chen},
  \citenamefont {Wang}, \citenamefont {Pan},\ and\ \citenamefont
  {Gong}}]{Chen:2010ws}%
  \BibitemOpen
  \bibfield  {author} {\bibinfo {author} {\bibfnamefont {X.}~\bibnamefont
  {Chen}}, \bibinfo {author} {\bibfnamefont {B.}~\bibnamefont {Wang}}, \bibinfo
  {author} {\bibfnamefont {N.}~\bibnamefont {Pan}}, \ and\ \bibinfo {author}
  {\bibfnamefont {Y.}~\bibnamefont {Gong}},\ }\href {\doibase
  10.1016/j.physletb.2010.11.032} {\bibfield  {journal} {\bibinfo  {journal}
  {Phys. Lett. B}\ }\textbf {\bibinfo {volume} {695}},\ \bibinfo {pages} {30}
  (\bibinfo {year} {2011})},\ \Eprint {http://arxiv.org/abs/1008.3455}
  {arXiv:1008.3455 [astro-ph.CO]} \BibitemShut {NoStop}%
\bibitem [{\citenamefont {Cao}\ \emph {et~al.}(2011)\citenamefont {Cao},
  \citenamefont {Liang},\ and\ \citenamefont {Zhu}}]{Cao:2010fb}%
  \BibitemOpen
  \bibfield  {author} {\bibinfo {author} {\bibfnamefont {S.}~\bibnamefont
  {Cao}}, \bibinfo {author} {\bibfnamefont {N.}~\bibnamefont {Liang}}, \ and\
  \bibinfo {author} {\bibfnamefont {Z.-H.}\ \bibnamefont {Zhu}},\ }\href
  {\doibase 10.1111/j.1365-2966.2011.19105.x} {\bibfield  {journal} {\bibinfo
  {journal} {Mon. Not. Roy. Astron. Soc.}\ }\textbf {\bibinfo {volume} {416}},\
  \bibinfo {pages} {1099} (\bibinfo {year} {2011})},\ \Eprint
  {http://arxiv.org/abs/1012.4879} {arXiv:1012.4879 [astro-ph.CO]} \BibitemShut
  {NoStop}%
\bibitem [{\citenamefont {Clemson}\ \emph {et~al.}(2012)\citenamefont
  {Clemson}, \citenamefont {Koyama}, \citenamefont {Zhao}, \citenamefont
  {Maartens},\ and\ \citenamefont {Valiviita}}]{Clemson:2011an}%
  \BibitemOpen
  \bibfield  {author} {\bibinfo {author} {\bibfnamefont {T.}~\bibnamefont
  {Clemson}}, \bibinfo {author} {\bibfnamefont {K.}~\bibnamefont {Koyama}},
  \bibinfo {author} {\bibfnamefont {G.-B.}\ \bibnamefont {Zhao}}, \bibinfo
  {author} {\bibfnamefont {R.}~\bibnamefont {Maartens}}, \ and\ \bibinfo
  {author} {\bibfnamefont {J.}~\bibnamefont {Valiviita}},\ }\href {\doibase
  10.1103/PhysRevD.85.043007} {\bibfield  {journal} {\bibinfo  {journal} {Phys.
  Rev. D}\ }\textbf {\bibinfo {volume} {85}},\ \bibinfo {pages} {043007}
  (\bibinfo {year} {2012})},\ \Eprint {http://arxiv.org/abs/1109.6234}
  {arXiv:1109.6234 [astro-ph.CO]} \BibitemShut {NoStop}%
\bibitem [{\citenamefont {He}\ \emph {et~al.}(2011)\citenamefont {He},
  \citenamefont {Wang},\ and\ \citenamefont {Abdalla}}]{He:2011qn}%
  \BibitemOpen
  \bibfield  {author} {\bibinfo {author} {\bibfnamefont {J.-H.}\ \bibnamefont
  {He}}, \bibinfo {author} {\bibfnamefont {B.}~\bibnamefont {Wang}}, \ and\
  \bibinfo {author} {\bibfnamefont {E.}~\bibnamefont {Abdalla}},\ }\href
  {\doibase 10.1103/PhysRevD.84.123526} {\bibfield  {journal} {\bibinfo
  {journal} {Phys. Rev. D}\ }\textbf {\bibinfo {volume} {84}},\ \bibinfo
  {pages} {123526} (\bibinfo {year} {2011})},\ \Eprint
  {http://arxiv.org/abs/1109.1730} {arXiv:1109.1730 [gr-qc]} \BibitemShut
  {NoStop}%
\bibitem [{\citenamefont {Harko}\ and\ \citenamefont
  {Lobo}(2013)}]{Harko:2012za}%
  \BibitemOpen
  \bibfield  {author} {\bibinfo {author} {\bibfnamefont {T.}~\bibnamefont
  {Harko}}\ and\ \bibinfo {author} {\bibfnamefont {F.~S.~N.}\ \bibnamefont
  {Lobo}},\ }\href {\doibase 10.1103/PhysRevD.87.044018} {\bibfield  {journal}
  {\bibinfo  {journal} {Phys. Rev. D}\ }\textbf {\bibinfo {volume} {87}},\
  \bibinfo {pages} {044018} (\bibinfo {year} {2013})},\ \Eprint
  {http://arxiv.org/abs/1210.3617} {arXiv:1210.3617 [gr-qc]} \BibitemShut
  {NoStop}%
\bibitem [{\citenamefont {Sun}\ and\ \citenamefont {Yue}(2013)}]{Sun:2013pda}%
  \BibitemOpen
  \bibfield  {author} {\bibinfo {author} {\bibfnamefont {C.-Y.}\ \bibnamefont
  {Sun}}\ and\ \bibinfo {author} {\bibfnamefont {R.-H.}\ \bibnamefont {Yue}},\
  }\href {\doibase 10.1088/1475-7516/2013/08/018} {\bibfield  {journal}
  {\bibinfo  {journal} {JCAP}\ }\textbf {\bibinfo {volume} {08}},\ \bibinfo
  {pages} {018} (\bibinfo {year} {2013})},\ \Eprint
  {http://arxiv.org/abs/1303.0684} {arXiv:1303.0684 [astro-ph.CO]} \BibitemShut
  {NoStop}%
\bibitem [{\citenamefont {Chimento}\ \emph {et~al.}(2013)\citenamefont
  {Chimento}, \citenamefont {Richarte},\ and\ \citenamefont
  {S\'anchez~Garc\'\i{}a}}]{Chimento:2013rya}%
  \BibitemOpen
  \bibfield  {author} {\bibinfo {author} {\bibfnamefont {L.~P.}\ \bibnamefont
  {Chimento}}, \bibinfo {author} {\bibfnamefont {M.~G.}\ \bibnamefont
  {Richarte}}, \ and\ \bibinfo {author} {\bibfnamefont {I.~E.}\ \bibnamefont
  {S\'anchez~Garc\'\i{}a}},\ }\href {\doibase 10.1103/PhysRevD.88.087301}
  {\bibfield  {journal} {\bibinfo  {journal} {Phys. Rev. D}\ }\textbf {\bibinfo
  {volume} {88}},\ \bibinfo {pages} {087301} (\bibinfo {year} {2013})},\
  \Eprint {http://arxiv.org/abs/1310.5335} {arXiv:1310.5335 [gr-qc]}
  \BibitemShut {NoStop}%
\bibitem [{\citenamefont {Li}\ and\ \citenamefont {Zhang}(2014)}]{Li:2013bya}%
  \BibitemOpen
  \bibfield  {author} {\bibinfo {author} {\bibfnamefont {Y.-H.}\ \bibnamefont
  {Li}}\ and\ \bibinfo {author} {\bibfnamefont {X.}~\bibnamefont {Zhang}},\
  }\href {\doibase 10.1103/PhysRevD.89.083009} {\bibfield  {journal} {\bibinfo
  {journal} {Phys. Rev. D}\ }\textbf {\bibinfo {volume} {89}},\ \bibinfo
  {pages} {083009} (\bibinfo {year} {2014})},\ \Eprint
  {http://arxiv.org/abs/1312.6328} {arXiv:1312.6328 [astro-ph.CO]} \BibitemShut
  {NoStop}%
\bibitem [{\citenamefont {Yang}\ and\ \citenamefont
  {Xu}(2014{\natexlab{a}})}]{Yang:2014gza}%
  \BibitemOpen
  \bibfield  {author} {\bibinfo {author} {\bibfnamefont {W.}~\bibnamefont
  {Yang}}\ and\ \bibinfo {author} {\bibfnamefont {L.}~\bibnamefont {Xu}},\
  }\href {\doibase 10.1103/PhysRevD.89.083517} {\bibfield  {journal} {\bibinfo
  {journal} {Phys. Rev. D}\ }\textbf {\bibinfo {volume} {89}},\ \bibinfo
  {pages} {083517} (\bibinfo {year} {2014}{\natexlab{a}})},\ \Eprint
  {http://arxiv.org/abs/1401.1286} {arXiv:1401.1286 [astro-ph.CO]} \BibitemShut
  {NoStop}%
\bibitem [{\citenamefont {Yang}\ and\ \citenamefont
  {Xu}(2014{\natexlab{b}})}]{Yang:2014okp}%
  \BibitemOpen
  \bibfield  {author} {\bibinfo {author} {\bibfnamefont {W.}~\bibnamefont
  {Yang}}\ and\ \bibinfo {author} {\bibfnamefont {L.}~\bibnamefont {Xu}},\
  }\href {\doibase 10.1088/1475-7516/2014/08/034} {\bibfield  {journal}
  {\bibinfo  {journal} {JCAP}\ }\textbf {\bibinfo {volume} {08}},\ \bibinfo
  {pages} {034} (\bibinfo {year} {2014}{\natexlab{b}})},\ \Eprint
  {http://arxiv.org/abs/1401.5177} {arXiv:1401.5177 [astro-ph.CO]} \BibitemShut
  {NoStop}%
\bibitem [{\citenamefont {Yang}\ and\ \citenamefont
  {Xu}(2014{\natexlab{c}})}]{Yang:2014hea}%
  \BibitemOpen
  \bibfield  {author} {\bibinfo {author} {\bibfnamefont {W.}~\bibnamefont
  {Yang}}\ and\ \bibinfo {author} {\bibfnamefont {L.}~\bibnamefont {Xu}},\
  }\href {\doibase 10.1103/PhysRevD.90.083532} {\bibfield  {journal} {\bibinfo
  {journal} {Phys. Rev. D}\ }\textbf {\bibinfo {volume} {90}},\ \bibinfo
  {pages} {083532} (\bibinfo {year} {2014}{\natexlab{c}})},\ \Eprint
  {http://arxiv.org/abs/1409.5533} {arXiv:1409.5533 [astro-ph.CO]} \BibitemShut
  {NoStop}%
\bibitem [{\citenamefont {Li}\ \emph {et~al.}(2014)\citenamefont {Li},
  \citenamefont {Zhang},\ and\ \citenamefont {Zhang}}]{Li:2014eha}%
  \BibitemOpen
  \bibfield  {author} {\bibinfo {author} {\bibfnamefont {Y.-H.}\ \bibnamefont
  {Li}}, \bibinfo {author} {\bibfnamefont {J.-F.}\ \bibnamefont {Zhang}}, \
  and\ \bibinfo {author} {\bibfnamefont {X.}~\bibnamefont {Zhang}},\ }\href
  {\doibase 10.1103/PhysRevD.90.063005} {\bibfield  {journal} {\bibinfo
  {journal} {Phys. Rev. D}\ }\textbf {\bibinfo {volume} {90}},\ \bibinfo
  {pages} {063005} (\bibinfo {year} {2014})},\ \Eprint
  {http://arxiv.org/abs/1404.5220} {arXiv:1404.5220 [astro-ph.CO]} \BibitemShut
  {NoStop}%
\bibitem [{\citenamefont {Tamanini}(2015)}]{Tamanini:2015iia}%
  \BibitemOpen
  \bibfield  {author} {\bibinfo {author} {\bibfnamefont {N.}~\bibnamefont
  {Tamanini}},\ }\href {\doibase 10.1103/PhysRevD.92.043524} {\bibfield
  {journal} {\bibinfo  {journal} {Phys. Rev. D}\ }\textbf {\bibinfo {volume}
  {92}},\ \bibinfo {pages} {043524} (\bibinfo {year} {2015})},\ \Eprint
  {http://arxiv.org/abs/1504.07397} {arXiv:1504.07397 [gr-qc]} \BibitemShut
  {NoStop}%
\bibitem [{\citenamefont {Goncalves}\ \emph {et~al.}(2015)\citenamefont
  {Goncalves}, \citenamefont {Carvalho},\ and\ \citenamefont
  {Alcaniz}}]{Goncalves:2015eaa}%
  \BibitemOpen
  \bibfield  {author} {\bibinfo {author} {\bibfnamefont {R.~S.}\ \bibnamefont
  {Goncalves}}, \bibinfo {author} {\bibfnamefont {G.~C.}\ \bibnamefont
  {Carvalho}}, \ and\ \bibinfo {author} {\bibfnamefont {J.~S.}\ \bibnamefont
  {Alcaniz}},\ }\href {\doibase 10.1103/PhysRevD.92.123504} {\bibfield
  {journal} {\bibinfo  {journal} {Phys. Rev. D}\ }\textbf {\bibinfo {volume}
  {92}},\ \bibinfo {pages} {123504} (\bibinfo {year} {2015})},\ \Eprint
  {http://arxiv.org/abs/1507.01921} {arXiv:1507.01921 [astro-ph.CO]}
  \BibitemShut {NoStop}%
\bibitem [{\citenamefont {Pan}\ \emph {et~al.}(2015)\citenamefont {Pan},
  \citenamefont {Bhattacharya},\ and\ \citenamefont
  {Chakraborty}}]{Pan:2012ki}%
  \BibitemOpen
  \bibfield  {author} {\bibinfo {author} {\bibfnamefont {S.}~\bibnamefont
  {Pan}}, \bibinfo {author} {\bibfnamefont {S.}~\bibnamefont {Bhattacharya}}, \
  and\ \bibinfo {author} {\bibfnamefont {S.}~\bibnamefont {Chakraborty}},\
  }\href {\doibase 10.1093/mnras/stv1495} {\bibfield  {journal} {\bibinfo
  {journal} {Mon. Not. Roy. Astron. Soc.}\ }\textbf {\bibinfo {volume} {452}},\
  \bibinfo {pages} {3038} (\bibinfo {year} {2015})},\ \Eprint
  {http://arxiv.org/abs/1210.0396} {arXiv:1210.0396 [gr-qc]} \BibitemShut
  {NoStop}%
\bibitem [{\citenamefont {Yang}\ \emph {et~al.}(2016)\citenamefont {Yang},
  \citenamefont {Li}, \citenamefont {Wu},\ and\ \citenamefont
  {Lu}}]{Yang:2016evp}%
  \BibitemOpen
  \bibfield  {author} {\bibinfo {author} {\bibfnamefont {W.}~\bibnamefont
  {Yang}}, \bibinfo {author} {\bibfnamefont {H.}~\bibnamefont {Li}}, \bibinfo
  {author} {\bibfnamefont {Y.}~\bibnamefont {Wu}}, \ and\ \bibinfo {author}
  {\bibfnamefont {J.}~\bibnamefont {Lu}},\ }\href {\doibase
  10.1088/1475-7516/2016/10/007} {\bibfield  {journal} {\bibinfo  {journal}
  {JCAP}\ }\textbf {\bibinfo {volume} {10}},\ \bibinfo {pages} {007} (\bibinfo
  {year} {2016})},\ \Eprint {http://arxiv.org/abs/1608.07039} {arXiv:1608.07039
  [astro-ph.CO]} \BibitemShut {NoStop}%
\bibitem [{\citenamefont {Nunes}\ \emph {et~al.}(2016)\citenamefont {Nunes},
  \citenamefont {Pan},\ and\ \citenamefont {Saridakis}}]{Nunes:2016dlj}%
  \BibitemOpen
  \bibfield  {author} {\bibinfo {author} {\bibfnamefont {R.~C.}\ \bibnamefont
  {Nunes}}, \bibinfo {author} {\bibfnamefont {S.}~\bibnamefont {Pan}}, \ and\
  \bibinfo {author} {\bibfnamefont {E.~N.}\ \bibnamefont {Saridakis}},\ }\href
  {\doibase 10.1103/PhysRevD.94.023508} {\bibfield  {journal} {\bibinfo
  {journal} {Phys. Rev. D}\ }\textbf {\bibinfo {volume} {94}},\ \bibinfo
  {pages} {023508} (\bibinfo {year} {2016})},\ \Eprint
  {http://arxiv.org/abs/1605.01712} {arXiv:1605.01712 [astro-ph.CO]}
  \BibitemShut {NoStop}%
\bibitem [{\citenamefont {Yang}\ \emph
  {et~al.}(2017{\natexlab{a}})\citenamefont {Yang}, \citenamefont {Banerjee},\
  and\ \citenamefont {Pan}}]{Yang:2017yme}%
  \BibitemOpen
  \bibfield  {author} {\bibinfo {author} {\bibfnamefont {W.}~\bibnamefont
  {Yang}}, \bibinfo {author} {\bibfnamefont {N.}~\bibnamefont {Banerjee}}, \
  and\ \bibinfo {author} {\bibfnamefont {S.}~\bibnamefont {Pan}},\ }\href
  {\doibase 10.1103/PhysRevD.95.123527} {\bibfield  {journal} {\bibinfo
  {journal} {Phys. Rev. D}\ }\textbf {\bibinfo {volume} {95}},\ \bibinfo
  {pages} {123527} (\bibinfo {year} {2017}{\natexlab{a}})},\ \bibinfo {note}
  {[Addendum: Phys.Rev.D 96, 089903 (2017)]},\ \Eprint
  {http://arxiv.org/abs/1705.09278} {arXiv:1705.09278 [astro-ph.CO]}
  \BibitemShut {NoStop}%
\bibitem [{\citenamefont {Dutta}\ \emph {et~al.}(2017)\citenamefont {Dutta},
  \citenamefont {Khyllep},\ and\ \citenamefont {Tamanini}}]{Dutta:2017kch}%
  \BibitemOpen
  \bibfield  {author} {\bibinfo {author} {\bibfnamefont {J.}~\bibnamefont
  {Dutta}}, \bibinfo {author} {\bibfnamefont {W.}~\bibnamefont {Khyllep}}, \
  and\ \bibinfo {author} {\bibfnamefont {N.}~\bibnamefont {Tamanini}},\ }\href
  {\doibase 10.1103/PhysRevD.95.023515} {\bibfield  {journal} {\bibinfo
  {journal} {Phys. Rev. D}\ }\textbf {\bibinfo {volume} {95}},\ \bibinfo
  {pages} {023515} (\bibinfo {year} {2017})},\ \Eprint
  {http://arxiv.org/abs/1701.00744} {arXiv:1701.00744 [gr-qc]} \BibitemShut
  {NoStop}%
\bibitem [{\citenamefont {Di~Valentino}\ \emph {et~al.}(2017)\citenamefont
  {Di~Valentino}, \citenamefont {Melchiorri},\ and\ \citenamefont
  {Mena}}]{DiValentino:2017iww}%
  \BibitemOpen
  \bibfield  {author} {\bibinfo {author} {\bibfnamefont {E.}~\bibnamefont
  {Di~Valentino}}, \bibinfo {author} {\bibfnamefont {A.}~\bibnamefont
  {Melchiorri}}, \ and\ \bibinfo {author} {\bibfnamefont {O.}~\bibnamefont
  {Mena}},\ }\href {\doibase 10.1103/PhysRevD.96.043503} {\bibfield  {journal}
  {\bibinfo  {journal} {Phys. Rev. D}\ }\textbf {\bibinfo {volume} {96}},\
  \bibinfo {pages} {043503} (\bibinfo {year} {2017})},\ \Eprint
  {http://arxiv.org/abs/1704.08342} {arXiv:1704.08342 [astro-ph.CO]}
  \BibitemShut {NoStop}%
\bibitem [{\citenamefont {Santos}\ \emph {et~al.}(2017)\citenamefont {Santos},
  \citenamefont {Zhao}, \citenamefont {Ferreira},\ and\ \citenamefont
  {Quintin}}]{Santos:2017bqm}%
  \BibitemOpen
  \bibfield  {author} {\bibinfo {author} {\bibfnamefont {L.}~\bibnamefont
  {Santos}}, \bibinfo {author} {\bibfnamefont {W.}~\bibnamefont {Zhao}},
  \bibinfo {author} {\bibfnamefont {E.~G.~M.}\ \bibnamefont {Ferreira}}, \ and\
  \bibinfo {author} {\bibfnamefont {J.}~\bibnamefont {Quintin}},\ }\href
  {\doibase 10.1103/PhysRevD.96.103529} {\bibfield  {journal} {\bibinfo
  {journal} {Phys. Rev. D}\ }\textbf {\bibinfo {volume} {96}},\ \bibinfo
  {pages} {103529} (\bibinfo {year} {2017})},\ \Eprint
  {http://arxiv.org/abs/1707.06827} {arXiv:1707.06827 [astro-ph.CO]}
  \BibitemShut {NoStop}%
\bibitem [{\citenamefont {Yang}\ \emph
  {et~al.}(2017{\natexlab{b}})\citenamefont {Yang}, \citenamefont {Pan},\ and\
  \citenamefont {Mota}}]{Yang:2017ccc}%
  \BibitemOpen
  \bibfield  {author} {\bibinfo {author} {\bibfnamefont {W.}~\bibnamefont
  {Yang}}, \bibinfo {author} {\bibfnamefont {S.}~\bibnamefont {Pan}}, \ and\
  \bibinfo {author} {\bibfnamefont {D.~F.}\ \bibnamefont {Mota}},\ }\href
  {\doibase 10.1103/PhysRevD.96.123508} {\bibfield  {journal} {\bibinfo
  {journal} {Phys. Rev. D}\ }\textbf {\bibinfo {volume} {96}},\ \bibinfo
  {pages} {123508} (\bibinfo {year} {2017}{\natexlab{b}})},\ \Eprint
  {http://arxiv.org/abs/1709.00006} {arXiv:1709.00006 [astro-ph.CO]}
  \BibitemShut {NoStop}%
\bibitem [{\citenamefont {Yang}\ \emph
  {et~al.}(2018{\natexlab{a}})\citenamefont {Yang}, \citenamefont {Pan},\ and\
  \citenamefont {Barrow}}]{Yang:2017zjs}%
  \BibitemOpen
  \bibfield  {author} {\bibinfo {author} {\bibfnamefont {W.}~\bibnamefont
  {Yang}}, \bibinfo {author} {\bibfnamefont {S.}~\bibnamefont {Pan}}, \ and\
  \bibinfo {author} {\bibfnamefont {J.~D.}\ \bibnamefont {Barrow}},\ }\href
  {\doibase 10.1103/PhysRevD.97.043529} {\bibfield  {journal} {\bibinfo
  {journal} {Phys. Rev. D}\ }\textbf {\bibinfo {volume} {97}},\ \bibinfo
  {pages} {043529} (\bibinfo {year} {2018}{\natexlab{a}})},\ \Eprint
  {http://arxiv.org/abs/1706.04953} {arXiv:1706.04953 [astro-ph.CO]}
  \BibitemShut {NoStop}%
\bibitem [{\citenamefont {Yang}\ \emph
  {et~al.}(2017{\natexlab{c}})\citenamefont {Yang}, \citenamefont {Xu},
  \citenamefont {Li}, \citenamefont {Wu},\ and\ \citenamefont
  {Lu}}]{Yang:2017iew}%
  \BibitemOpen
  \bibfield  {author} {\bibinfo {author} {\bibfnamefont {W.}~\bibnamefont
  {Yang}}, \bibinfo {author} {\bibfnamefont {L.}~\bibnamefont {Xu}}, \bibinfo
  {author} {\bibfnamefont {H.}~\bibnamefont {Li}}, \bibinfo {author}
  {\bibfnamefont {Y.}~\bibnamefont {Wu}}, \ and\ \bibinfo {author}
  {\bibfnamefont {J.}~\bibnamefont {Lu}},\ }\href {\doibase 10.3390/e19070327}
  {\bibfield  {journal} {\bibinfo  {journal} {Entropy}\ }\textbf {\bibinfo
  {volume} {19}},\ \bibinfo {pages} {327} (\bibinfo {year}
  {2017}{\natexlab{c}})}\BibitemShut {NoStop}%
\bibitem [{\citenamefont {Yang}\ \emph
  {et~al.}(2018{\natexlab{b}})\citenamefont {Yang}, \citenamefont {Pan},
  \citenamefont {Di~Valentino}, \citenamefont {Nunes}, \citenamefont
  {Vagnozzi},\ and\ \citenamefont {Mota}}]{Yang:2018euj}%
  \BibitemOpen
  \bibfield  {author} {\bibinfo {author} {\bibfnamefont {W.}~\bibnamefont
  {Yang}}, \bibinfo {author} {\bibfnamefont {S.}~\bibnamefont {Pan}}, \bibinfo
  {author} {\bibfnamefont {E.}~\bibnamefont {Di~Valentino}}, \bibinfo {author}
  {\bibfnamefont {R.~C.}\ \bibnamefont {Nunes}}, \bibinfo {author}
  {\bibfnamefont {S.}~\bibnamefont {Vagnozzi}}, \ and\ \bibinfo {author}
  {\bibfnamefont {D.~F.}\ \bibnamefont {Mota}},\ }\href {\doibase
  10.1088/1475-7516/2018/09/019} {\bibfield  {journal} {\bibinfo  {journal}
  {JCAP}\ }\textbf {\bibinfo {volume} {09}},\ \bibinfo {pages} {019} (\bibinfo
  {year} {2018}{\natexlab{b}})},\ \Eprint {http://arxiv.org/abs/1805.08252}
  {arXiv:1805.08252 [astro-ph.CO]} \BibitemShut {NoStop}%
\bibitem [{\citenamefont {Yang}\ \emph
  {et~al.}(2018{\natexlab{c}})\citenamefont {Yang}, \citenamefont {Mukherjee},
  \citenamefont {Di~Valentino},\ and\ \citenamefont {Pan}}]{Yang:2018uae}%
  \BibitemOpen
  \bibfield  {author} {\bibinfo {author} {\bibfnamefont {W.}~\bibnamefont
  {Yang}}, \bibinfo {author} {\bibfnamefont {A.}~\bibnamefont {Mukherjee}},
  \bibinfo {author} {\bibfnamefont {E.}~\bibnamefont {Di~Valentino}}, \ and\
  \bibinfo {author} {\bibfnamefont {S.}~\bibnamefont {Pan}},\ }\href {\doibase
  10.1103/PhysRevD.98.123527} {\bibfield  {journal} {\bibinfo  {journal} {Phys.
  Rev. D}\ }\textbf {\bibinfo {volume} {98}},\ \bibinfo {pages} {123527}
  (\bibinfo {year} {2018}{\natexlab{c}})},\ \Eprint
  {http://arxiv.org/abs/1809.06883} {arXiv:1809.06883 [astro-ph.CO]}
  \BibitemShut {NoStop}%
\bibitem [{\citenamefont {Li}\ \emph {et~al.}(2020)\citenamefont {Li},
  \citenamefont {Ren}, \citenamefont {Khurshudyan},\ and\ \citenamefont
  {Cai}}]{Li:2019loh}%
  \BibitemOpen
  \bibfield  {author} {\bibinfo {author} {\bibfnamefont {C.}~\bibnamefont
  {Li}}, \bibinfo {author} {\bibfnamefont {X.}~\bibnamefont {Ren}}, \bibinfo
  {author} {\bibfnamefont {M.}~\bibnamefont {Khurshudyan}}, \ and\ \bibinfo
  {author} {\bibfnamefont {Y.-F.}\ \bibnamefont {Cai}},\ }\href {\doibase
  10.1016/j.physletb.2019.135141} {\bibfield  {journal} {\bibinfo  {journal}
  {Phys. Lett. B}\ }\textbf {\bibinfo {volume} {801}},\ \bibinfo {pages}
  {135141} (\bibinfo {year} {2020})},\ \Eprint
  {http://arxiv.org/abs/1904.02458} {arXiv:1904.02458 [astro-ph.CO]}
  \BibitemShut {NoStop}%
\bibitem [{\citenamefont {Pan}\ \emph {et~al.}(2019)\citenamefont {Pan},
  \citenamefont {Yang}, \citenamefont {Di~Valentino}, \citenamefont
  {Saridakis},\ and\ \citenamefont {Chakraborty}}]{Pan:2019gop}%
  \BibitemOpen
  \bibfield  {author} {\bibinfo {author} {\bibfnamefont {S.}~\bibnamefont
  {Pan}}, \bibinfo {author} {\bibfnamefont {W.}~\bibnamefont {Yang}}, \bibinfo
  {author} {\bibfnamefont {E.}~\bibnamefont {Di~Valentino}}, \bibinfo {author}
  {\bibfnamefont {E.~N.}\ \bibnamefont {Saridakis}}, \ and\ \bibinfo {author}
  {\bibfnamefont {S.}~\bibnamefont {Chakraborty}},\ }\href {\doibase
  10.1103/PhysRevD.100.103520} {\bibfield  {journal} {\bibinfo  {journal}
  {Phys. Rev. D}\ }\textbf {\bibinfo {volume} {100}},\ \bibinfo {pages}
  {103520} (\bibinfo {year} {2019})},\ \Eprint
  {http://arxiv.org/abs/1907.07540} {arXiv:1907.07540 [astro-ph.CO]}
  \BibitemShut {NoStop}%
\bibitem [{\citenamefont {Di~Valentino}\ \emph
  {et~al.}(2020{\natexlab{a}})\citenamefont {Di~Valentino}, \citenamefont
  {Melchiorri}, \citenamefont {Mena},\ and\ \citenamefont
  {Vagnozzi}}]{DiValentino:2019ffd}%
  \BibitemOpen
  \bibfield  {author} {\bibinfo {author} {\bibfnamefont {E.}~\bibnamefont
  {Di~Valentino}}, \bibinfo {author} {\bibfnamefont {A.}~\bibnamefont
  {Melchiorri}}, \bibinfo {author} {\bibfnamefont {O.}~\bibnamefont {Mena}}, \
  and\ \bibinfo {author} {\bibfnamefont {S.}~\bibnamefont {Vagnozzi}},\ }\href
  {\doibase 10.1016/j.dark.2020.100666} {\bibfield  {journal} {\bibinfo
  {journal} {Phys. Dark Univ.}\ }\textbf {\bibinfo {volume} {30}},\ \bibinfo
  {pages} {100666} (\bibinfo {year} {2020}{\natexlab{a}})},\ \Eprint
  {http://arxiv.org/abs/1908.04281} {arXiv:1908.04281 [astro-ph.CO]}
  \BibitemShut {NoStop}%
\bibitem [{\citenamefont {Cheng}\ \emph {et~al.}(2020)\citenamefont {Cheng},
  \citenamefont {Ma}, \citenamefont {Wu}, \citenamefont {Zhang},\ and\
  \citenamefont {Chen}}]{Cheng:2019bkh}%
  \BibitemOpen
  \bibfield  {author} {\bibinfo {author} {\bibfnamefont {G.}~\bibnamefont
  {Cheng}}, \bibinfo {author} {\bibfnamefont {Y.-Z.}\ \bibnamefont {Ma}},
  \bibinfo {author} {\bibfnamefont {F.}~\bibnamefont {Wu}}, \bibinfo {author}
  {\bibfnamefont {J.}~\bibnamefont {Zhang}}, \ and\ \bibinfo {author}
  {\bibfnamefont {X.}~\bibnamefont {Chen}},\ }\href {\doibase
  10.1103/PhysRevD.102.043517} {\bibfield  {journal} {\bibinfo  {journal}
  {Phys. Rev. D}\ }\textbf {\bibinfo {volume} {102}},\ \bibinfo {pages}
  {043517} (\bibinfo {year} {2020})},\ \Eprint
  {http://arxiv.org/abs/1911.04520} {arXiv:1911.04520 [astro-ph.CO]}
  \BibitemShut {NoStop}%
\bibitem [{\citenamefont {Di~Valentino}\ \emph
  {et~al.}(2020{\natexlab{b}})\citenamefont {Di~Valentino}, \citenamefont
  {Melchiorri}, \citenamefont {Mena},\ and\ \citenamefont
  {Vagnozzi}}]{DiValentino:2019jae}%
  \BibitemOpen
  \bibfield  {author} {\bibinfo {author} {\bibfnamefont {E.}~\bibnamefont
  {Di~Valentino}}, \bibinfo {author} {\bibfnamefont {A.}~\bibnamefont
  {Melchiorri}}, \bibinfo {author} {\bibfnamefont {O.}~\bibnamefont {Mena}}, \
  and\ \bibinfo {author} {\bibfnamefont {S.}~\bibnamefont {Vagnozzi}},\ }\href
  {\doibase 10.1103/PhysRevD.101.063502} {\bibfield  {journal} {\bibinfo
  {journal} {Phys. Rev. D}\ }\textbf {\bibinfo {volume} {101}},\ \bibinfo
  {pages} {063502} (\bibinfo {year} {2020}{\natexlab{b}})},\ \Eprint
  {http://arxiv.org/abs/1910.09853} {arXiv:1910.09853 [astro-ph.CO]}
  \BibitemShut {NoStop}%
\bibitem [{\citenamefont {Pan}\ \emph {et~al.}(2020{\natexlab{a}})\citenamefont
  {Pan}, \citenamefont {Sharov},\ and\ \citenamefont {Yang}}]{Pan:2020zza}%
  \BibitemOpen
  \bibfield  {author} {\bibinfo {author} {\bibfnamefont {S.}~\bibnamefont
  {Pan}}, \bibinfo {author} {\bibfnamefont {G.~S.}\ \bibnamefont {Sharov}}, \
  and\ \bibinfo {author} {\bibfnamefont {W.}~\bibnamefont {Yang}},\ }\href
  {\doibase 10.1103/PhysRevD.101.103533} {\bibfield  {journal} {\bibinfo
  {journal} {Phys. Rev. D}\ }\textbf {\bibinfo {volume} {101}},\ \bibinfo
  {pages} {103533} (\bibinfo {year} {2020}{\natexlab{a}})},\ \Eprint
  {http://arxiv.org/abs/2001.03120} {arXiv:2001.03120 [astro-ph.CO]}
  \BibitemShut {NoStop}%
\bibitem [{\citenamefont {Di~Valentino}\ and\ \citenamefont
  {Mena}(2020)}]{DiValentino:2020leo}%
  \BibitemOpen
  \bibfield  {author} {\bibinfo {author} {\bibfnamefont {E.}~\bibnamefont
  {Di~Valentino}}\ and\ \bibinfo {author} {\bibfnamefont {O.}~\bibnamefont
  {Mena}},\ }\href {\doibase 10.1093/mnrasl/slaa175} {\bibfield  {journal}
  {\bibinfo  {journal} {Mon. Not. Roy. Astron. Soc.}\ }\textbf {\bibinfo
  {volume} {500}},\ \bibinfo {pages} {L22} (\bibinfo {year} {2020})},\ \Eprint
  {http://arxiv.org/abs/2009.12620} {arXiv:2009.12620 [astro-ph.CO]}
  \BibitemShut {NoStop}%
\bibitem [{\citenamefont {Pan}\ \emph {et~al.}(2020{\natexlab{b}})\citenamefont
  {Pan}, \citenamefont {de~Haro}, \citenamefont {Yang},\ and\ \citenamefont
  {Amor\'os}}]{Pan:2020mst}%
  \BibitemOpen
  \bibfield  {author} {\bibinfo {author} {\bibfnamefont {S.}~\bibnamefont
  {Pan}}, \bibinfo {author} {\bibfnamefont {J.}~\bibnamefont {de~Haro}},
  \bibinfo {author} {\bibfnamefont {W.}~\bibnamefont {Yang}}, \ and\ \bibinfo
  {author} {\bibfnamefont {J.}~\bibnamefont {Amor\'os}},\ }\href {\doibase
  10.1103/PhysRevD.101.123506} {\bibfield  {journal} {\bibinfo  {journal}
  {Phys. Rev. D}\ }\textbf {\bibinfo {volume} {101}},\ \bibinfo {pages}
  {123506} (\bibinfo {year} {2020}{\natexlab{b}})},\ \Eprint
  {http://arxiv.org/abs/2001.09885} {arXiv:2001.09885 [gr-qc]} \BibitemShut
  {NoStop}%
\bibitem [{\citenamefont {Di~Valentino}\ \emph
  {et~al.}(2021{\natexlab{d}})\citenamefont {Di~Valentino}, \citenamefont
  {Melchiorri}, \citenamefont {Mena}, \citenamefont {Pan},\ and\ \citenamefont
  {Yang}}]{DiValentino:2020kpf}%
  \BibitemOpen
  \bibfield  {author} {\bibinfo {author} {\bibfnamefont {E.}~\bibnamefont
  {Di~Valentino}}, \bibinfo {author} {\bibfnamefont {A.}~\bibnamefont
  {Melchiorri}}, \bibinfo {author} {\bibfnamefont {O.}~\bibnamefont {Mena}},
  \bibinfo {author} {\bibfnamefont {S.}~\bibnamefont {Pan}}, \ and\ \bibinfo
  {author} {\bibfnamefont {W.}~\bibnamefont {Yang}},\ }\href {\doibase
  10.1093/mnrasl/slaa207} {\bibfield  {journal} {\bibinfo  {journal} {Mon. Not.
  Roy. Astron. Soc.}\ }\textbf {\bibinfo {volume} {502}},\ \bibinfo {pages}
  {L23} (\bibinfo {year} {2021}{\natexlab{d}})},\ \Eprint
  {http://arxiv.org/abs/2011.00283} {arXiv:2011.00283 [astro-ph.CO]}
  \BibitemShut {NoStop}%
\bibitem [{\citenamefont {Yang}\ \emph {et~al.}(2021)\citenamefont {Yang},
  \citenamefont {Pan}, \citenamefont {Di~Valentino}, \citenamefont {Mena},\
  and\ \citenamefont {Melchiorri}}]{Yang:2021hxg}%
  \BibitemOpen
  \bibfield  {author} {\bibinfo {author} {\bibfnamefont {W.}~\bibnamefont
  {Yang}}, \bibinfo {author} {\bibfnamefont {S.}~\bibnamefont {Pan}}, \bibinfo
  {author} {\bibfnamefont {E.}~\bibnamefont {Di~Valentino}}, \bibinfo {author}
  {\bibfnamefont {O.}~\bibnamefont {Mena}}, \ and\ \bibinfo {author}
  {\bibfnamefont {A.}~\bibnamefont {Melchiorri}},\ }\href {\doibase
  10.1088/1475-7516/2021/10/008} {\bibfield  {journal} {\bibinfo  {journal}
  {JCAP}\ }\textbf {\bibinfo {volume} {10}},\ \bibinfo {pages} {008} (\bibinfo
  {year} {2021})},\ \Eprint {http://arxiv.org/abs/2101.03129} {arXiv:2101.03129
  [astro-ph.CO]} \BibitemShut {NoStop}%
\bibitem [{\citenamefont {Gariazzo}\ \emph {et~al.}(2022)\citenamefont
  {Gariazzo}, \citenamefont {Di~Valentino}, \citenamefont {Mena},\ and\
  \citenamefont {Nunes}}]{Gariazzo:2021qtg}%
  \BibitemOpen
  \bibfield  {author} {\bibinfo {author} {\bibfnamefont {S.}~\bibnamefont
  {Gariazzo}}, \bibinfo {author} {\bibfnamefont {E.}~\bibnamefont
  {Di~Valentino}}, \bibinfo {author} {\bibfnamefont {O.}~\bibnamefont {Mena}},
  \ and\ \bibinfo {author} {\bibfnamefont {R.~C.}\ \bibnamefont {Nunes}},\
  }\href {\doibase 10.1103/PhysRevD.106.023530} {\bibfield  {journal} {\bibinfo
   {journal} {Phys. Rev. D}\ }\textbf {\bibinfo {volume} {106}},\ \bibinfo
  {pages} {023530} (\bibinfo {year} {2022})},\ \Eprint
  {http://arxiv.org/abs/2111.03152} {arXiv:2111.03152 [astro-ph.CO]}
  \BibitemShut {NoStop}%
\bibitem [{\citenamefont {Gao}\ \emph {et~al.}(2021)\citenamefont {Gao},
  \citenamefont {Zhao}, \citenamefont {Xue},\ and\ \citenamefont
  {Zhang}}]{Gao:2021xnk}%
  \BibitemOpen
  \bibfield  {author} {\bibinfo {author} {\bibfnamefont {L.-Y.}\ \bibnamefont
  {Gao}}, \bibinfo {author} {\bibfnamefont {Z.-W.}\ \bibnamefont {Zhao}},
  \bibinfo {author} {\bibfnamefont {S.-S.}\ \bibnamefont {Xue}}, \ and\
  \bibinfo {author} {\bibfnamefont {X.}~\bibnamefont {Zhang}},\ }\href
  {\doibase 10.1088/1475-7516/2021/07/005} {\bibfield  {journal} {\bibinfo
  {journal} {JCAP}\ }\textbf {\bibinfo {volume} {07}},\ \bibinfo {pages} {005}
  (\bibinfo {year} {2021})},\ \Eprint {http://arxiv.org/abs/2101.10714}
  {arXiv:2101.10714 [astro-ph.CO]} \BibitemShut {NoStop}%
\bibitem [{\citenamefont {Guo}\ \emph {et~al.}(2021)\citenamefont {Guo},
  \citenamefont {Feng}, \citenamefont {Yao},\ and\ \citenamefont
  {Chen}}]{Guo:2021rrz}%
  \BibitemOpen
  \bibfield  {author} {\bibinfo {author} {\bibfnamefont {R.-Y.}\ \bibnamefont
  {Guo}}, \bibinfo {author} {\bibfnamefont {L.}~\bibnamefont {Feng}}, \bibinfo
  {author} {\bibfnamefont {T.-Y.}\ \bibnamefont {Yao}}, \ and\ \bibinfo
  {author} {\bibfnamefont {X.-Y.}\ \bibnamefont {Chen}},\ }\href {\doibase
  10.1088/1475-7516/2021/12/036} {\bibfield  {journal} {\bibinfo  {journal}
  {JCAP}\ }\textbf {\bibinfo {volume} {12}},\ \bibinfo {pages} {036} (\bibinfo
  {year} {2021})},\ \Eprint {http://arxiv.org/abs/2110.02536} {arXiv:2110.02536
  [gr-qc]} \BibitemShut {NoStop}%
\bibitem [{\citenamefont {Chatzidakis}\ \emph {et~al.}(2022)\citenamefont
  {Chatzidakis}, \citenamefont {Giacomini}, \citenamefont {Leach},
  \citenamefont {Leon}, \citenamefont {Paliathanasis},\ and\ \citenamefont
  {Pan}}]{Chatzidakis:2022mpf}%
  \BibitemOpen
  \bibfield  {author} {\bibinfo {author} {\bibfnamefont {S.}~\bibnamefont
  {Chatzidakis}}, \bibinfo {author} {\bibfnamefont {A.}~\bibnamefont
  {Giacomini}}, \bibinfo {author} {\bibfnamefont {P.~G.~L.}\ \bibnamefont
  {Leach}}, \bibinfo {author} {\bibfnamefont {G.}~\bibnamefont {Leon}},
  \bibinfo {author} {\bibfnamefont {A.}~\bibnamefont {Paliathanasis}}, \ and\
  \bibinfo {author} {\bibfnamefont {S.}~\bibnamefont {Pan}},\ }\href {\doibase
  10.1016/j.jheap.2022.10.001} {\bibfield  {journal} {\bibinfo  {journal}
  {JHEAp}\ }\textbf {\bibinfo {volume} {36}},\ \bibinfo {pages} {141} (\bibinfo
  {year} {2022})},\ \Eprint {http://arxiv.org/abs/2206.06639} {arXiv:2206.06639
  [gr-qc]} \BibitemShut {NoStop}%
\bibitem [{\citenamefont {Zhao}\ \emph {et~al.}(2023)\citenamefont {Zhao},
  \citenamefont {Liu}, \citenamefont {Liao}, \citenamefont {Zhang},
  \citenamefont {Liu},\ and\ \citenamefont {Du}}]{Zhao:2022ycr}%
  \BibitemOpen
  \bibfield  {author} {\bibinfo {author} {\bibfnamefont {Y.}~\bibnamefont
  {Zhao}}, \bibinfo {author} {\bibfnamefont {Y.}~\bibnamefont {Liu}}, \bibinfo
  {author} {\bibfnamefont {S.}~\bibnamefont {Liao}}, \bibinfo {author}
  {\bibfnamefont {J.}~\bibnamefont {Zhang}}, \bibinfo {author} {\bibfnamefont
  {X.}~\bibnamefont {Liu}}, \ and\ \bibinfo {author} {\bibfnamefont
  {W.}~\bibnamefont {Du}},\ }\href {\doibase 10.1093/mnras/stad1814} {\bibfield
   {journal} {\bibinfo  {journal} {Mon. Not. Roy. Astron. Soc.}\ }\textbf
  {\bibinfo {volume} {523}},\ \bibinfo {pages} {5962} (\bibinfo {year}
  {2023})},\ \Eprint {http://arxiv.org/abs/2212.02050} {arXiv:2212.02050
  [astro-ph.CO]} \BibitemShut {NoStop}%
\bibitem [{\citenamefont {Gao}\ \emph {et~al.}(2022)\citenamefont {Gao},
  \citenamefont {Xue},\ and\ \citenamefont {Zhang}}]{Gao:2022ahg}%
  \BibitemOpen
  \bibfield  {author} {\bibinfo {author} {\bibfnamefont {L.-Y.}\ \bibnamefont
  {Gao}}, \bibinfo {author} {\bibfnamefont {S.-S.}\ \bibnamefont {Xue}}, \ and\
  \bibinfo {author} {\bibfnamefont {X.}~\bibnamefont {Zhang}},\ }\href@noop {}
  {\bibfield  {journal} {\bibinfo  {journal} {2212.13146}\ } (\bibinfo {year}
  {2022})}\BibitemShut {NoStop}%
\bibitem [{\citenamefont {Hou}\ \emph {et~al.}(2023)\citenamefont {Hou},
  \citenamefont {Qi}, \citenamefont {Han}, \citenamefont {Zhang}, \citenamefont
  {Cao},\ and\ \citenamefont {Zhang}}]{Hou:2022rvk}%
  \BibitemOpen
  \bibfield  {author} {\bibinfo {author} {\bibfnamefont {W.-T.}\ \bibnamefont
  {Hou}}, \bibinfo {author} {\bibfnamefont {J.-Z.}\ \bibnamefont {Qi}},
  \bibinfo {author} {\bibfnamefont {T.}~\bibnamefont {Han}}, \bibinfo {author}
  {\bibfnamefont {J.-F.}\ \bibnamefont {Zhang}}, \bibinfo {author}
  {\bibfnamefont {S.}~\bibnamefont {Cao}}, \ and\ \bibinfo {author}
  {\bibfnamefont {X.}~\bibnamefont {Zhang}},\ }\href {\doibase
  10.1088/1475-7516/2023/05/017} {\bibfield  {journal} {\bibinfo  {journal}
  {JCAP}\ }\textbf {\bibinfo {volume} {05}},\ \bibinfo {pages} {017} (\bibinfo
  {year} {2023})},\ \Eprint {http://arxiv.org/abs/2211.10087} {arXiv:2211.10087
  [astro-ph.CO]} \BibitemShut {NoStop}%
\bibitem [{\citenamefont {Nunes}\ \emph {et~al.}(2022)\citenamefont {Nunes},
  \citenamefont {Vagnozzi}, \citenamefont {Kumar}, \citenamefont
  {Di~Valentino},\ and\ \citenamefont {Mena}}]{Nunes:2022bhn}%
  \BibitemOpen
  \bibfield  {author} {\bibinfo {author} {\bibfnamefont {R.~C.}\ \bibnamefont
  {Nunes}}, \bibinfo {author} {\bibfnamefont {S.}~\bibnamefont {Vagnozzi}},
  \bibinfo {author} {\bibfnamefont {S.}~\bibnamefont {Kumar}}, \bibinfo
  {author} {\bibfnamefont {E.}~\bibnamefont {Di~Valentino}}, \ and\ \bibinfo
  {author} {\bibfnamefont {O.}~\bibnamefont {Mena}},\ }\href {\doibase
  10.1103/PhysRevD.105.123506} {\bibfield  {journal} {\bibinfo  {journal}
  {Phys. Rev. D}\ }\textbf {\bibinfo {volume} {105}},\ \bibinfo {pages}
  {123506} (\bibinfo {year} {2022})},\ \Eprint
  {http://arxiv.org/abs/2203.08093} {arXiv:2203.08093 [astro-ph.CO]}
  \BibitemShut {NoStop}%
\bibitem [{\citenamefont {Zhai}\ \emph {et~al.}(2023)\citenamefont {Zhai},
  \citenamefont {Giar\`e}, \citenamefont {van~de Bruck}, \citenamefont
  {Di~Valentino}, \citenamefont {Mena},\ and\ \citenamefont
  {Nunes}}]{Zhai:2023yny}%
  \BibitemOpen
  \bibfield  {author} {\bibinfo {author} {\bibfnamefont {Y.}~\bibnamefont
  {Zhai}}, \bibinfo {author} {\bibfnamefont {W.}~\bibnamefont {Giar\`e}},
  \bibinfo {author} {\bibfnamefont {C.}~\bibnamefont {van~de Bruck}}, \bibinfo
  {author} {\bibfnamefont {E.}~\bibnamefont {Di~Valentino}}, \bibinfo {author}
  {\bibfnamefont {O.}~\bibnamefont {Mena}}, \ and\ \bibinfo {author}
  {\bibfnamefont {R.~C.}\ \bibnamefont {Nunes}},\ }\href {\doibase
  10.1088/1475-7516/2023/07/032} {\bibfield  {journal} {\bibinfo  {journal}
  {JCAP}\ }\textbf {\bibinfo {volume} {07}},\ \bibinfo {pages} {032} (\bibinfo
  {year} {2023})},\ \Eprint {http://arxiv.org/abs/2303.08201} {arXiv:2303.08201
  [astro-ph.CO]} \BibitemShut {NoStop}%
\bibitem [{\citenamefont {Li}\ and\ \citenamefont {Zhang}(2023)}]{Li:2023fdk}%
  \BibitemOpen
  \bibfield  {author} {\bibinfo {author} {\bibfnamefont {Y.-H.}\ \bibnamefont
  {Li}}\ and\ \bibinfo {author} {\bibfnamefont {X.}~\bibnamefont {Zhang}},\
  }\href {\doibase 10.1088/1475-7516/2023/09/046} {\bibfield  {journal}
  {\bibinfo  {journal} {JCAP}\ }\textbf {\bibinfo {volume} {09}},\ \bibinfo
  {pages} {046} (\bibinfo {year} {2023})},\ \Eprint
  {http://arxiv.org/abs/2306.01593} {arXiv:2306.01593 [astro-ph.CO]}
  \BibitemShut {NoStop}%
\bibitem [{\citenamefont {Benisty}\ \emph {et~al.}(2024)\citenamefont
  {Benisty}, \citenamefont {Pan}, \citenamefont {Staicova}, \citenamefont
  {Di~Valentino},\ and\ \citenamefont {Nunes}}]{Benisty:2024lmj}%
  \BibitemOpen
  \bibfield  {author} {\bibinfo {author} {\bibfnamefont {D.}~\bibnamefont
  {Benisty}}, \bibinfo {author} {\bibfnamefont {S.}~\bibnamefont {Pan}},
  \bibinfo {author} {\bibfnamefont {D.}~\bibnamefont {Staicova}}, \bibinfo
  {author} {\bibfnamefont {E.}~\bibnamefont {Di~Valentino}}, \ and\ \bibinfo
  {author} {\bibfnamefont {R.~C.}\ \bibnamefont {Nunes}},\ }\href@noop {}
  {\bibfield  {journal} {\bibinfo  {journal} {2403.00056}\ } (\bibinfo {year}
  {2024})}\BibitemShut {NoStop}%
\bibitem [{\citenamefont {Halder}\ \emph
  {et~al.}(2024{\natexlab{a}})\citenamefont {Halder}, \citenamefont {de~Haro},
  \citenamefont {Saha},\ and\ \citenamefont {Pan}}]{Halder:2024uao}%
  \BibitemOpen
  \bibfield  {author} {\bibinfo {author} {\bibfnamefont {S.}~\bibnamefont
  {Halder}}, \bibinfo {author} {\bibfnamefont {J.}~\bibnamefont {de~Haro}},
  \bibinfo {author} {\bibfnamefont {T.}~\bibnamefont {Saha}}, \ and\ \bibinfo
  {author} {\bibfnamefont {S.}~\bibnamefont {Pan}},\ }\href@noop {} {\bibfield
  {journal} {\bibinfo  {journal} {2403.01397}\ } (\bibinfo {year}
  {2024}{\natexlab{a}})}\BibitemShut {NoStop}%
\bibitem [{\citenamefont {Giar\`e}\ \emph
  {et~al.}(2024{\natexlab{a}})\citenamefont {Giar\`e}, \citenamefont {Sabogal},
  \citenamefont {Nunes},\ and\ \citenamefont {Di~Valentino}}]{Giare:2024smz}%
  \BibitemOpen
  \bibfield  {author} {\bibinfo {author} {\bibfnamefont {W.}~\bibnamefont
  {Giar\`e}}, \bibinfo {author} {\bibfnamefont {M.~A.}\ \bibnamefont
  {Sabogal}}, \bibinfo {author} {\bibfnamefont {R.~C.}\ \bibnamefont {Nunes}},
  \ and\ \bibinfo {author} {\bibfnamefont {E.}~\bibnamefont {Di~Valentino}},\
  }\href@noop {} {\bibfield  {journal} {\bibinfo  {journal} {2404.15232}\ }
  (\bibinfo {year} {2024}{\natexlab{a}})}\BibitemShut {NoStop}%
\bibitem [{\citenamefont {Giar\`e}\ \emph
  {et~al.}(2024{\natexlab{b}})\citenamefont {Giar\`e}, \citenamefont {Zhai},
  \citenamefont {Pan}, \citenamefont {Di~Valentino}, \citenamefont {Nunes},\
  and\ \citenamefont {van~de Bruck}}]{Giare:2024ytc}%
  \BibitemOpen
  \bibfield  {author} {\bibinfo {author} {\bibfnamefont {W.}~\bibnamefont
  {Giar\`e}}, \bibinfo {author} {\bibfnamefont {Y.}~\bibnamefont {Zhai}},
  \bibinfo {author} {\bibfnamefont {S.}~\bibnamefont {Pan}}, \bibinfo {author}
  {\bibfnamefont {E.}~\bibnamefont {Di~Valentino}}, \bibinfo {author}
  {\bibfnamefont {R.~C.}\ \bibnamefont {Nunes}}, \ and\ \bibinfo {author}
  {\bibfnamefont {C.}~\bibnamefont {van~de Bruck}},\ }\href {\doibase
  10.1103/PhysRevD.110.063527} {\bibfield  {journal} {\bibinfo  {journal}
  {Phys. Rev. D}\ }\textbf {\bibinfo {volume} {110}},\ \bibinfo {pages}
  {063527} (\bibinfo {year} {2024}{\natexlab{b}})},\ \Eprint
  {http://arxiv.org/abs/2404.02110} {arXiv:2404.02110 [astro-ph.CO]}
  \BibitemShut {NoStop}%
\bibitem [{\citenamefont {Halder}\ \emph
  {et~al.}(2024{\natexlab{b}})\citenamefont {Halder}, \citenamefont {Pan},
  \citenamefont {S\'a},\ and\ \citenamefont {Saha}}]{Halder:2024gag}%
  \BibitemOpen
  \bibfield  {author} {\bibinfo {author} {\bibfnamefont {S.}~\bibnamefont
  {Halder}}, \bibinfo {author} {\bibfnamefont {S.}~\bibnamefont {Pan}},
  \bibinfo {author} {\bibfnamefont {P.~M.}\ \bibnamefont {S\'a}}, \ and\
  \bibinfo {author} {\bibfnamefont {T.}~\bibnamefont {Saha}},\ }\href {\doibase
  10.1103/PhysRevD.110.063529} {\bibfield  {journal} {\bibinfo  {journal}
  {Phys. Rev. D}\ }\textbf {\bibinfo {volume} {110}},\ \bibinfo {pages}
  {063529} (\bibinfo {year} {2024}{\natexlab{b}})},\ \Eprint
  {http://arxiv.org/abs/2407.15804} {arXiv:2407.15804 [gr-qc]} \BibitemShut
  {NoStop}%
\bibitem [{\citenamefont {Sabogal}\ \emph {et~al.}(2024)\citenamefont
  {Sabogal}, \citenamefont {Silva}, \citenamefont {Nunes}, \citenamefont
  {Kumar}, \citenamefont {Di~Valentino},\ and\ \citenamefont
  {Giar\`e}}]{Sabogal:2024yha}%
  \BibitemOpen
  \bibfield  {author} {\bibinfo {author} {\bibfnamefont {M.~A.}\ \bibnamefont
  {Sabogal}}, \bibinfo {author} {\bibfnamefont {E.}~\bibnamefont {Silva}},
  \bibinfo {author} {\bibfnamefont {R.~C.}\ \bibnamefont {Nunes}}, \bibinfo
  {author} {\bibfnamefont {S.}~\bibnamefont {Kumar}}, \bibinfo {author}
  {\bibfnamefont {E.}~\bibnamefont {Di~Valentino}}, \ and\ \bibinfo {author}
  {\bibfnamefont {W.}~\bibnamefont {Giar\`e}},\ }\href@noop {} {\bibfield
  {journal} {\bibinfo  {journal} {2408.12403}\ } (\bibinfo {year}
  {2024})}\BibitemShut {NoStop}%
\bibitem [{\citenamefont {Ghedini}\ \emph {et~al.}(2024)\citenamefont
  {Ghedini}, \citenamefont {Hajjar},\ and\ \citenamefont
  {Mena}}]{Ghedini:2024mdu}%
  \BibitemOpen
  \bibfield  {author} {\bibinfo {author} {\bibfnamefont {P.}~\bibnamefont
  {Ghedini}}, \bibinfo {author} {\bibfnamefont {R.}~\bibnamefont {Hajjar}}, \
  and\ \bibinfo {author} {\bibfnamefont {O.}~\bibnamefont {Mena}},\ }\href@noop
  {} {\bibfield  {journal} {\bibinfo  {journal} {2409.02700}\ } (\bibinfo
  {year} {2024})}\BibitemShut {NoStop}%
\bibitem [{\citenamefont {Bolotin}\ \emph {et~al.}(2014)\citenamefont
  {Bolotin}, \citenamefont {Kostenko}, \citenamefont {Lemets},\ and\
  \citenamefont {Yerokhin}}]{Bolotin:2013jpa}%
  \BibitemOpen
  \bibfield  {author} {\bibinfo {author} {\bibfnamefont {Y.~L.}\ \bibnamefont
  {Bolotin}}, \bibinfo {author} {\bibfnamefont {A.}~\bibnamefont {Kostenko}},
  \bibinfo {author} {\bibfnamefont {O.~A.}\ \bibnamefont {Lemets}}, \ and\
  \bibinfo {author} {\bibfnamefont {D.~A.}\ \bibnamefont {Yerokhin}},\ }\href
  {\doibase 10.1142/S0218271815300074} {\bibfield  {journal} {\bibinfo
  {journal} {Int. J. Mod. Phys. D}\ }\textbf {\bibinfo {volume} {24}},\
  \bibinfo {pages} {1530007} (\bibinfo {year} {2014})},\ \Eprint
  {http://arxiv.org/abs/1310.0085} {arXiv:1310.0085 [astro-ph.CO]} \BibitemShut
  {NoStop}%
\bibitem [{\citenamefont {Wang}\ \emph {et~al.}(2016)\citenamefont {Wang},
  \citenamefont {Abdalla}, \citenamefont {Atrio-Barandela},\ and\ \citenamefont
  {Pavon}}]{Wang:2016lxa}%
  \BibitemOpen
  \bibfield  {author} {\bibinfo {author} {\bibfnamefont {B.}~\bibnamefont
  {Wang}}, \bibinfo {author} {\bibfnamefont {E.}~\bibnamefont {Abdalla}},
  \bibinfo {author} {\bibfnamefont {F.}~\bibnamefont {Atrio-Barandela}}, \ and\
  \bibinfo {author} {\bibfnamefont {D.}~\bibnamefont {Pavon}},\ }\href
  {\doibase 10.1088/0034-4885/79/9/096901} {\bibfield  {journal} {\bibinfo
  {journal} {Rept. Prog. Phys.}\ }\textbf {\bibinfo {volume} {79}},\ \bibinfo
  {pages} {096901} (\bibinfo {year} {2016})},\ \Eprint
  {http://arxiv.org/abs/1603.08299} {arXiv:1603.08299 [astro-ph.CO]}
  \BibitemShut {NoStop}%
\bibitem [{\citenamefont {Wang}\ \emph {et~al.}(2024)\citenamefont {Wang},
  \citenamefont {Abdalla}, \citenamefont {Atrio-Barandela},\ and\ \citenamefont
  {Pav\'on}}]{Wang:2024vmw}%
  \BibitemOpen
  \bibfield  {author} {\bibinfo {author} {\bibfnamefont {B.}~\bibnamefont
  {Wang}}, \bibinfo {author} {\bibfnamefont {E.}~\bibnamefont {Abdalla}},
  \bibinfo {author} {\bibfnamefont {F.}~\bibnamefont {Atrio-Barandela}}, \ and\
  \bibinfo {author} {\bibfnamefont {D.}~\bibnamefont {Pav\'on}},\ }\href@noop
  {} {\bibfield  {journal} {\bibinfo  {journal} {2402.00819}\ } (\bibinfo
  {year} {2024})}\BibitemShut {NoStop}%
\bibitem [{\citenamefont {Pavon}\ and\ \citenamefont
  {Zimdahl}(2005)}]{Pavon:2005yx}%
  \BibitemOpen
  \bibfield  {author} {\bibinfo {author} {\bibfnamefont {D.}~\bibnamefont
  {Pavon}}\ and\ \bibinfo {author} {\bibfnamefont {W.}~\bibnamefont
  {Zimdahl}},\ }\href {\doibase 10.1016/j.physletb.2005.08.134} {\bibfield
  {journal} {\bibinfo  {journal} {Phys. Lett. B}\ }\textbf {\bibinfo {volume}
  {628}},\ \bibinfo {pages} {206} (\bibinfo {year} {2005})},\ \Eprint
  {http://arxiv.org/abs/gr-qc/0505020} {arXiv:gr-qc/0505020} \BibitemShut
  {NoStop}%
\bibitem [{\citenamefont {Huey}\ and\ \citenamefont
  {Wandelt}(2006)}]{Huey:2004qv}%
  \BibitemOpen
  \bibfield  {author} {\bibinfo {author} {\bibfnamefont {G.}~\bibnamefont
  {Huey}}\ and\ \bibinfo {author} {\bibfnamefont {B.~D.}\ \bibnamefont
  {Wandelt}},\ }\href {\doibase 10.1103/PhysRevD.74.023519} {\bibfield
  {journal} {\bibinfo  {journal} {Phys. Rev. D}\ }\textbf {\bibinfo {volume}
  {74}},\ \bibinfo {pages} {023519} (\bibinfo {year} {2006})},\ \Eprint
  {http://arxiv.org/abs/astro-ph/0407196} {arXiv:astro-ph/0407196} \BibitemShut
  {NoStop}%
\bibitem [{\citenamefont {del Campo}\ \emph {et~al.}(2008)\citenamefont {del
  Campo}, \citenamefont {Herrera},\ and\ \citenamefont
  {Pavon}}]{delCampo:2008sr}%
  \BibitemOpen
  \bibfield  {author} {\bibinfo {author} {\bibfnamefont {S.}~\bibnamefont {del
  Campo}}, \bibinfo {author} {\bibfnamefont {R.}~\bibnamefont {Herrera}}, \
  and\ \bibinfo {author} {\bibfnamefont {D.}~\bibnamefont {Pavon}},\ }\href
  {\doibase 10.1103/PhysRevD.78.021302} {\bibfield  {journal} {\bibinfo
  {journal} {Phys. Rev. D}\ }\textbf {\bibinfo {volume} {78}},\ \bibinfo
  {pages} {021302} (\bibinfo {year} {2008})},\ \Eprint
  {http://arxiv.org/abs/0806.2116} {arXiv:0806.2116 [astro-ph]} \BibitemShut
  {NoStop}%
\bibitem [{\citenamefont {del Campo}\ \emph {et~al.}(2009)\citenamefont {del
  Campo}, \citenamefont {Herrera},\ and\ \citenamefont
  {Pavon}}]{delCampo:2008jx}%
  \BibitemOpen
  \bibfield  {author} {\bibinfo {author} {\bibfnamefont {S.}~\bibnamefont {del
  Campo}}, \bibinfo {author} {\bibfnamefont {R.}~\bibnamefont {Herrera}}, \
  and\ \bibinfo {author} {\bibfnamefont {D.}~\bibnamefont {Pavon}},\ }\href
  {\doibase 10.1088/1475-7516/2009/01/020} {\bibfield  {journal} {\bibinfo
  {journal} {JCAP}\ }\textbf {\bibinfo {volume} {01}},\ \bibinfo {pages} {020}
  (\bibinfo {year} {2009})},\ \Eprint {http://arxiv.org/abs/0812.2210}
  {arXiv:0812.2210 [gr-qc]} \BibitemShut {NoStop}%
\bibitem [{\citenamefont {Wang}\ \emph {et~al.}(2005)\citenamefont {Wang},
  \citenamefont {Gong},\ and\ \citenamefont {Abdalla}}]{Wang:2005jx}%
  \BibitemOpen
  \bibfield  {author} {\bibinfo {author} {\bibfnamefont {B.}~\bibnamefont
  {Wang}}, \bibinfo {author} {\bibfnamefont {Y.-g.}\ \bibnamefont {Gong}}, \
  and\ \bibinfo {author} {\bibfnamefont {E.}~\bibnamefont {Abdalla}},\ }\href
  {\doibase 10.1016/j.physletb.2005.08.008} {\bibfield  {journal} {\bibinfo
  {journal} {Phys. Lett. B}\ }\textbf {\bibinfo {volume} {624}},\ \bibinfo
  {pages} {141} (\bibinfo {year} {2005})},\ \Eprint
  {http://arxiv.org/abs/hep-th/0506069} {arXiv:hep-th/0506069} \BibitemShut
  {NoStop}%
\bibitem [{\citenamefont {Sadjadi}\ and\ \citenamefont
  {Honardoost}(2007)}]{Sadjadi:2006qb}%
  \BibitemOpen
  \bibfield  {author} {\bibinfo {author} {\bibfnamefont {H.~M.}\ \bibnamefont
  {Sadjadi}}\ and\ \bibinfo {author} {\bibfnamefont {M.}~\bibnamefont
  {Honardoost}},\ }\href {\doibase 10.1016/j.physletb.2007.02.016} {\bibfield
  {journal} {\bibinfo  {journal} {Phys. Lett. B}\ }\textbf {\bibinfo {volume}
  {647}},\ \bibinfo {pages} {231} (\bibinfo {year} {2007})},\ \Eprint
  {http://arxiv.org/abs/gr-qc/0609076} {arXiv:gr-qc/0609076} \BibitemShut
  {NoStop}%
\bibitem [{\citenamefont {Pan}\ and\ \citenamefont
  {Chakraborty}(2014)}]{Pan:2014afa}%
  \BibitemOpen
  \bibfield  {author} {\bibinfo {author} {\bibfnamefont {S.}~\bibnamefont
  {Pan}}\ and\ \bibinfo {author} {\bibfnamefont {S.}~\bibnamefont
  {Chakraborty}},\ }\href {\doibase 10.1142/S0218271814500928} {\bibfield
  {journal} {\bibinfo  {journal} {Int. J. Mod. Phys. D}\ }\textbf {\bibinfo
  {volume} {23}},\ \bibinfo {pages} {1450092} (\bibinfo {year} {2014})},\
  \Eprint {http://arxiv.org/abs/1410.8281} {arXiv:1410.8281 [gr-qc]}
  \BibitemShut {NoStop}%
\bibitem [{\citenamefont {Pan}\ and\ \citenamefont {Yang}(2023)}]{Pan:2023mie}%
  \BibitemOpen
  \bibfield  {author} {\bibinfo {author} {\bibfnamefont {S.}~\bibnamefont
  {Pan}}\ and\ \bibinfo {author} {\bibfnamefont {W.}~\bibnamefont {Yang}},\
  }\href@noop {} {\bibfield  {journal} {\bibinfo  {journal} {2310.07260}\ }
  (\bibinfo {year} {2023})}\BibitemShut {NoStop}%
\bibitem [{\citenamefont {Yang}\ \emph {et~al.}(2023)\citenamefont {Yang},
  \citenamefont {Pan}, \citenamefont {Mena},\ and\ \citenamefont
  {Di~Valentino}}]{Yang:2022csz}%
  \BibitemOpen
  \bibfield  {author} {\bibinfo {author} {\bibfnamefont {W.}~\bibnamefont
  {Yang}}, \bibinfo {author} {\bibfnamefont {S.}~\bibnamefont {Pan}}, \bibinfo
  {author} {\bibfnamefont {O.}~\bibnamefont {Mena}}, \ and\ \bibinfo {author}
  {\bibfnamefont {E.}~\bibnamefont {Di~Valentino}},\ }\href {\doibase
  10.1016/j.jheap.2023.09.001} {\bibfield  {journal} {\bibinfo  {journal}
  {JHEAp}\ }\textbf {\bibinfo {volume} {40}},\ \bibinfo {pages} {19} (\bibinfo
  {year} {2023})},\ \Eprint {http://arxiv.org/abs/2209.14816} {arXiv:2209.14816
  [astro-ph.CO]} \BibitemShut {NoStop}%
\bibitem [{\citenamefont {Pan}\ \emph {et~al.}(2020{\natexlab{c}})\citenamefont
  {Pan}, \citenamefont {Yang},\ and\ \citenamefont
  {Paliathanasis}}]{Pan:2020bur}%
  \BibitemOpen
  \bibfield  {author} {\bibinfo {author} {\bibfnamefont {S.}~\bibnamefont
  {Pan}}, \bibinfo {author} {\bibfnamefont {W.}~\bibnamefont {Yang}}, \ and\
  \bibinfo {author} {\bibfnamefont {A.}~\bibnamefont {Paliathanasis}},\ }\href
  {\doibase 10.1093/mnras/staa213} {\bibfield  {journal} {\bibinfo  {journal}
  {Mon. Not. Roy. Astron. Soc.}\ }\textbf {\bibinfo {volume} {493}},\ \bibinfo
  {pages} {3114} (\bibinfo {year} {2020}{\natexlab{c}})},\ \Eprint
  {http://arxiv.org/abs/2002.03408} {arXiv:2002.03408 [astro-ph.CO]}
  \BibitemShut {NoStop}%
\bibitem [{\citenamefont {Yang}\ \emph
  {et~al.}(2019{\natexlab{a}})\citenamefont {Yang}, \citenamefont {Mena},
  \citenamefont {Pan},\ and\ \citenamefont {Di~Valentino}}]{Yang:2019uzo}%
  \BibitemOpen
  \bibfield  {author} {\bibinfo {author} {\bibfnamefont {W.}~\bibnamefont
  {Yang}}, \bibinfo {author} {\bibfnamefont {O.}~\bibnamefont {Mena}}, \bibinfo
  {author} {\bibfnamefont {S.}~\bibnamefont {Pan}}, \ and\ \bibinfo {author}
  {\bibfnamefont {E.}~\bibnamefont {Di~Valentino}},\ }\href {\doibase
  10.1103/PhysRevD.100.083509} {\bibfield  {journal} {\bibinfo  {journal}
  {Phys. Rev. D}\ }\textbf {\bibinfo {volume} {100}},\ \bibinfo {pages}
  {083509} (\bibinfo {year} {2019}{\natexlab{a}})},\ \Eprint
  {http://arxiv.org/abs/1906.11697} {arXiv:1906.11697 [astro-ph.CO]}
  \BibitemShut {NoStop}%
\bibitem [{\citenamefont {Pourtsidou}\ and\ \citenamefont
  {Tram}(2016)}]{Pourtsidou:2016ico}%
  \BibitemOpen
  \bibfield  {author} {\bibinfo {author} {\bibfnamefont {A.}~\bibnamefont
  {Pourtsidou}}\ and\ \bibinfo {author} {\bibfnamefont {T.}~\bibnamefont
  {Tram}},\ }\href {\doibase 10.1103/PhysRevD.94.043518} {\bibfield  {journal}
  {\bibinfo  {journal} {Phys. Rev. D}\ }\textbf {\bibinfo {volume} {94}},\
  \bibinfo {pages} {043518} (\bibinfo {year} {2016})},\ \Eprint
  {http://arxiv.org/abs/1604.04222} {arXiv:1604.04222 [astro-ph.CO]}
  \BibitemShut {NoStop}%
\bibitem [{\citenamefont {An}\ \emph {et~al.}(2018)\citenamefont {An},
  \citenamefont {Feng},\ and\ \citenamefont {Wang}}]{An:2017crg}%
  \BibitemOpen
  \bibfield  {author} {\bibinfo {author} {\bibfnamefont {R.}~\bibnamefont
  {An}}, \bibinfo {author} {\bibfnamefont {C.}~\bibnamefont {Feng}}, \ and\
  \bibinfo {author} {\bibfnamefont {B.}~\bibnamefont {Wang}},\ }\href {\doibase
  10.1088/1475-7516/2018/02/038} {\bibfield  {journal} {\bibinfo  {journal}
  {JCAP}\ }\textbf {\bibinfo {volume} {02}},\ \bibinfo {pages} {038} (\bibinfo
  {year} {2018})},\ \Eprint {http://arxiv.org/abs/1711.06799} {arXiv:1711.06799
  [astro-ph.CO]} \BibitemShut {NoStop}%
\bibitem [{\citenamefont {Kumar}\ \emph {et~al.}(2019)\citenamefont {Kumar},
  \citenamefont {Nunes},\ and\ \citenamefont {Yadav}}]{Kumar:2019wfs}%
  \BibitemOpen
  \bibfield  {author} {\bibinfo {author} {\bibfnamefont {S.}~\bibnamefont
  {Kumar}}, \bibinfo {author} {\bibfnamefont {R.~C.}\ \bibnamefont {Nunes}}, \
  and\ \bibinfo {author} {\bibfnamefont {S.~K.}\ \bibnamefont {Yadav}},\ }\href
  {\doibase 10.1140/epjc/s10052-019-7087-7} {\bibfield  {journal} {\bibinfo
  {journal} {Eur. Phys. J. C}\ }\textbf {\bibinfo {volume} {79}},\ \bibinfo
  {pages} {576} (\bibinfo {year} {2019})},\ \Eprint
  {http://arxiv.org/abs/1903.04865} {arXiv:1903.04865 [astro-ph.CO]}
  \BibitemShut {NoStop}%
\bibitem [{\citenamefont {Tsujikawa}(2013)}]{Tsujikawa:2013fta}%
  \BibitemOpen
  \bibfield  {author} {\bibinfo {author} {\bibfnamefont {S.}~\bibnamefont
  {Tsujikawa}},\ }\href {\doibase 10.1088/0264-9381/30/21/214003} {\bibfield
  {journal} {\bibinfo  {journal} {Class. Quant. Grav.}\ }\textbf {\bibinfo
  {volume} {30}},\ \bibinfo {pages} {214003} (\bibinfo {year} {2013})},\
  \Eprint {http://arxiv.org/abs/1304.1961} {arXiv:1304.1961 [gr-qc]}
  \BibitemShut {NoStop}%
\bibitem [{\citenamefont {Steinhardt}\ \emph {et~al.}(1999)\citenamefont
  {Steinhardt}, \citenamefont {Wang},\ and\ \citenamefont
  {Zlatev}}]{Steinhardt:1999nw}%
  \BibitemOpen
  \bibfield  {author} {\bibinfo {author} {\bibfnamefont {P.~J.}\ \bibnamefont
  {Steinhardt}}, \bibinfo {author} {\bibfnamefont {L.-M.}\ \bibnamefont
  {Wang}}, \ and\ \bibinfo {author} {\bibfnamefont {I.}~\bibnamefont
  {Zlatev}},\ }\href {\doibase 10.1103/PhysRevD.59.123504} {\bibfield
  {journal} {\bibinfo  {journal} {Phys. Rev. D}\ }\textbf {\bibinfo {volume}
  {59}},\ \bibinfo {pages} {123504} (\bibinfo {year} {1999})},\ \Eprint
  {http://arxiv.org/abs/astro-ph/9812313} {arXiv:astro-ph/9812313} \BibitemShut
  {NoStop}%
\bibitem [{\citenamefont {Kiselev}(2008)}]{Kiselev:2006jy}%
  \BibitemOpen
  \bibfield  {author} {\bibinfo {author} {\bibfnamefont {V.~V.}\ \bibnamefont
  {Kiselev}},\ }\href {\doibase 10.1088/1475-7516/2008/01/019} {\bibfield
  {journal} {\bibinfo  {journal} {JCAP}\ }\textbf {\bibinfo {volume} {01}},\
  \bibinfo {pages} {019} (\bibinfo {year} {2008})},\ \Eprint
  {http://arxiv.org/abs/gr-qc/0611064} {arXiv:gr-qc/0611064} \BibitemShut
  {NoStop}%
\bibitem [{\citenamefont {Roy}\ and\ \citenamefont
  {Banerjee}(2014)}]{Roy:2013wqa}%
  \BibitemOpen
  \bibfield  {author} {\bibinfo {author} {\bibfnamefont {N.}~\bibnamefont
  {Roy}}\ and\ \bibinfo {author} {\bibfnamefont {N.}~\bibnamefont {Banerjee}},\
  }\href {\doibase 10.1007/s10714-013-1651-5} {\bibfield  {journal} {\bibinfo
  {journal} {Gen. Rel. Grav.}\ }\textbf {\bibinfo {volume} {46}},\ \bibinfo
  {pages} {1651} (\bibinfo {year} {2014})},\ \Eprint
  {http://arxiv.org/abs/1312.2670} {arXiv:1312.2670 [gr-qc]} \BibitemShut
  {NoStop}%
\bibitem [{\citenamefont {Gong}(2014)}]{Gong:2014dia}%
  \BibitemOpen
  \bibfield  {author} {\bibinfo {author} {\bibfnamefont {Y.}~\bibnamefont
  {Gong}},\ }\href {\doibase 10.1016/j.physletb.2014.03.013} {\bibfield
  {journal} {\bibinfo  {journal} {Phys. Lett. B}\ }\textbf {\bibinfo {volume}
  {731}},\ \bibinfo {pages} {342} (\bibinfo {year} {2014})},\ \Eprint
  {http://arxiv.org/abs/1401.1959} {arXiv:1401.1959 [gr-qc]} \BibitemShut
  {NoStop}%
\bibitem [{\citenamefont {Haro}\ \emph {et~al.}(2020)\citenamefont {Haro},
  \citenamefont {Amor\'os},\ and\ \citenamefont {Pan}}]{Haro:2019peq}%
  \BibitemOpen
  \bibfield  {author} {\bibinfo {author} {\bibfnamefont {J.}~\bibnamefont
  {Haro}}, \bibinfo {author} {\bibfnamefont {J.}~\bibnamefont {Amor\'os}}, \
  and\ \bibinfo {author} {\bibfnamefont {S.}~\bibnamefont {Pan}},\ }\href
  {\doibase 10.1140/epjc/s10052-020-7950-6} {\bibfield  {journal} {\bibinfo
  {journal} {Eur. Phys. J. C}\ }\textbf {\bibinfo {volume} {80}},\ \bibinfo
  {pages} {404} (\bibinfo {year} {2020})},\ \Eprint
  {http://arxiv.org/abs/1908.01516} {arXiv:1908.01516 [gr-qc]} \BibitemShut
  {NoStop}%
\bibitem [{\citenamefont {Ure\~na L\'opez}\ and\ \citenamefont
  {Roy}(2020)}]{Urena-Lopez:2020npg}%
  \BibitemOpen
  \bibfield  {author} {\bibinfo {author} {\bibfnamefont {L.~A.}\ \bibnamefont
  {Ure\~na L\'opez}}\ and\ \bibinfo {author} {\bibfnamefont {N.}~\bibnamefont
  {Roy}},\ }\href {\doibase 10.1103/PhysRevD.102.063510} {\bibfield  {journal}
  {\bibinfo  {journal} {Phys. Rev. D}\ }\textbf {\bibinfo {volume} {102}},\
  \bibinfo {pages} {063510} (\bibinfo {year} {2020})},\ \Eprint
  {http://arxiv.org/abs/2007.08873} {arXiv:2007.08873 [astro-ph.CO]}
  \BibitemShut {NoStop}%
\bibitem [{\citenamefont {Alho}\ \emph {et~al.}(2024)\citenamefont {Alho},
  \citenamefont {Uggla},\ and\ \citenamefont {Wainwright}}]{Alho:2023xel}%
  \BibitemOpen
  \bibfield  {author} {\bibinfo {author} {\bibfnamefont {A.}~\bibnamefont
  {Alho}}, \bibinfo {author} {\bibfnamefont {C.}~\bibnamefont {Uggla}}, \ and\
  \bibinfo {author} {\bibfnamefont {J.}~\bibnamefont {Wainwright}},\ }\href
  {\doibase 10.1016/j.dark.2024.101433} {\bibfield  {journal} {\bibinfo
  {journal} {Phys. Dark Univ.}\ }\textbf {\bibinfo {volume} {44}},\ \bibinfo
  {pages} {101433} (\bibinfo {year} {2024})},\ \Eprint
  {http://arxiv.org/abs/2311.16775} {arXiv:2311.16775 [gr-qc]} \BibitemShut
  {NoStop}%
\bibitem [{\citenamefont {Xia}(2009)}]{Xia:2009zzb}%
  \BibitemOpen
  \bibfield  {author} {\bibinfo {author} {\bibfnamefont {J.-Q.}\ \bibnamefont
  {Xia}},\ }\href {\doibase 10.1103/PhysRevD.80.103514} {\bibfield  {journal}
  {\bibinfo  {journal} {Phys. Rev. D}\ }\textbf {\bibinfo {volume} {80}},\
  \bibinfo {pages} {103514} (\bibinfo {year} {2009})},\ \Eprint
  {http://arxiv.org/abs/0911.4820} {arXiv:0911.4820 [astro-ph.CO]} \BibitemShut
  {NoStop}%
\bibitem [{\citenamefont {van~de Bruck}\ \emph {et~al.}(2017)\citenamefont
  {van~de Bruck}, \citenamefont {Mifsud},\ and\ \citenamefont
  {Morrice}}]{vandeBruck:2016hpz}%
  \BibitemOpen
  \bibfield  {author} {\bibinfo {author} {\bibfnamefont {C.}~\bibnamefont
  {van~de Bruck}}, \bibinfo {author} {\bibfnamefont {J.}~\bibnamefont
  {Mifsud}}, \ and\ \bibinfo {author} {\bibfnamefont {J.}~\bibnamefont
  {Morrice}},\ }\href {\doibase 10.1103/PhysRevD.95.043513} {\bibfield
  {journal} {\bibinfo  {journal} {Phys. Rev. D}\ }\textbf {\bibinfo {volume}
  {95}},\ \bibinfo {pages} {043513} (\bibinfo {year} {2017})},\ \Eprint
  {http://arxiv.org/abs/1609.09855} {arXiv:1609.09855 [astro-ph.CO]}
  \BibitemShut {NoStop}%
\bibitem [{\citenamefont {Mifsud}\ and\ \citenamefont {Van
  De~Bruck}(2017)}]{Mifsud:2017fsy}%
  \BibitemOpen
  \bibfield  {author} {\bibinfo {author} {\bibfnamefont {J.}~\bibnamefont
  {Mifsud}}\ and\ \bibinfo {author} {\bibfnamefont {C.}~\bibnamefont {Van
  De~Bruck}},\ }\href {\doibase 10.1088/1475-7516/2017/11/001} {\bibfield
  {journal} {\bibinfo  {journal} {JCAP}\ }\textbf {\bibinfo {volume} {11}},\
  \bibinfo {pages} {001} (\bibinfo {year} {2017})},\ \Eprint
  {http://arxiv.org/abs/1707.07667} {arXiv:1707.07667 [astro-ph.CO]}
  \BibitemShut {NoStop}%
\bibitem [{\citenamefont {Van De~Bruck}\ and\ \citenamefont
  {Mifsud}(2018)}]{VanDeBruck:2017mua}%
  \BibitemOpen
  \bibfield  {author} {\bibinfo {author} {\bibfnamefont {C.}~\bibnamefont {Van
  De~Bruck}}\ and\ \bibinfo {author} {\bibfnamefont {J.}~\bibnamefont
  {Mifsud}},\ }\href {\doibase 10.1103/PhysRevD.97.023506} {\bibfield
  {journal} {\bibinfo  {journal} {Phys. Rev. D}\ }\textbf {\bibinfo {volume}
  {97}},\ \bibinfo {pages} {023506} (\bibinfo {year} {2018})},\ \Eprint
  {http://arxiv.org/abs/1709.04882} {arXiv:1709.04882 [astro-ph.CO]}
  \BibitemShut {NoStop}%
\bibitem [{\citenamefont {Liu}\ \emph {et~al.}(2020)\citenamefont {Liu},
  \citenamefont {Heneka},\ and\ \citenamefont {Amendola}}]{Liu:2019ygl}%
  \BibitemOpen
  \bibfield  {author} {\bibinfo {author} {\bibfnamefont {X.-W.}\ \bibnamefont
  {Liu}}, \bibinfo {author} {\bibfnamefont {C.}~\bibnamefont {Heneka}}, \ and\
  \bibinfo {author} {\bibfnamefont {L.}~\bibnamefont {Amendola}},\ }\href
  {\doibase 10.1088/1475-7516/2020/05/038} {\bibfield  {journal} {\bibinfo
  {journal} {JCAP}\ }\textbf {\bibinfo {volume} {05}},\ \bibinfo {pages} {038}
  (\bibinfo {year} {2020})},\ \Eprint {http://arxiv.org/abs/1910.02763}
  {arXiv:1910.02763 [astro-ph.CO]} \BibitemShut {NoStop}%
\bibitem [{\citenamefont {da~Fonseca}\ \emph {et~al.}(2022)\citenamefont
  {da~Fonseca}, \citenamefont {Barreiro},\ and\ \citenamefont
  {Nunes}}]{daFonseca:2021imp}%
  \BibitemOpen
  \bibfield  {author} {\bibinfo {author} {\bibfnamefont {V.}~\bibnamefont
  {da~Fonseca}}, \bibinfo {author} {\bibfnamefont {T.}~\bibnamefont
  {Barreiro}}, \ and\ \bibinfo {author} {\bibfnamefont {N.~J.}\ \bibnamefont
  {Nunes}},\ }\href {\doibase 10.1016/j.dark.2021.100940} {\bibfield  {journal}
  {\bibinfo  {journal} {Phys. Dark Univ.}\ }\textbf {\bibinfo {volume} {35}},\
  \bibinfo {pages} {100940} (\bibinfo {year} {2022})},\ \Eprint
  {http://arxiv.org/abs/2104.14889} {arXiv:2104.14889 [astro-ph.CO]}
  \BibitemShut {NoStop}%
\bibitem [{\citenamefont {Barros}\ \emph {et~al.}(2023)\citenamefont {Barros},
  \citenamefont {Castel\~ao}, \citenamefont {da~Fonseca}, \citenamefont
  {Barreiro}, \citenamefont {Nunes},\ and\ \citenamefont
  {Tereno}}]{Barros:2022bdv}%
  \BibitemOpen
  \bibfield  {author} {\bibinfo {author} {\bibfnamefont {B.~J.}\ \bibnamefont
  {Barros}}, \bibinfo {author} {\bibfnamefont {D.}~\bibnamefont {Castel\~ao}},
  \bibinfo {author} {\bibfnamefont {V.}~\bibnamefont {da~Fonseca}}, \bibinfo
  {author} {\bibfnamefont {T.}~\bibnamefont {Barreiro}}, \bibinfo {author}
  {\bibfnamefont {N.~J.}\ \bibnamefont {Nunes}}, \ and\ \bibinfo {author}
  {\bibfnamefont {I.}~\bibnamefont {Tereno}},\ }\href {\doibase
  10.1088/1475-7516/2023/01/013} {\bibfield  {journal} {\bibinfo  {journal}
  {JCAP}\ }\textbf {\bibinfo {volume} {01}},\ \bibinfo {pages} {013} (\bibinfo
  {year} {2023})},\ \Eprint {http://arxiv.org/abs/2209.04468} {arXiv:2209.04468
  [astro-ph.CO]} \BibitemShut {NoStop}%
\bibitem [{\citenamefont {Teixeira}\ \emph {et~al.}(2023)\citenamefont
  {Teixeira}, \citenamefont {Daniel}, \citenamefont {Frusciante},\ and\
  \citenamefont {van~de Bruck}}]{Teixeira:2023zjt}%
  \BibitemOpen
  \bibfield  {author} {\bibinfo {author} {\bibfnamefont {E.~M.}\ \bibnamefont
  {Teixeira}}, \bibinfo {author} {\bibfnamefont {R.}~\bibnamefont {Daniel}},
  \bibinfo {author} {\bibfnamefont {N.}~\bibnamefont {Frusciante}}, \ and\
  \bibinfo {author} {\bibfnamefont {C.}~\bibnamefont {van~de Bruck}},\ }\href
  {\doibase 10.1103/PhysRevD.108.084070} {\bibfield  {journal} {\bibinfo
  {journal} {Phys. Rev. D}\ }\textbf {\bibinfo {volume} {108}},\ \bibinfo
  {pages} {084070} (\bibinfo {year} {2023})},\ \Eprint
  {http://arxiv.org/abs/2309.06544} {arXiv:2309.06544 [astro-ph.CO]}
  \BibitemShut {NoStop}%
\bibitem [{\citenamefont {Poulin}\ \emph {et~al.}(2019)\citenamefont {Poulin},
  \citenamefont {Smith}, \citenamefont {Karwal},\ and\ \citenamefont
  {Kamionkowski}}]{Poulin:2018cxd}%
  \BibitemOpen
  \bibfield  {author} {\bibinfo {author} {\bibfnamefont {V.}~\bibnamefont
  {Poulin}}, \bibinfo {author} {\bibfnamefont {T.~L.}\ \bibnamefont {Smith}},
  \bibinfo {author} {\bibfnamefont {T.}~\bibnamefont {Karwal}}, \ and\ \bibinfo
  {author} {\bibfnamefont {M.}~\bibnamefont {Kamionkowski}},\ }\href {\doibase
  10.1103/PhysRevLett.122.221301} {\bibfield  {journal} {\bibinfo  {journal}
  {Phys. Rev. Lett.}\ }\textbf {\bibinfo {volume} {122}},\ \bibinfo {pages}
  {221301} (\bibinfo {year} {2019})},\ \Eprint
  {http://arxiv.org/abs/1811.04083} {arXiv:1811.04083 [astro-ph.CO]}
  \BibitemShut {NoStop}%
\bibitem [{\citenamefont {Sakstein}\ and\ \citenamefont
  {Trodden}(2020)}]{Sakstein:2019fmf}%
  \BibitemOpen
  \bibfield  {author} {\bibinfo {author} {\bibfnamefont {J.}~\bibnamefont
  {Sakstein}}\ and\ \bibinfo {author} {\bibfnamefont {M.}~\bibnamefont
  {Trodden}},\ }\href {\doibase 10.1103/PhysRevLett.124.161301} {\bibfield
  {journal} {\bibinfo  {journal} {Phys. Rev. Lett.}\ }\textbf {\bibinfo
  {volume} {124}},\ \bibinfo {pages} {161301} (\bibinfo {year} {2020})},\
  \Eprint {http://arxiv.org/abs/1911.11760} {arXiv:1911.11760 [astro-ph.CO]}
  \BibitemShut {NoStop}%
\bibitem [{\citenamefont {Niedermann}\ and\ \citenamefont
  {Sloth}(2021)}]{Niedermann:2019olb}%
  \BibitemOpen
  \bibfield  {author} {\bibinfo {author} {\bibfnamefont {F.}~\bibnamefont
  {Niedermann}}\ and\ \bibinfo {author} {\bibfnamefont {M.~S.}\ \bibnamefont
  {Sloth}},\ }\href {\doibase 10.1103/PhysRevD.103.L041303} {\bibfield
  {journal} {\bibinfo  {journal} {Phys. Rev. D}\ }\textbf {\bibinfo {volume}
  {103}},\ \bibinfo {pages} {L041303} (\bibinfo {year} {2021})},\ \Eprint
  {http://arxiv.org/abs/1910.10739} {arXiv:1910.10739 [astro-ph.CO]}
  \BibitemShut {NoStop}%
\bibitem [{\citenamefont {Chudaykin}\ \emph {et~al.}(2021)\citenamefont
  {Chudaykin}, \citenamefont {Gorbunov},\ and\ \citenamefont
  {Nedelko}}]{Chudaykin:2020igl}%
  \BibitemOpen
  \bibfield  {author} {\bibinfo {author} {\bibfnamefont {A.}~\bibnamefont
  {Chudaykin}}, \bibinfo {author} {\bibfnamefont {D.}~\bibnamefont {Gorbunov}},
  \ and\ \bibinfo {author} {\bibfnamefont {N.}~\bibnamefont {Nedelko}},\ }\href
  {\doibase 10.1103/PhysRevD.103.043529} {\bibfield  {journal} {\bibinfo
  {journal} {Phys. Rev. D}\ }\textbf {\bibinfo {volume} {103}},\ \bibinfo
  {pages} {043529} (\bibinfo {year} {2021})},\ \Eprint
  {http://arxiv.org/abs/2011.04682} {arXiv:2011.04682 [astro-ph.CO]}
  \BibitemShut {NoStop}%
\bibitem [{\citenamefont {Niedermann}\ and\ \citenamefont
  {Sloth}(2020)}]{Niedermann:2020dwg}%
  \BibitemOpen
  \bibfield  {author} {\bibinfo {author} {\bibfnamefont {F.}~\bibnamefont
  {Niedermann}}\ and\ \bibinfo {author} {\bibfnamefont {M.~S.}\ \bibnamefont
  {Sloth}},\ }\href {\doibase 10.1103/PhysRevD.102.063527} {\bibfield
  {journal} {\bibinfo  {journal} {Phys. Rev. D}\ }\textbf {\bibinfo {volume}
  {102}},\ \bibinfo {pages} {063527} (\bibinfo {year} {2020})},\ \Eprint
  {http://arxiv.org/abs/2006.06686} {arXiv:2006.06686 [astro-ph.CO]}
  \BibitemShut {NoStop}%
\bibitem [{\citenamefont {Smith}\ \emph {et~al.}(2021)\citenamefont {Smith},
  \citenamefont {Poulin}, \citenamefont {Bernal}, \citenamefont {Boddy},
  \citenamefont {Kamionkowski},\ and\ \citenamefont {Murgia}}]{Smith:2020rxx}%
  \BibitemOpen
  \bibfield  {author} {\bibinfo {author} {\bibfnamefont {T.~L.}\ \bibnamefont
  {Smith}}, \bibinfo {author} {\bibfnamefont {V.}~\bibnamefont {Poulin}},
  \bibinfo {author} {\bibfnamefont {J.~L.}\ \bibnamefont {Bernal}}, \bibinfo
  {author} {\bibfnamefont {K.~K.}\ \bibnamefont {Boddy}}, \bibinfo {author}
  {\bibfnamefont {M.}~\bibnamefont {Kamionkowski}}, \ and\ \bibinfo {author}
  {\bibfnamefont {R.}~\bibnamefont {Murgia}},\ }\href {\doibase
  10.1103/PhysRevD.103.123542} {\bibfield  {journal} {\bibinfo  {journal}
  {Phys. Rev. D}\ }\textbf {\bibinfo {volume} {103}},\ \bibinfo {pages}
  {123542} (\bibinfo {year} {2021})},\ \Eprint
  {http://arxiv.org/abs/2009.10740} {arXiv:2009.10740 [astro-ph.CO]}
  \BibitemShut {NoStop}%
\bibitem [{\citenamefont {Seto}\ and\ \citenamefont
  {Toda}(2021)}]{Seto:2021xua}%
  \BibitemOpen
  \bibfield  {author} {\bibinfo {author} {\bibfnamefont {O.}~\bibnamefont
  {Seto}}\ and\ \bibinfo {author} {\bibfnamefont {Y.}~\bibnamefont {Toda}},\
  }\href {\doibase 10.1103/PhysRevD.103.123501} {\bibfield  {journal} {\bibinfo
   {journal} {Phys. Rev. D}\ }\textbf {\bibinfo {volume} {103}},\ \bibinfo
  {pages} {123501} (\bibinfo {year} {2021})},\ \Eprint
  {http://arxiv.org/abs/2101.03740} {arXiv:2101.03740 [astro-ph.CO]}
  \BibitemShut {NoStop}%
\bibitem [{\citenamefont {Agrawal}\ \emph {et~al.}(2023)\citenamefont
  {Agrawal}, \citenamefont {Cyr-Racine}, \citenamefont {Pinner},\ and\
  \citenamefont {Randall}}]{Agrawal:2019lmo}%
  \BibitemOpen
  \bibfield  {author} {\bibinfo {author} {\bibfnamefont {P.}~\bibnamefont
  {Agrawal}}, \bibinfo {author} {\bibfnamefont {F.-Y.}\ \bibnamefont
  {Cyr-Racine}}, \bibinfo {author} {\bibfnamefont {D.}~\bibnamefont {Pinner}},
  \ and\ \bibinfo {author} {\bibfnamefont {L.}~\bibnamefont {Randall}},\ }\href
  {\doibase 10.1016/j.dark.2023.101347} {\bibfield  {journal} {\bibinfo
  {journal} {Phys. Dark Univ.}\ }\textbf {\bibinfo {volume} {42}},\ \bibinfo
  {pages} {101347} (\bibinfo {year} {2023})},\ \Eprint
  {http://arxiv.org/abs/1904.01016} {arXiv:1904.01016 [astro-ph.CO]}
  \BibitemShut {NoStop}%
\bibitem [{\citenamefont {Karwal}\ and\ \citenamefont
  {Kamionkowski}(2016)}]{Karwal:2016vyq}%
  \BibitemOpen
  \bibfield  {author} {\bibinfo {author} {\bibfnamefont {T.}~\bibnamefont
  {Karwal}}\ and\ \bibinfo {author} {\bibfnamefont {M.}~\bibnamefont
  {Kamionkowski}},\ }\href {\doibase 10.1103/PhysRevD.94.103523} {\bibfield
  {journal} {\bibinfo  {journal} {Phys. Rev. D}\ }\textbf {\bibinfo {volume}
  {94}},\ \bibinfo {pages} {103523} (\bibinfo {year} {2016})},\ \Eprint
  {http://arxiv.org/abs/1608.01309} {arXiv:1608.01309 [astro-ph.CO]}
  \BibitemShut {NoStop}%
\bibitem [{\citenamefont {Poulin}\ \emph {et~al.}(2023)\citenamefont {Poulin},
  \citenamefont {Smith},\ and\ \citenamefont {Karwal}}]{Poulin:2023lkg}%
  \BibitemOpen
  \bibfield  {author} {\bibinfo {author} {\bibfnamefont {V.}~\bibnamefont
  {Poulin}}, \bibinfo {author} {\bibfnamefont {T.~L.}\ \bibnamefont {Smith}}, \
  and\ \bibinfo {author} {\bibfnamefont {T.}~\bibnamefont {Karwal}},\ }\href
  {\doibase 10.1016/j.dark.2023.101348} {\bibfield  {journal} {\bibinfo
  {journal} {Phys. Dark Univ.}\ }\textbf {\bibinfo {volume} {42}},\ \bibinfo
  {pages} {101348} (\bibinfo {year} {2023})},\ \Eprint
  {http://arxiv.org/abs/2302.09032} {arXiv:2302.09032 [astro-ph.CO]}
  \BibitemShut {NoStop}%
\bibitem [{\citenamefont {Knox}\ and\ \citenamefont
  {Millea}(2020)}]{Knox:2019rjx}%
  \BibitemOpen
  \bibfield  {author} {\bibinfo {author} {\bibfnamefont {L.}~\bibnamefont
  {Knox}}\ and\ \bibinfo {author} {\bibfnamefont {M.}~\bibnamefont {Millea}},\
  }\href {\doibase 10.1103/PhysRevD.101.043533} {\bibfield  {journal} {\bibinfo
   {journal} {Phys. Rev. D}\ }\textbf {\bibinfo {volume} {101}},\ \bibinfo
  {pages} {043533} (\bibinfo {year} {2020})},\ \Eprint
  {http://arxiv.org/abs/1908.03663} {arXiv:1908.03663 [astro-ph.CO]}
  \BibitemShut {NoStop}%
\bibitem [{\citenamefont {G\'omez-Valent}\ \emph {et~al.}(2020)\citenamefont
  {G\'omez-Valent}, \citenamefont {Pettorino},\ and\ \citenamefont
  {Amendola}}]{Gomez-Valent:2020mqn}%
  \BibitemOpen
  \bibfield  {author} {\bibinfo {author} {\bibfnamefont {A.}~\bibnamefont
  {G\'omez-Valent}}, \bibinfo {author} {\bibfnamefont {V.}~\bibnamefont
  {Pettorino}}, \ and\ \bibinfo {author} {\bibfnamefont {L.}~\bibnamefont
  {Amendola}},\ }\href {\doibase 10.1103/PhysRevD.101.123513} {\bibfield
  {journal} {\bibinfo  {journal} {Phys. Rev. D}\ }\textbf {\bibinfo {volume}
  {101}},\ \bibinfo {pages} {123513} (\bibinfo {year} {2020})},\ \Eprint
  {http://arxiv.org/abs/2004.00610} {arXiv:2004.00610 [astro-ph.CO]}
  \BibitemShut {NoStop}%
\bibitem [{\citenamefont {Wu}\ \emph {et~al.}(2020)\citenamefont {Wu},
  \citenamefont {Motloch}, \citenamefont {Hu},\ and\ \citenamefont
  {Raveri}}]{Wu:2020nxz}%
  \BibitemOpen
  \bibfield  {author} {\bibinfo {author} {\bibfnamefont {W.~L.~K.}\
  \bibnamefont {Wu}}, \bibinfo {author} {\bibfnamefont {P.}~\bibnamefont
  {Motloch}}, \bibinfo {author} {\bibfnamefont {W.}~\bibnamefont {Hu}}, \ and\
  \bibinfo {author} {\bibfnamefont {M.}~\bibnamefont {Raveri}},\ }\href
  {\doibase 10.1103/PhysRevD.102.023510} {\bibfield  {journal} {\bibinfo
  {journal} {Phys. Rev. D}\ }\textbf {\bibinfo {volume} {102}},\ \bibinfo
  {pages} {023510} (\bibinfo {year} {2020})},\ \Eprint
  {http://arxiv.org/abs/2004.10207} {arXiv:2004.10207 [astro-ph.CO]}
  \BibitemShut {NoStop}%
\bibitem [{\citenamefont {Jedamzik}\ \emph {et~al.}(2021)\citenamefont
  {Jedamzik}, \citenamefont {Pogosian},\ and\ \citenamefont
  {Zhao}}]{Jedamzik:2020zmd}%
  \BibitemOpen
  \bibfield  {author} {\bibinfo {author} {\bibfnamefont {K.}~\bibnamefont
  {Jedamzik}}, \bibinfo {author} {\bibfnamefont {L.}~\bibnamefont {Pogosian}},
  \ and\ \bibinfo {author} {\bibfnamefont {G.-B.}\ \bibnamefont {Zhao}},\
  }\href {\doibase 10.1038/s42005-021-00628-x} {\bibfield  {journal} {\bibinfo
  {journal} {Commun. in Phys.}\ }\textbf {\bibinfo {volume} {4}},\ \bibinfo
  {pages} {123} (\bibinfo {year} {2021})},\ \Eprint
  {http://arxiv.org/abs/2010.04158} {arXiv:2010.04158 [astro-ph.CO]}
  \BibitemShut {NoStop}%
\bibitem [{\citenamefont {Maziashvili}\ and\ \citenamefont
  {Tsintsabadze}(2024)}]{Maziashvili:2023rjr}%
  \BibitemOpen
  \bibfield  {author} {\bibinfo {author} {\bibfnamefont {M.}~\bibnamefont
  {Maziashvili}}\ and\ \bibinfo {author} {\bibfnamefont {V.}~\bibnamefont
  {Tsintsabadze}},\ }\href {\doibase 10.1016/j.astropartphys.2023.102901}
  {\bibfield  {journal} {\bibinfo  {journal} {Astropart. Phys.}\ }\textbf
  {\bibinfo {volume} {154}},\ \bibinfo {pages} {102901} (\bibinfo {year}
  {2024})},\ \Eprint {http://arxiv.org/abs/2302.00380} {arXiv:2302.00380
  [astro-ph.CO]} \BibitemShut {NoStop}%
\bibitem [{\citenamefont {Ratra}\ and\ \citenamefont
  {Peebles}(1988)}]{Ratra:1987rm}%
  \BibitemOpen
  \bibfield  {author} {\bibinfo {author} {\bibfnamefont {B.}~\bibnamefont
  {Ratra}}\ and\ \bibinfo {author} {\bibfnamefont {P.~J.~E.}\ \bibnamefont
  {Peebles}},\ }\href {\doibase 10.1103/PhysRevD.37.3406} {\bibfield  {journal}
  {\bibinfo  {journal} {Phys. Rev. D}\ }\textbf {\bibinfo {volume} {37}},\
  \bibinfo {pages} {3406} (\bibinfo {year} {1988})}\BibitemShut {NoStop}%
\bibitem [{\citenamefont {Kesden}\ and\ \citenamefont
  {Kamionkowski}(2006)}]{Kesden_2006}%
  \BibitemOpen
  \bibfield  {author} {\bibinfo {author} {\bibfnamefont {M.}~\bibnamefont
  {Kesden}}\ and\ \bibinfo {author} {\bibfnamefont {M.}~\bibnamefont
  {Kamionkowski}},\ }\href {\doibase 10.1103/physrevd.74.083007} {\bibfield
  {journal} {\bibinfo  {journal} {Physical Review D}\ }\textbf {\bibinfo
  {volume} {74}} (\bibinfo {year} {2006}),\
  10.1103/physrevd.74.083007}\BibitemShut {NoStop}%
\bibitem [{\citenamefont {Khoury}\ and\ \citenamefont
  {Weltman}(2004)}]{Khoury_2004}%
  \BibitemOpen
  \bibfield  {author} {\bibinfo {author} {\bibfnamefont {J.}~\bibnamefont
  {Khoury}}\ and\ \bibinfo {author} {\bibfnamefont {A.}~\bibnamefont
  {Weltman}},\ }\href {\doibase 10.1103/physrevd.69.044026} {\bibfield
  {journal} {\bibinfo  {journal} {Physical Review D}\ }\textbf {\bibinfo
  {volume} {69}} (\bibinfo {year} {2004}),\
  10.1103/physrevd.69.044026}\BibitemShut {NoStop}%
\bibitem [{\citenamefont {Vainshtein}(1972)}]{VAINSHTEIN1972393}%
  \BibitemOpen
  \bibfield  {author} {\bibinfo {author} {\bibfnamefont {A.}~\bibnamefont
  {Vainshtein}},\ }\href {\doibase
  https://doi.org/10.1016/0370-2693(72)90147-5} {\bibfield  {journal} {\bibinfo
   {journal} {Physics Letters B}\ }\textbf {\bibinfo {volume} {39}},\ \bibinfo
  {pages} {393} (\bibinfo {year} {1972})}\BibitemShut {NoStop}%
\bibitem [{\citenamefont {Hu}\ and\ \citenamefont
  {Sugiyama}(1996)}]{Hu:1995en}%
  \BibitemOpen
  \bibfield  {author} {\bibinfo {author} {\bibfnamefont {W.}~\bibnamefont
  {Hu}}\ and\ \bibinfo {author} {\bibfnamefont {N.}~\bibnamefont {Sugiyama}},\
  }\href {\doibase 10.1086/177989} {\bibfield  {journal} {\bibinfo  {journal}
  {Astrophys. J.}\ }\textbf {\bibinfo {volume} {471}},\ \bibinfo {pages} {542}
  (\bibinfo {year} {1996})},\ \Eprint {http://arxiv.org/abs/astro-ph/9510117}
  {arXiv:astro-ph/9510117} \BibitemShut {NoStop}%
\bibitem [{\citenamefont {Ma}\ and\ \citenamefont
  {Bertschinger}(1995)}]{Ma:1995ey}%
  \BibitemOpen
  \bibfield  {author} {\bibinfo {author} {\bibfnamefont {C.-P.}\ \bibnamefont
  {Ma}}\ and\ \bibinfo {author} {\bibfnamefont {E.}~\bibnamefont
  {Bertschinger}},\ }\href {\doibase 10.1086/176550} {\bibfield  {journal}
  {\bibinfo  {journal} {Astrophys. J.}\ }\textbf {\bibinfo {volume} {455}},\
  \bibinfo {pages} {7} (\bibinfo {year} {1995})},\ \Eprint
  {http://arxiv.org/abs/astro-ph/9506072} {arXiv:astro-ph/9506072} \BibitemShut
  {NoStop}%
\bibitem [{\citenamefont {Weller}\ and\ \citenamefont
  {Lewis}(2003)}]{Weller_2003}%
  \BibitemOpen
  \bibfield  {author} {\bibinfo {author} {\bibfnamefont {J.}~\bibnamefont
  {Weller}}\ and\ \bibinfo {author} {\bibfnamefont {A.~M.}\ \bibnamefont
  {Lewis}},\ }\href {\doibase 10.1111/j.1365-2966.2003.07144.x} {\bibfield
  {journal} {\bibinfo  {journal} {Monthly Notices of the Royal Astronomical
  Society}\ }\textbf {\bibinfo {volume} {346}},\ \bibinfo {pages} {987–993}
  (\bibinfo {year} {2003})}\BibitemShut {NoStop}%
\bibitem [{\citenamefont {Aghanim}\ \emph
  {et~al.}(2020{\natexlab{b}})\citenamefont {Aghanim} \emph
  {et~al.}}]{Aghanim:2019ame}%
  \BibitemOpen
  \bibfield  {author} {\bibinfo {author} {\bibfnamefont {N.}~\bibnamefont
  {Aghanim}} \emph {et~al.} (\bibinfo {collaboration} {Planck}),\ }\href
  {\doibase 10.1051/0004-6361/201936386} {\bibfield  {journal} {\bibinfo
  {journal} {Astron. Astrophys.}\ }\textbf {\bibinfo {volume} {641}},\ \bibinfo
  {pages} {A5} (\bibinfo {year} {2020}{\natexlab{b}})},\ \Eprint
  {http://arxiv.org/abs/1907.12875} {arXiv:1907.12875 [astro-ph.CO]}
  \BibitemShut {NoStop}%
\bibitem [{\citenamefont {Beutler}\ \emph {et~al.}(2011)\citenamefont
  {Beutler}, \citenamefont {Blake}, \citenamefont {Colless}, \citenamefont
  {Jones}, \citenamefont {Staveley-Smith}, \citenamefont {Campbell},
  \citenamefont {Parker}, \citenamefont {Saunders},\ and\ \citenamefont
  {Watson}}]{Beutler:2011hx}%
  \BibitemOpen
  \bibfield  {author} {\bibinfo {author} {\bibfnamefont {F.}~\bibnamefont
  {Beutler}}, \bibinfo {author} {\bibfnamefont {C.}~\bibnamefont {Blake}},
  \bibinfo {author} {\bibfnamefont {M.}~\bibnamefont {Colless}}, \bibinfo
  {author} {\bibfnamefont {D.}~\bibnamefont {Jones}}, \bibinfo {author}
  {\bibfnamefont {L.}~\bibnamefont {Staveley-Smith}}, \bibinfo {author}
  {\bibfnamefont {L.}~\bibnamefont {Campbell}}, \bibinfo {author}
  {\bibfnamefont {Q.}~\bibnamefont {Parker}}, \bibinfo {author} {\bibfnamefont
  {W.}~\bibnamefont {Saunders}}, \ and\ \bibinfo {author} {\bibfnamefont
  {F.}~\bibnamefont {Watson}},\ }\href {\doibase
  10.1111/j.1365-2966.2011.19250.x} {\bibfield  {journal} {\bibinfo  {journal}
  {Mon. Not. Roy. Astron. Soc.}\ }\textbf {\bibinfo {volume} {416}},\ \bibinfo
  {pages} {3017} (\bibinfo {year} {2011})},\ \Eprint
  {http://arxiv.org/abs/1106.3366} {arXiv:1106.3366 [astro-ph.CO]} \BibitemShut
  {NoStop}%
\bibitem [{\citenamefont {Ross}\ \emph {et~al.}(2015)\citenamefont {Ross},
  \citenamefont {Samushia}, \citenamefont {Howlett}, \citenamefont {Percival},
  \citenamefont {Burden},\ and\ \citenamefont {Manera}}]{Ross:2014qpa}%
  \BibitemOpen
  \bibfield  {author} {\bibinfo {author} {\bibfnamefont {A.~J.}\ \bibnamefont
  {Ross}}, \bibinfo {author} {\bibfnamefont {L.}~\bibnamefont {Samushia}},
  \bibinfo {author} {\bibfnamefont {C.}~\bibnamefont {Howlett}}, \bibinfo
  {author} {\bibfnamefont {W.~J.}\ \bibnamefont {Percival}}, \bibinfo {author}
  {\bibfnamefont {A.}~\bibnamefont {Burden}}, \ and\ \bibinfo {author}
  {\bibfnamefont {M.}~\bibnamefont {Manera}},\ }\href {\doibase
  10.1093/mnras/stv154} {\bibfield  {journal} {\bibinfo  {journal} {Mon. Not.
  Roy. Astron. Soc.}\ }\textbf {\bibinfo {volume} {449}},\ \bibinfo {pages}
  {835} (\bibinfo {year} {2015})},\ \Eprint {http://arxiv.org/abs/1409.3242}
  {arXiv:1409.3242 [astro-ph.CO]} \BibitemShut {NoStop}%
\bibitem [{\citenamefont {Alam}\ \emph {et~al.}(2017)\citenamefont {Alam} \emph
  {et~al.}}]{Alam:2016hwk}%
  \BibitemOpen
  \bibfield  {author} {\bibinfo {author} {\bibfnamefont {S.}~\bibnamefont
  {Alam}} \emph {et~al.} (\bibinfo {collaboration} {BOSS}),\ }\href {\doibase
  10.1093/mnras/stx721} {\bibfield  {journal} {\bibinfo  {journal} {Mon. Not.
  Roy. Astron. Soc.}\ }\textbf {\bibinfo {volume} {470}},\ \bibinfo {pages}
  {2617} (\bibinfo {year} {2017})},\ \Eprint {http://arxiv.org/abs/1607.03155}
  {arXiv:1607.03155 [astro-ph.CO]} \BibitemShut {NoStop}%
\bibitem [{\citenamefont {Scolnic}\ \emph {et~al.}(2018)\citenamefont {Scolnic}
  \emph {et~al.}}]{Scolnic:2017caz}%
  \BibitemOpen
  \bibfield  {author} {\bibinfo {author} {\bibfnamefont {D.}~\bibnamefont
  {Scolnic}} \emph {et~al.},\ }\href {\doibase 10.3847/1538-4357/aab9bb}
  {\bibfield  {journal} {\bibinfo  {journal} {Astrophys. J.}\ }\textbf
  {\bibinfo {volume} {859}},\ \bibinfo {pages} {101} (\bibinfo {year}
  {2018})},\ \Eprint {http://arxiv.org/abs/1710.00845} {arXiv:1710.00845
  [astro-ph.CO]} \BibitemShut {NoStop}%
\bibitem [{\citenamefont {Lewis}\ \emph {et~al.}(2000)\citenamefont {Lewis},
  \citenamefont {Challinor},\ and\ \citenamefont {Lasenby}}]{Lewis:1999bs}%
  \BibitemOpen
  \bibfield  {author} {\bibinfo {author} {\bibfnamefont {A.}~\bibnamefont
  {Lewis}}, \bibinfo {author} {\bibfnamefont {A.}~\bibnamefont {Challinor}}, \
  and\ \bibinfo {author} {\bibfnamefont {A.}~\bibnamefont {Lasenby}},\ }\href
  {\doibase 10.1086/309179} {\bibfield  {journal} {\bibinfo  {journal}
  {Astrophys. J.}\ }\textbf {\bibinfo {volume} {538}},\ \bibinfo {pages} {473}
  (\bibinfo {year} {2000})},\ \Eprint {http://arxiv.org/abs/astro-ph/9911177}
  {arXiv:astro-ph/9911177} \BibitemShut {NoStop}%
\bibitem [{\citenamefont {Lewis}\ and\ \citenamefont
  {Bridle}(2002)}]{Lewis:2002ah}%
  \BibitemOpen
  \bibfield  {author} {\bibinfo {author} {\bibfnamefont {A.}~\bibnamefont
  {Lewis}}\ and\ \bibinfo {author} {\bibfnamefont {S.}~\bibnamefont {Bridle}},\
  }\href {\doibase 10.1103/PhysRevD.66.103511} {\bibfield  {journal} {\bibinfo
  {journal} {Phys. Rev. D}\ }\textbf {\bibinfo {volume} {66}},\ \bibinfo
  {pages} {103511} (\bibinfo {year} {2002})},\ \Eprint
  {http://arxiv.org/abs/astro-ph/0205436} {arXiv:astro-ph/0205436} \BibitemShut
  {NoStop}%
\bibitem [{\citenamefont {Lewis}(2013)}]{Lewis:2013hha}%
  \BibitemOpen
  \bibfield  {author} {\bibinfo {author} {\bibfnamefont {A.}~\bibnamefont
  {Lewis}},\ }\href {\doibase 10.1103/PhysRevD.87.103529} {\bibfield  {journal}
  {\bibinfo  {journal} {Phys. Rev. D}\ }\textbf {\bibinfo {volume} {87}},\
  \bibinfo {pages} {103529} (\bibinfo {year} {2013})},\ \Eprint
  {http://arxiv.org/abs/1304.4473} {arXiv:1304.4473 [astro-ph.CO]} \BibitemShut
  {NoStop}%
\bibitem [{\citenamefont {Gelman}\ and\ \citenamefont
  {Rubin}(1992)}]{Gelman:1992zz}%
  \BibitemOpen
  \bibfield  {author} {\bibinfo {author} {\bibfnamefont {A.}~\bibnamefont
  {Gelman}}\ and\ \bibinfo {author} {\bibfnamefont {D.~B.}\ \bibnamefont
  {Rubin}},\ }\href {\doibase 10.1214/ss/1177011136} {\bibfield  {journal}
  {\bibinfo  {journal} {Statist. Sci.}\ }\textbf {\bibinfo {volume} {7}},\
  \bibinfo {pages} {457} (\bibinfo {year} {1992})}\BibitemShut {NoStop}%
\bibitem [{\citenamefont {Heavens}\ \emph
  {et~al.}(2017{\natexlab{a}})\citenamefont {Heavens}, \citenamefont {Fantaye},
  \citenamefont {Sellentin}, \citenamefont {Eggers}, \citenamefont {Hosenie},
  \citenamefont {Kroon},\ and\ \citenamefont {Mootoovaloo}}]{Heavens:2017hkr}%
  \BibitemOpen
  \bibfield  {author} {\bibinfo {author} {\bibfnamefont {A.}~\bibnamefont
  {Heavens}}, \bibinfo {author} {\bibfnamefont {Y.}~\bibnamefont {Fantaye}},
  \bibinfo {author} {\bibfnamefont {E.}~\bibnamefont {Sellentin}}, \bibinfo
  {author} {\bibfnamefont {H.}~\bibnamefont {Eggers}}, \bibinfo {author}
  {\bibfnamefont {Z.}~\bibnamefont {Hosenie}}, \bibinfo {author} {\bibfnamefont
  {S.}~\bibnamefont {Kroon}}, \ and\ \bibinfo {author} {\bibfnamefont
  {A.}~\bibnamefont {Mootoovaloo}},\ }\href {\doibase
  10.1103/PhysRevLett.119.101301} {\bibfield  {journal} {\bibinfo  {journal}
  {Phys. Rev. Lett.}\ }\textbf {\bibinfo {volume} {119}},\ \bibinfo {pages}
  {101301} (\bibinfo {year} {2017}{\natexlab{a}})},\ \Eprint
  {http://arxiv.org/abs/1704.03467} {arXiv:1704.03467 [astro-ph.CO]}
  \BibitemShut {NoStop}%
\bibitem [{\citenamefont {Heavens}\ \emph
  {et~al.}(2017{\natexlab{b}})\citenamefont {Heavens}, \citenamefont {Fantaye},
  \citenamefont {Mootoovaloo}, \citenamefont {Eggers}, \citenamefont {Hosenie},
  \citenamefont {Kroon},\ and\ \citenamefont {Sellentin}}]{Heavens:2017afc}%
  \BibitemOpen
  \bibfield  {author} {\bibinfo {author} {\bibfnamefont {A.}~\bibnamefont
  {Heavens}}, \bibinfo {author} {\bibfnamefont {Y.}~\bibnamefont {Fantaye}},
  \bibinfo {author} {\bibfnamefont {A.}~\bibnamefont {Mootoovaloo}}, \bibinfo
  {author} {\bibfnamefont {H.}~\bibnamefont {Eggers}}, \bibinfo {author}
  {\bibfnamefont {Z.}~\bibnamefont {Hosenie}}, \bibinfo {author} {\bibfnamefont
  {S.}~\bibnamefont {Kroon}}, \ and\ \bibinfo {author} {\bibfnamefont
  {E.}~\bibnamefont {Sellentin}},\ }\href@noop {} {\  (\bibinfo {year}
  {2017}{\natexlab{b}})},\ \Eprint {http://arxiv.org/abs/1704.03472}
  {arXiv:1704.03472 [stat.CO]} \BibitemShut {NoStop}%
\bibitem [{\citenamefont {Pan}\ \emph {et~al.}(2018)\citenamefont {Pan},
  \citenamefont {Saridakis},\ and\ \citenamefont {Yang}}]{Pan_2018}%
  \BibitemOpen
  \bibfield  {author} {\bibinfo {author} {\bibfnamefont {S.}~\bibnamefont
  {Pan}}, \bibinfo {author} {\bibfnamefont {E.~N.}\ \bibnamefont {Saridakis}},
  \ and\ \bibinfo {author} {\bibfnamefont {W.}~\bibnamefont {Yang}},\ }\href
  {\doibase 10.1103/physrevd.98.063510} {\bibfield  {journal} {\bibinfo
  {journal} {Physical Review D}\ }\textbf {\bibinfo {volume} {98}} (\bibinfo
  {year} {2018}),\ 10.1103/physrevd.98.063510}\BibitemShut {NoStop}%
\bibitem [{\citenamefont {Yang}\ \emph
  {et~al.}(2019{\natexlab{b}})\citenamefont {Yang}, \citenamefont {Shahalam},
  \citenamefont {Pal}, \citenamefont {Pan},\ and\ \citenamefont
  {Wang}}]{Yang_2019}%
  \BibitemOpen
  \bibfield  {author} {\bibinfo {author} {\bibfnamefont {W.}~\bibnamefont
  {Yang}}, \bibinfo {author} {\bibfnamefont {M.}~\bibnamefont {Shahalam}},
  \bibinfo {author} {\bibfnamefont {B.}~\bibnamefont {Pal}}, \bibinfo {author}
  {\bibfnamefont {S.}~\bibnamefont {Pan}}, \ and\ \bibinfo {author}
  {\bibfnamefont {A.}~\bibnamefont {Wang}},\ }\href {\doibase
  10.1103/physrevd.100.023522} {\bibfield  {journal} {\bibinfo  {journal}
  {Physical Review D}\ }\textbf {\bibinfo {volume} {100}} (\bibinfo {year}
  {2019}{\natexlab{b}}),\ 10.1103/physrevd.100.023522}\BibitemShut {NoStop}%
\bibitem [{\citenamefont {Kass}\ and\ \citenamefont
  {Raftery}(1995)}]{Kass:1995loi}%
  \BibitemOpen
  \bibfield  {author} {\bibinfo {author} {\bibfnamefont {R.~E.}\ \bibnamefont
  {Kass}}\ and\ \bibinfo {author} {\bibfnamefont {A.~E.}\ \bibnamefont
  {Raftery}},\ }\href {\doibase 10.1080/01621459.1995.10476572} {\bibfield
  {journal} {\bibinfo  {journal} {J. Am. Statist. Assoc.}\ }\textbf {\bibinfo
  {volume} {90}},\ \bibinfo {pages} {773} (\bibinfo {year} {1995})}\BibitemShut
  {NoStop}%
\end{thebibliography}%
\end{document}